\documentclass[aps,prb,twocolumn,10pt,floatfix,flushbottom]{revtex4-2}
\usepackage{adjustbox}
\usepackage[utf8]{inputenc}
\usepackage[colorlinks=true,linkcolor=magenta,urlcolor=blue,citecolor=blue]{hyperref}
\usepackage{mathtools,amsmath,amssymb,amsfonts,graphicx,tabularx,dcolumn,bm,xcolor,times,float,algorithm,algorithmic,tikz,pgf}
\allowdisplaybreaks
\usepackage{dsfont}
\usepackage{capt-of}
\usepackage{multirow}
\usepackage{CJKutf8}
\usepackage{subfig} 
\usepackage[justification=justifying]{ragged2e}

\makeatletter

\newcommand{\skipthis}[1]{}
\newcommand{\II}{\mathbb{I}\,}
\newcommand{\BB}{\mathbb{B}}

\newcommand{\PP}{\mathbb{P}}
\newcommand{\QQ}{\mathbb{Q}}
\newcommand{\WW}{\mathbb{W}}
\renewcommand{\Re}{\operatorname{Re}}
\newcommand{\ZZ}{\mathbb{Z}}
\makeatother

\DeclareMathOperator{\tr}{tr}
\usetikzlibrary{arrows.meta, decorations.markings,calc}
\usetikzlibrary{positioning}
\tikzstyle{vecArrow} = [thick, decoration={markings,mark=at position
   1 with {\arrow[semithick]{open triangle 60}}},
   double distance=1.4pt, shorten >= 5.5pt,
   preaction = {decorate},
   postaction = {draw,line width=1.4pt, white,shorten >= 4.5pt}]
\tikzstyle{innerWhite} = [semithick, white,line width=1.4pt, shorten >= 4.5pt]
\tikzset{middlearrow/.style={
        decoration={markings,
            mark= at position 0.5 with {\arrow{#1}} ,
        },
        postaction={decorate}
    }
}

\begin{document}

\begin{CJK*}{UTF8}{gbsn} 
\title{Operator Spreading in Random Unitary Circuits with Orthogonal/Symplectic-invariant Gate Distributions}
\author{Zhiyang Tan (谭志阳)}
\author{Piet W.\ Brouwer}%
\affiliation{Dahlem Center for Complex Quantum Systems, Physics Department, Halle-Berlin-Regensburg Cluster of Excellence CCE, and Freie Universit\"at Berlin, Arnimallee 14, 14195 Berlin, Germany}
\date{\today}

\begin{abstract}
We investigate operator spreading in random quantum circuits with gates drawn from orthogonal-invariant or symplectic-invariant ensembles, revealing several key distinctions from the well-studied unitary-invariant case. We find that the ensemble-averaged Pauli-string weights relax to a ternary-valued structure, instead of the binary structure of unitary-invariant circuits. For orthogonal- or symplectic-invariant circuits, the domain wall separating trivial and scrambled regions has a finite width even for Haar-random gates, whereas domain walls are sharp for Haar-distributed random unitary circuits. We further find a fundamental dichotomy between random circuits with two-qubit gates from the two disconnected components of the orthogonal group: While the butterfly velocity for the special orthogonal ensemble lies between zero and the Haar value, the negative-determinant sector exhibits a non-zero lower bound for any gate distribution. Moreover, for qudit size $q=2$, the butterfly velocity can exceed that of the Haar-random ensemble.
\end{abstract}

\maketitle
\end{CJK*}
\section{Introduction}

\begin{figure}[tbp]
\vspace{-0.6cm}
\centering
\includegraphics[scale=0.8,trim={0cm 0cm 0cm 0cm},clip]{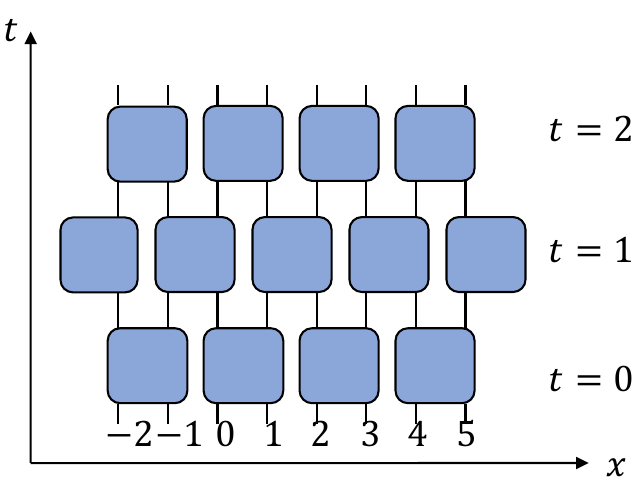}
\caption{\justifying \small
Schematic of a random unitary circuit with a brickwork structure. The random circuit is composed of two-qudit gates (shown in blue) uncorrelated in space and time.}
\label{Fig:circuit}
\end{figure}

Random unitary circuits are minimal models for studying fundamental aspects of many-body quantum dynamics, including information spreading, quantum chaos, and thermalization \cite{arnaud2007distribution,arnaud2008efficiency,brown2013scrambling,nahum2017quantum,chan2018solution,nahum_operator_2018, von_keyserlingk_operator_2018, rakovszky_diffusive_2018, khemani_operator_2018, Fisher2023random, skinner2023lecture}. These circuits consist of a sequence of gate operations acting on pairs of $q$-dimensional quantum degrees of freedom (``qudits''), whereby the gate operations are drawn from a statistical distribution \cite{nahum_operator_2018}. Typically, the gate operations follow a brickwork structure, in which qudits are acted upon pairwise by two-qudit gates, with a pattern that alternates in space and time, see Fig. \ref{Fig:circuit}.

An initially local operator $O$ acquires nonlocal correlations under the circuit evolution 
\cite{xu_locality_2019, styliaris_information_2021,xu_scrambling_2022}. 
{\em Operator spreading} is the increase in the spatial support of the set of Pauli-string basis operators required to represent $O(t)$ with time $t$ \cite{xu_scrambling_2022,mi_information_2021}.
The decay of the four-point out-of-time-order correlator (OTOC) provides a direct probe of this spreading \cite{nahum_operator_2018,von_keyserlingk_operator_2018, rakovszky_diffusive_2018, khemani_operator_2018}, and recent work has generalized this to higher-point OTOCs as measures of operator ergodicity in the full operator space \cite{vardhan2025freemutualinformationhigherpoint,abanin2025observation}. 

For a one-dimensional random unitary circuit, operator spreading can
be described in terms of a front propagating according to a
drift-diffusion model \cite{nahum_operator_2018,
von_keyserlingk_operator_2018, rakovszky_diffusive_2018,
khemani_operator_2018}. The drift velocity, which is known as the
``butterfly velocity'', and the diffusion constant were first
calculated for the circuit with maximal randomness, which has its
two-qudit gate operators drawn from a Haar distribution
\cite{nahum_operator_2018, von_keyserlingk_operator_2018,
rakovszky_diffusive_2018, khemani_operator_2018}. For
Haar-distributed gates, the randomness of the local gate operation is
maximal. This raises the question to what extent the properties of
operator spreading found in this case are universal: which of them
persist for a broader class of gate ensembles, and which new features
emerge once the local randomness is reduced? Ensembles whose
probability distribution is invariant under a group of basis
transformations provide a natural setting in which this question can
be addressed in a controlled manner.

In a recent publication, we
addressed this question for random unitary circuits with an arbitrary
basis-independent probability distribution for the two-qudit gates
\cite{tan2025operatorspreadingrandomunitary}, {\em i.e.}, probability
distributions in which the two-qudit gate operator ${\cal U}$
satisfies the invariance property
\begin{equation}
  P({\cal U}) = P({\cal V} {\cal U} {\cal V}^{\dagger}),
\label{eq:unitary_invariance}
\end{equation}
where ${\cal V}$ is an arbitrary unitary matrix. 
 The Haar-random circuit is a special case of this larger class of unitary circuits, but the invariance property (\ref{eq:unitary_invariance}) also admits, {\em e.g.}, circuits that continuously interpolate between the trivial circuit and the Haar-random one, such as continuous Brownian circuits \cite{xu_locality_2019,xu_scrambling_2022}.
Apart from a
different drift velocity and diffusion constant, we found two
qualitative differences between a fully random Haar-distributed
circuit and a circuit with the more general unitary invariance
property of Eq.~\eqref{eq:unitary_invariance}: (i) whereas
Haar-distributed unitary circuits have a sharp domain wall between
the trivial and maximally random regions, general unitary-invariant
circuits have a finite domain-wall width; (ii) for the Haar-random
circuit, the probability of finding a given generalized Pauli
operator $\sigma_{p_x}$ in the Pauli-string expansion of ${\cal
O}(t)$ depends only on whether $\sigma_{p_x}$ is the identity,
whereas for a general unitary-invariant circuit, such a ``binary''
distribution emerges only after a finite ``thermalization time''.

Antiunitary symmetries of the underlying Hamiltonian, such as
time-reversal or particle-hole symmetry, endow the two-qudit gates
with additional structure. In their presence, the natural basis
transformations are no longer arbitrary unitary matrices, but
orthogonal or symplectic ones. The orthogonal- and symplectic-Haar
ensembles are therefore natural symmetry-constrained counterparts of
the Haar-unitary circuit \cite{haferkamp2021improved,shaya2025}.  However, because the Haar measure combines
the relevant basis invariance with maximal randomness, it does not
allow one to distinguish which features of operator spreading
originate from the symmetry constraint itself and which rely on the
full Haar distribution. This motivates the study of the full class
of orthogonal- and symplectic-invariant gate ensembles.

In this work, we consider operator spreading in random circuits in
which the two-qudit gate operators satisfy an invariance property of
the form \eqref{eq:unitary_invariance}, but with ${\cal V}$ an
arbitrary orthogonal or symplectic matrix. Beyond this invariance
condition, we impose no restriction on the probability distribution
of the gates. This provides a setting in which the effects of the symmetry
constraint can be disentangled from those of the gate randomness:
properties shared by all gate distributions with a given invariance
originate from the symmetry constraint alone, whereas properties
that vary within the invariant class reflect the randomness of the
gate distribution.
{\color{black} Moreover, as in the unitary case, the invariance property (\ref{eq:unitary_invariance}) allows us to study continuous interpolations between trivial and Haar-random circuits as well as continuous Brownian circuits.}

{\color{black} An invariance property with respect to orthogonal or symplectic basis transformations} is natural if the two-qudit gate operator
${\cal U}$ itself satisfies an additional symmetry, making it an
orthogonal or symplectic matrix itself. However, there also exist
random matrix ensembles in which ${\cal U}$ is unitary, but its
probability distribution satisfies
Eq.~\eqref{eq:unitary_invariance} with ${\cal V}$ restricted to the
orthogonal or symplectic group \cite{forrester2010,mehta2004random}.
Random circuits with orthogonal, unitary-symmetric, or symplectic
two-qudit gates have recently been considered in
Refs.~\cite{arnaud2007distribution,hashagen2018,
haferkamp2021improved, west2024symplectic,schuster2025random,
grevink2025glue,sauliere2025universality,magni2025anticoncentration,
West_2025,shaya2025,khanna2026random,hunterjones2026lxeb}.
Symplectic-invariant circuits require that the qudit array itself has
an even-odd structure, as we discuss in further detail below.

As we show in this article, the drift velocity and the diffusion
constant for random circuits with the orthogonal or symplectic
invariance property differ quantitatively from their counterparts
with unitary invariance. In addition, we also find qualitative
differences. For orthogonal- or symplectic-invariant circuits, we
find that the domain-wall width remains finite even for
Haar-distributed random circuits. We also find that the probability
distribution of generalized Pauli operators approaches a ``ternary''
form, in which it depends not only on whether or not the generalized
Pauli operator equals the identity operator, but also on whether it
is symmetric or antisymmetric. Because the invariant gate
distribution need not be Haar, the butterfly velocity can attain
values inaccessible to the corresponding Haar ensemble: in the
$\mathrm{SO}^-(4)$ sector, its maximum exceeds the Haar-orthogonal
value. Moreover, in this sector the minimum butterfly velocity is
strictly positive for any gate distribution.

The remainder of this article is organized as follows: In Sec.~\ref{sec:2}, we briefly review the formulation of random unitary circuits with a brickwork structure, closely following Ref.~\cite{nahum_operator_2018}, and we review the expansion of an operator $O$ in terms of Pauli strings. In Sec.~\ref{sec:3}, we elaborate on the connection between the gate ensemble's invariance property, Eq.~\eqref{eq:unitary_invariance}, and the underlying symmetries of the Hamiltonian. We point out that a symplectic structure occurs naturally only if the qudits in the circuit have an even-odd structure, such that there is a natural involution squaring to $-1$ for every second qudit. We derive the governing equations for the time dependence of the Pauli-string density matrix in Sec.\ \ref{sec:4}.
%
The core of our analysis is presented in Sec.~\ref{sec:5}, where we specialize to the case of qudit dimension $q=2^m$. By mapping the quantum evolution to a classical stochastic growth model, we derive the butterfly velocity $v_{\rm B}$ and diffusion constant $\mathcal{D}$ for circuits with orthogonal and symplectic-invariant gate distributions. For circuits composed of gates that are themselves orthogonal or symplectic, we show that the Pauli-string weights relax to a ternary form after a finite thermalization time $\tau_{\rm t}$. In Sec.~\ref{sec:qgeneral}, we generalize these results from $q=2^m$ to an arbitrary qudit dimension $q$. We illustrate our findings with three specific examples in Sec.~\ref{sec:7}. Finally, we conclude in Sec.~\ref{sec:8}. Additional details of our calculations are provided in the appendices.

\section{Operator spreading in random unitary circuits}
\label{sec:2}

The non-local correlations that an initially local operator $O$ acquires under the circuit evolution can be described by expressing the Heisenberg-picture operator $O(t)$ (with $O(0) = O$) in terms of a set of Pauli-string basis operators $O_p = \otimes_x \sigma_{p_{x}}$, which are products of generalized Pauli operators $\sigma_{p_x}$ associated with each qudit,
\begin{equation}
  O(t) = \sum_{p} \gamma_p(t) O_p, \label{eq:Ot}
\end{equation}
where the expansion coefficient $\gamma_p(t)$ denotes the amplitude of the basis operator $O_p$. The Pauli string index $p = [\ldots,p_{x-1},p_{x},p_{x+1},\ldots]$.
Its support is the set of sites $x$ for which $\sigma_{p_x}$ is different from the identity operator. We define the front position of a Pauli string as the position of its rightmost nonidentity operator; all sites to the right of the front position therefore carry the identity operator.
In App.\ \ref{app:1}, we review the Weyl-Heisenberg choice for the generalized Pauli matrices, as well as a construction of generalized Pauli matrices from tensor products of conventional $2 \times 2$ Pauli matrices for special values $q = 2^m$ of the qudit dimension. In either case, the generalized Pauli operators $\sigma_{p_x}$ form a complete, orthonormal basis of operators for the qudit at the site $x$, satisfying
\begin{equation}
  \frac{1}{q} \mbox{tr}\, \sigma_{p_x} \sigma_{q_x}^{\dagger} = \delta_{p_x,q_x}.
  \label{eq:orthonormality}
\end{equation}
Correspondingly, the Pauli string operators $O_p$ satisfy the orthonormality relation
\begin{equation}
  \frac{1}{q^L} \mbox{tr}\, O_{p} O_{q}^{\dagger} = \delta_{p,q} ,
\end{equation}
where $L$ is the number of sites of the circuit.
The Pauli-string amplitude $\gamma_p(t)$ satisfies the normalization condition
\begin{equation}
  \sum_{p} |\gamma_{p}(t)|^2 = 1
  \label{eq:gamma_normalization}
\end{equation}
for all $t$.

The time evolution of an operator $O$ in the Heisenberg picture is given by the relation
\begin{equation}
  O(t) = U^\dagger(t,t') O(t') U(t,t').
  \label{eq:UO}
\end{equation}
Following Refs.\ \cite{nahum_operator_2018, von_keyserlingk_operator_2018, rakovszky_diffusive_2018}, the evolution matrix $U$ is taken to be the
product of two-qudit gates in a brickwork pattern,
\begin{equation}
\label{eq:gates}
U(t,t-1) =
\begin{cases}  
\bigotimes_{\text{even } x} \mathcal{U}_{x,x+1}, & \text{if } t \text{ is even}, \\  
\bigotimes_{\text{even } x} \mathcal{U}_{x-1,x}, & \text{if } t \text{ is odd},
\end{cases}
\end{equation}
Here, $\mathcal{U}_{x,y}$ is a $q^2 \times q^2$ unitary matrix acting on the qudits at sites $x$ and $y$. 
We take the two-qudit gate operators $U_{x,y}$ from a probability distribution, which satisfies the invariance condition
\begin{equation}
P({\cal U}_{x,y}) = P({\cal V}_{x,y} {\cal U}_{x,y} {\cal V}_{x,y}^{\dagger}),
  \label{eq:unitary_invariance_xy}
\end{equation}
with ${\cal V}_{x,y}$ an arbitrary unitary matrix with the symmetry constraint
\begin{equation}
  {\cal V}_{x,y}^{\dagger} = \Omega_{x,y} {\cal V}_{x,y}^{\rm T} \Omega_{x,y}^{\dagger},
  \label{eq:Vorthosympl}
\end{equation}
where $\Omega_{x,y}$ is a symmetric or an antisymmetric unitary matrix. Since one may always choose the basis such that $\Omega_{x,y} = \openone_{q^2}$ if $\Omega_{x,y}$ is symmetric and $\Omega_{x,y} = \openone_{q^2/2} \otimes \sigma_2$ if $\Omega_{x,y}$ is antisymmetric, we refer to matrices ${\cal V}_{x,y}$ satisfying the constraint (\ref{eq:Vorthosympl}) as orthogonal and symplectic, respectively.
The gate operator ${\cal U}_{x,y}$ may itself also be subject to additional symmetries, {\em e.g.}, ${\cal U}_{x,y}$ may be orthogonal or symplectic, too, but this is not a necessary condition in our general analysis. We discuss random circuits for which the two-qudit gate operators ${\cal U}_{x,y}$ have an additional symmetry constraint in the next Section.

The time evolution of the Pauli-string coefficients $\gamma_p$ follows from that of the operator $O$, see Eq.\ (\ref{eq:UO}). For a single time step, we have
\begin{align}
  \gamma_{p}(t)  =&\,
  \sum_{a} W_{ap}(t,t-1) \gamma_a(t-1)\,
  \label{eq:gammat}
\end{align}
where the summation variable $a$ labels Pauli strings and
\begin{align}
  W_{ap}(t,t-1) = \frac{1}{q^L} 
  \mbox{tr}\, U^\dagger(t,t-1) O_{a} U(t,t-1) O_p^{\dagger}.
  \nonumber
\end{align}
For an evolution matrix of the form (\ref{eq:gates}), this evaluates to
\begin{equation}
  \label{eq:Wap}
  W_{ap} =
  \left\{ \begin{array}{ll}
    \prod_{\textrm{$x$ even}}
    {\cal W}_{ap;x,x+1}
    & \mbox{$t$ even}, \\
    \prod_{\textrm{$x$ even}}
   {\cal W}_{ap;x-1,x} & \mbox{$t$ odd},
  \end{array} \right.
\end{equation}
Here, the factors ${\cal W}_{ap;x,y}$ do not depend on the full Pauli strings $a$ and $p$, but only on the labels $a_x$, $a_y$, $p_x$, and $p_y$ that belong to the sites $x$ and $y$.
Its explicit expression is given in Appendix \ref{app:3}. 

\section{orthogonal and symplectic-invariant ensembles}
\label{sec:3}
The invariance condition (\ref{eq:unitary_invariance}), with ${\cal V}_{x,y}$ orthogonal or symplectic, can be seen as a purely mathematical condition on the pair-qudit gate operation. We here consider symmetry conditions for the evolution of single qudits and of qudit pairs, for which such a symmetry restriction on the matrix ${\cal V}_{x,y}$
appears naturally. We consider qudits with $q$ degrees of freedom and gate operations that arise from the operation of a Hamiltonian for a finite period of time, whereby the randomness appears because these Hamiltonians differ from time step to time step.

The restriction to invariance with orthogonal or symplectic basis transformations appears naturally if there exists an antiunitary involution 
\begin{equation}
  \psi_x \to Z_x \psi_x^*
  \label{eq:involution}
\end{equation}
on the $q$-dimensional single-qudit spinors, where $Z_x$ is a $q \times q$ unitary matrix with $Z_x Z_x^* = \pm 1$, and the Hamiltonian that is applied to a single qudit or a pair of qudits is symmetric or antisymmetric under the involution. Specifically, we consider involutions corresponding to time-reversal symmetry and particle-hole symmetry. For time-reversal symmetry, the qudit Hamiltonian is symmetric under the involution, whereby $Z_x$ is symmetric if the qudit has integer spin and antisymmetric if the qudit spin is half-integer. In the language of the tenfold-way classification \cite{altland1997nonstandard,schnyder2008classification,kitaev2009periodic}, these two cases correspond to the Cartan classes AI and AII, respectively. For particle-hole symmetry, the qudit Hamiltonian is antisymmetric under the involution. The cases of symmetric and antisymmetric $Z_x$ then correspond to Cartan classes D and C, respectively. 
For integer spin, we may choose the qudit basis such that 
\begin{equation}
  Z_x = \openone.
  \label{eq:Zx1}
\end{equation}
For half-integer spin, $q$ must be even, and we can choose the qudit basis such that
\begin{equation}
  Z_x = \sigma_2 \otimes \openone_{q/2}.
  \label{eq:Zx2}
\end{equation}

For a single-qudit gate operation described by the Hamiltonian $H_x$, one has the constraint
\begin{equation}
  H_x = Z_x H_x^* Z_x^{\dagger}\ \ \mbox{or}\ \ H_x = -Z_x H_x^* Z_x^{\dagger}
  \label{eq:Hconstraint}
\end{equation}
for the case of time-reversal symmetry or particle-hole symmetry, respectively. Correspondingly, for the special choice of basis corresponding to Eqs.\ (\ref{eq:Zx1}) and (\ref{eq:Zx2}), the single-qudit gate operations are unitary-symmetric, unitary-antisymmetric, orthogonal, and symplectic for Cartan classes AI, AII, D, and C, respectively. Whereas these are well-defined symmetry properties for the evolution matrix of a single time step, the properties of being unitary symmetric or unitary antisymmetric are not preserved if multiple time steps are taken and one ends up with evolution matrices that are simply unitary, without further symmetry constraints.

The generalized Pauli matrices $\sigma_{p_x}$ form a basis for operators on the qudit $x$. If the qudit dimension $q = 2^m$, we can take tensor products of the standard $2 \times 2$ Pauli matrices for the generalized Pauli matrices. In this case, the generalized Pauli matrices $\sigma_{p_x}$ can be chosen such that they have a well-defined parity $s_{p_x} \in \{1,-1\}$ under the involution (\ref{eq:involution}),
\begin{equation}
  \sigma_{p_x} = s_{p_x} Z_x \sigma_{p_x}^{\rm T} Z_x^{\dagger}.
  \label{eq:sigmaparity}
\end{equation}
For general $q$, a generalization of Eq.\ (\ref{eq:sigmaparity}) exists, which will be discussed in Sec.\ \ref{sec:qgeneral}.

The invariance property (\ref{eq:unitary_invariance}) refers to two-qudit gates. The two-qudit Hamiltonian $H_{x,x+1}$ satisfies a constraint analogous to Eq.\ (\ref{eq:Hconstraint}), but with 
\begin{equation}
  \label{eq:OmegaxxDef}
\Omega_{x,x+1} = Z_x \otimes Z_{x+1}
\end{equation}
instead of $Z_x$. The corresponding two-qudit gate operator ${\cal U}_{x,x+1}$ then satisfies the symmetry relation
\begin{equation}
  {\cal U}_{x,x+1} = \Omega_{x,x+1} {\cal U}_{x,x+1}^{\rm T} \Omega_{x,x+1}^{\dagger}
\end{equation}
for time-reversal symmetry and
\begin{equation}
  {\cal U}_{x,x+1} = \Omega_{x,x+1} {\cal U}_{x,x+1}^{*} \Omega_{x,x+1}^{\dagger}
\end{equation}
for particle-hole symmetry. Since $\Omega_{x,x+1}$ is symmetric if $Z_{x}$ and $Z_{x+1}$ are both symmetric as well as if they are both antisymmetric, the corresponding evolution operator ${\cal U}_{x,x+1}$ is unitary-symmetric or orthogonal (in the sense described below Eq.\ (\ref{eq:Vorthosympl})), 
regardless of whether the spin of the individual qudits is half-integer or integer. This is to be expected, since the spin of two neighboring qudits is always an integer, regardless of whether the individual spins are of half-integer or integer type. Evolution operators of unitary-self-dual or symplectic type appear if the qudits in the circuit are alternatingly of half-integer and integer spin, {\em e.g.}, if the $Z_x$ are symmetric for even $x$ and antisymmetric for odd $x$, so that $\Omega_{x,x+1} = Z_x \otimes Z_{x+1}$ is an antisymmetric matrix.

If the two-qudit gate evolution matrix ${\cal U}_{x,x+1}$ is orthogonal or symplectic, the transition factors ${\cal W}_{ap;x,x+1}$ of Eq.\ (\ref{eq:Wap}) are parity preserving for $q = 2^{m}$, {\em i.e.},
\begin{equation}
  {\cal W}_{ap;x,x+1} = 0\ \
  \mbox{if} s_{p_x} s_{p_{x+1}} s_{a_x} s_{a_{x+1}} = -1.
  \label{eq:Wparity}
\end{equation}
A generalization to arbitrary $q$ will be discussed in Sec.\ \ref{sec:qgeneral}.

\section{Time evolution of Pauli-string density matrix}
\label{sec:4}

The Pauli-string coefficient $\gamma_p(t)$ contains the full
information about the evolution of the operator $O$ under the unitary
dynamics. We introduce the second moment of the Pauli-string
amplitudes,
\begin{equation}
  \rho_{pq}(t)
  =
  \left\langle \gamma_p(t)\gamma_q^*(t) \right\rangle,
\end{equation}
where the average is taken over the ensemble of two-qudit gates.
The matrix $\rho(t)$ is Hermitian and positive semidefinite, and the
normalization condition in Eq.~\eqref{eq:gamma_normalization} implies
$\operatorname{tr}\rho(t)=1$. We therefore refer to it as the
\emph{Pauli-string density matrix}. Its diagonal elements
$\rho_{pp}(t)$ may be interpreted as a classical probability
distribution over the set of Pauli strings $p$
\cite{nahum_operator_2018}.

With the help of Eq.~(\ref{eq:gammat}), the Pauli-string density
matrix at time $t$ can be expressed in terms of that at time $t-1$
and the covariance of the transition amplitudes,
\begin{equation}
  \rho_{pq}(t)
  =
  \sum_{a,b}
  \left\langle W_{ap}W_{bq}^{*}\right\rangle
  \rho_{ab}(t-1).
  \label{eq:gammasqavg0}
\end{equation}
Using unitarity and the invariance of the identity operator under conjugation, one verifies that the evolution equation (\ref{eq:gammasqavg0}) admits the maximally random steady-state solution
\begin{equation}
  \label{eq:rhomax}
  \rho^{\infty}_{pq}(t) = q^{-2 L} \prod_x \delta_{p_x,q_x}
\end{equation}
as well as the trivial solution
\begin{equation}
  \label{eq:rhotrivial}
  \rho^0_{pq}(t) =
  \prod_{x} \delta_{p_x,0} \delta_{q_x,0}.
\end{equation}
The butterfly velocity $v_{\rm B}$ corresponds to the speed at which a domain wall between these two solutions propagates through the qudit array. For definiteness, we here will consider a domain wall separating a maximally random region to the left (smaller $x$) from a trivial region to the right (larger $x$). Such a domain wall propagates to the right (in the positive $x$ direction).

For circuits with orthogonal or symplectic gates, the parity-conservation condition (\ref{eq:Wparity}) implies that the total parity of a Pauli string is preserved under time evolution. In principle, this allows for two maximally random steady-state solutions, each with well-defined global parity. A global parity constraint has no effect on how the steady-state solution looks locally. Hence, for the problem of operator spreading, it is sufficient to consider a domain wall between steady-state solutions that locally have the forms of Eqs.\ (\ref{eq:rhomax}) and (\ref{eq:rhotrivial}).

For a Haar-random unitary circuit, the expectation value
\begin{equation}
  \langle W_{ap} W_{bq}^* \rangle = 0\ \mbox{if $a \neq b$ or $p \neq q$}
  \label{eq:WW1}.
\end{equation}
Moreover, $\langle |W_{ap}|^2 \rangle$ only depends on whether the indices $a$ and $p$ are trivial or nontrivial. This means that the Pauli-string density matrix $\rho(t)$ acquires a
``binary'' form after the first time step: its only nonzero elements are the diagonal elements $\rho_{pp}(t)$, which depend only on whether $p_x$ is trivial or nontrivial at each site $x$.

For a generic unitary-invariant circuit, Eq.\ (\ref{eq:WW1}) no longer holds, but there is a weaker constraint \cite{tan2025operatorspreadingrandomunitary},
\begin{equation}
  \langle W_{ap} W_{bq}^* \rangle = 0\ \mbox{if $a \neq b$ and $p=q$ or if
  $p \neq q$ and $a = b$} \label{eq:WW}.
\end{equation}
The constraint (\ref{eq:WW}) is sufficient to show that the evolution of the diagonal elements $\rho_{pp}$ can be solved separately from that of the off-diagonal elements, {\em i.e.}, 
to find diagonal elements $\rho_{pp}(t)$ it is sufficient to know the diagonal elements $\rho_{aa}(t-1)$ at the preceding times,
\begin{align}
  \rho_{pp}(t)
  =&\,
  \sum_{a}\
  \rho_{aa}(t-1)
  \langle |W_{ap}|^2 \rangle.
  \label{eq:gammasqavg}
\end{align}
In Ref.\ \cite{tan2025operatorspreadingrandomunitary}, we showed that the diagonal elements $\rho_{pp}$ exponentially relax to the binary form after a finite relaxation time.

In an orthogonal-invariant or symplectic-invariant circuit, Eq.\ (\ref{eq:WW}) holds only for a basis of generalized Pauli matrices that have well-defined parity $s_{p_x}$, see Eq.\ (\ref{eq:sigmaparity}). Such a basis exists for qudit size $q = 2^m$, but not for other qudit sizes.
We therefore first consider the case $q = 2^m$ and postpone the discussion of general $q$ to Sec.\ \ref{sec:qgeneral}.

\section{The case $q = 2^m$}
\label{sec:5}
\subsection{Mapping to a stochastic growth model}
\label{sec:5a}

The transition probabilities $\langle |W_{ap}|^2 \rangle$ in the stochastic growth model (\ref{eq:gammasqavg}) factorize in pair contributions from the two-qudit gates,
\begin{align}
  \label{eq:Wevenodd}
  \langle |W_{ap}|^2 \rangle
  =&\,
  \prod_{x\, {\rm even}}
  \left\{ \begin{array}{ll}
    \langle |{\cal W}_{ap;x,x+1}|^2 \rangle & \mbox{for $t$ even}, \\
    \langle |{\cal W}_{ap;x-1,x}|^2 \rangle & \mbox{for $t$ odd}.
  \end{array} \right.
\end{align}
In Ref. \cite{tan2025operatorspreadingrandomunitary} we showed that for a unitary-invariant circuit, the pair transition probability $\langle |{\cal W}_{ap;x,x+1}|^2 \rangle$ is of the form
\begin{align}
  \langle |{\cal W}_{ap;x,y}|^2 \rangle =&\, \delta_{a,0} \delta_{p,0}
  + (1 - \delta_{a,0})(1 - \delta_{p,0}) 
  \nonumber \\ &\, \mbox{} \times
  \left[ \mathcal{P}_1 + \mathcal{P}_2 \varphi(a,p) + \mathcal{P}_3 \delta_{a,p} \right],
  \label{eq:Wgeneral}
\end{align}
where $\mathcal{P}_1$, $\mathcal{P}_2$, and $\mathcal{P}_3$ are coefficients that depend on moments of the pair-qudit evolution matrix ${\cal U}_{x,x+1}$ involving up to four factors of ${\cal U}_{x,x+1}$ or ${\cal U}^{\dagger}_{x,x+1}$ and the function $\varphi(a,p)$ is defined by
\begin{equation}
  \label{eq:varphi}
  \varphi(a,p) = \left\{ \begin{array}{rl} 1 & \mbox{if $\sigma_{p_x} \otimes \sigma_{p_{x+1}}$ and $\sigma_{a_x} \otimes \sigma_{a_{x+1}}$ commute}, \\
    - 1 & \mbox{if $\sigma_{p_x} \otimes \sigma_{p_{x+1}}$ and $\sigma_{a_x} \otimes \sigma_{a_{x+1}}$ anti-}\\ & \ \ \ \ \mbox{commute}.
  \end{array} \right.
\end{equation}
For an orthogonal-invariant or symplectic-invariant circuit, $\langle |{\cal W}_{ap;x,y}|^2 \rangle$ is also of the form of Eq.\ (\ref{eq:Wgeneral}), but the coefficients $\mathcal{P}_1$, $\mathcal{P}_2$, and $\mathcal{P}_3$ additionally depend on the products $s_{a_x} s_{a_{y}}$ and $s_{p_x} s_{p_{y}}$ of the parities of the generalized Pauli matrices.
Explicit expressions for the coefficients $\mathcal{P}_1$, $\mathcal{P}_2$, and $\mathcal{P}_3$ are given in App.\ \ref{app:3}.

For the Haar orthogonal and symplectic ensembles, the averages can be
evaluated directly, without invoking the machinery required for the more general orthogonal-invariant and symplectic-invariant ensembles \cite{hunter-jones_operator_2018}. In this case, the coefficient $\mathcal P_1$ reads
\begin{align}
\mathcal P_1
={}&\,
\frac{2}{q^2\mp1}
\left\{
\begin{array}{ll}
\displaystyle \frac{1}{q^2\pm2},
& \text{if }
s_{p_x}s_{p_{x+1}}
=
s_{a_x}s_{a_{x+1}}
=
1,
\\[6pt]
\displaystyle \frac{1}{q^2},
& \text{if }
s_{p_x}s_{p_{x+1}}
=
s_{a_x}s_{a_{x+1}}
=
-1,
\\[6pt]
0,
& \text{if }
s_{p_x}s_{p_{x+1}}
\neq
s_{a_x}s_{a_{x+1}}.
\end{array}
\right.
\label{eq:P1Orthogonal}
\end{align}
(Recall that $s_{p_x}$, $s_{p_{x+1}}$, $s_{a_x}$, and $s_{a_{x+1}}$ are the parities of the generalized Pauli matrices.)
In Eq.\ (\ref{eq:P1Orthogonal}), the upper sign applies to the orthogonal case and the
lower sign to the symplectic case. As in the Haar-random unitary case,
one has
\begin{equation}
\mathcal P_2=\mathcal P_3=0.
\label{eq:P23Orthogonal}
\end{equation}

For the COE and CSE ensembles, standard Weingarten calculus gives
\begin{align}
\label{eq:PCOECSE}
\mathcal P_1
={}&\,
\frac{q^4\pm4q^2+2}
{q^4(q^2\pm1)(q^2\pm3)},
\nonumber\\
\mathcal P_2
={}&\,
-\frac{2}
{q^4(q^2\pm1)(q^2\pm3)},
\nonumber\\
\mathcal P_3
={}&\,
\frac{2(q^2\pm2)}
{q^2(q^2\pm1)(q^2\pm3)}.
\end{align}
These coefficients are independent of the parities of the generalized
Pauli matrices, and hence do not depend on the signs of
$s_{p_x}s_{p_{x+1}}$ and $s_{a_x}s_{a_{x+1}}$. In these expressions,
the upper sign corresponds to the COE and the lower sign to the CSE.

We give a simple derivation of Eq.\ (\ref{eq:P1Orthogonal}) that does not use the full machinery for orthogonal- and symplectic-invariant circuits in App.\ \ref{app:3}. For the corresponding derivation of Eq.\ (\ref{eq:PCOECSE}) we refer to Ref.\ \cite{hunter-jones_operator_2018}.

\subsection{Projected ternary strings}
\label{subsec:3D}

Since the stochastic growth process defined by Eqs.\ (\ref{eq:gammasqavg})--(\ref{eq:Wevenodd}) refers to the basis of qudit operators --- the generalized Pauli matrices $\sigma_{p_x}$ ---, it has $q^2$ degrees of freedom per qudit. This obstructs an efficient simulation of the stochastic growth process for a large qudit dimension $q$. In Ref.\ \cite{tan2025operatorspreadingrandomunitary} we showed that for a unitary-invariant circuit, by appropriate summation the growth process may be simplified to a growth process with only two degrees of freedom per qudit, specifying whether the index $p_x$ is trivial ($p_x = 0$) or nontrivial ($p_x \neq 0$). Hereto, we defined the {\em binary Pauli string} $\bar p$ associated with $p$ as the string $[\bar p_0, \bar p_1, \ldots, \bar p_{L-1}]$, where $\bar p_x = 0$ if $p_x= 0$ and $\bar p_x = 1$ if $p_x \neq 0$. Writing $p \to \bar p$ to denote the binary string $\bar p$ corresponding to the Pauli string $p$, the ``projected binary-string distribution'' $\bar \rho_{\bar p}$ corresponding to $\rho_{pp}$ is then defined as
\begin{equation}
  \bar \rho_{\bar p} = \sum_{p \to \bar p} \rho_{pp},
  \label{eq:binarydef}
\end{equation}
{\em i.e.}, $\bar \rho_{\bar p}$ is the sum of all diagonal elements $\rho_{pp}$ with the same binary Pauli string $\bar p$.
The evolution equation (\ref{eq:gammasqavg0}) closes when applied to the projected binary-string distribution, {\em i.e.}, $\bar \rho_{\bar p}(t)$ depends only on $\bar \rho_{\bar p}(t-1)$ \cite{tan2025operatorspreadingrandomunitary}. Since the projected binary-string distribution has only two degrees of freedom per qubit, its stochastic evolution equation can be easily simulated classically, independent of the actual qudit dimension $q$. Moreover, it is a good starting point for analytical approximation schemes.

A similar simplification is possible for an orthogonal-invariant or symplectic-invariant circuit, but with three degrees of freedom per qudit instead of two.
In order to achieve this simplification, we consider a projection of the classical probabilities $\rho_p(t)$ onto a ``ternary'' probability string $\bar \rho_{\bar p}$, where the string index $\bar p$ is a list of ternaries $[\bar p_0,\bar p_1,\ldots,\bar p_{L-1}]$ with $\bar p_x \in \{-1,0,1 \}$, where $\bar p_x = 0$ if $\sigma_{p_x} = \openone$ and $\bar p_x = s_{p_x}$ if $\sigma_{p_x} \neq \openone$. (The parity $s_{p_x}$ is defined in Eq.\ (\ref{eq:sigmaparity}).)
In analogy to Eq.~(\ref{eq:binarydef}), we write $p\to\bar p$ for the
ternary string associated with the Pauli string $p$ and define the {\em projected ternary-string distribution} $\bar\rho_{\bar p}$ from
$\rho_{pp}$ by the same equation.

In App.\ \ref{app:4} we show that there is a closed Markovian evolution equation for the projected ternary distribution function $\bar \rho_{\bar p}$,
\begin{equation}
  \bar \rho_{\bar p}(t) = \sum_{\bar a} \bar \rho_{\bar a}(t-1) T_{\bar a \bar p},
  \label{eq:rhobinaryevolution}
\end{equation}
with transition probabilities $T_{\bar a\bar p}$ that are products of pair-transition probabilities,
\begin{equation}
  T_{\bar a\bar p} =   
  \prod_{x\, {\rm even}}
  \left\{ \begin{array}{ll}
    T_{\bar a_x \bar a_{x+1};\bar p_x,\bar p_{x+1}} & \mbox{for $t$ even}, \\
    T_{\bar a_{x-1} \bar a_x;\bar p_{x-1} \bar p_x} & \mbox{for $t$ odd}.
  \end{array} \right.
  \label{eq:Tpair}
\end{equation}
We refer to App.\ \ref{app:4} for explicit expressions.

The projected binary– or ternary–string distributions corresponding to the maximally random steady state \eqref{eq:rhomax} factorize sitewise,
\begin{equation}
  \bar \rho^{\infty}_{\bar p}(t) = \prod_x \bar r^{\infty}_{\bar p_x},
  \label{eq:barrhosteady}
\end{equation}
with 
\begin{equation}
  \label{eq:rinftyU}
  \bar r^{\infty}_{\bar p_x} =
  \frac{1}{q^2}\!\left[ \delta_{\bar p_x,0} +
  (q^2-1)\,\delta_{\bar p_x,1} \right]
\end{equation}
for the unitary case. For orthogonal- or symplectic-invariant random circuits, the maximally random steady state also has the form (\ref{eq:barrhosteady}), but with
\begin{equation}
  \label{eq:rinftyO}
  \bar r^{\infty}_{\bar p_x} =
  \frac{1}{q^2}\!\left[ \delta_{\bar p_x,0} +
  \frac{q-1}{2}\!\left((q+2)\,\delta_{\bar p_x,1}
  + q\,\delta_{\bar p_x,-1} \right) \right],
\end{equation}
for qudits with integer spin ({\em i.e.}, qudits that have an involution of the form (\ref{eq:involution}) with symmetric $Z_x$) and 
\begin{equation}
  \label{eq:rinftyS}
  \bar r^{\infty}_{\bar p_x} =
  \frac{1}{q^2}\!\left[ \delta_{\bar p_x,0} +
  \frac{q+1}{2}\!\left((q-2)\,\delta_{\bar p_x,1}
  + q\,\delta_{\bar p_x,-1} \right) \right],
\end{equation}
for qudits with half-integer spin ({\em i.e.}, qudits that have an involution with antisymmetric $Z_x$). We recall that orthogonal-invariant two-gate distributions appear if all qudits have the same spin, whereas symplectic-invariant two-gate distributions appear if the qudits alternate between integer and half-integer spin between even and odd $x$. The projected ternary-string distribution corresponding to the trivial solution (\ref{eq:rhotrivial}) is
\begin{equation}
  \label{eq:rhobinarytrivial}
  \bar \rho^0_{\bar p}(t) =
  \prod_{x} \delta_{\bar p_x,0}
\end{equation}
for all cases.

\subsection{Right-propagating $n$-point density}
\label{sec:3e}

The next steps in the analysis of operator spreading in orthogonal-invariant or symplectic-invariant random circuits closely follow those of the analysis of operator spreading in unitary-invariant circuits in Ref.\ \cite{tan2025operatorspreadingrandomunitary}. We say that the ternary string $\bar p$ has front position $x$, or equivalently ``ends at $x$'' if $x$ is the rightmost qudit position with $\bar p_x \neq 0$. (This implies that $\bar p_y = 0$ for all $y > x$.) Following Ref.\ \cite{nahum_operator_2018}, we define the ``right-propagating density'' $\bar \rho_{\rm }^{(0)}(\Delta x;t;\bar a_0)$ as the fraction of ternary Pauli strings whose front position is $x = t + \Delta x$ and whose last nontrivial element is equal to $\bar a_0 \in \{-1,1\}$,
\begin{equation}
  \bar \rho_{\rm }^{(0)}(\Delta x;t;\bar a_0) = \sum_{\textrm{$\bar p$ ends at $t + \Delta x$}} \bar \rho_{\bar p}(t) \delta_{\bar p_{t + \Delta x},\bar a_0}.
\end{equation}
(Note that the argument $\Delta x$ is measured with respect to the ballistic propagation at unit velocity.) We also define the ``right-propagating $n$-point density''
\begin{widetext}
\begin{equation}
  \label{eq:rhoRdef}
    \bar\rho_{\rm }^{(n)}(\Delta x;t;\bar a_n,\ldots,\bar a_0)
 =\,
  \sum_{\textrm{$\bar p$ ends at $t + \Delta x$}} \bar \rho_{\bar p}(t)\,
  \delta_{\bar{p}_{t+\Delta x},\bar a_0}\delta_{\bar p_{t+\Delta x-1},\bar a_{1}} \ldots \delta_{\bar p_{t+\Delta x-n},\bar a_{n}},
\end{equation}
which is the fraction of ternary Pauli strings that end at $t + \Delta x$ and that have the sequence $\bar a_n,\ldots,\bar a_0$ leading up to the end of the string at $\Delta x$.

From the stochastic evolution equation for the full ternary probability density $\bar \rho_{\bar p}(t)$, we may deduce evolution equations for the right-propagating density and for the right-propagating $n$-point densities. These are
\begin{align}
  \label{eq:rhoevol1}
  \bar \rho_{\rm }^{(n)}(\Delta x;t;\bar p_n,\ldots,\bar p_0) =&\,
  \sum_{\bar a_0,\ldots,\bar a_{n}} 
  T_{\bar a_{n} \bar a_{n-1};\bar p_{n} \bar p_{n-1}} \ldots T_{\bar a_2,\bar a_1;\bar p_2,\bar p_1}
  T_{\bar a_00;\bar p_00} \bar \rho_{\rm }^{(n)}(\Delta x+1;t-1;\bar a_n,\ldots,\bar a_0)
  \\ \nonumber &\, \mbox{}
  + \sum_{\bar a_0,\ldots,\bar a_{n+1}}
  T_{\bar a_{n+1}\bar a_{n};\bar p_{n}\bar p_{n-1}}
  \ldots T_{\bar a_3,\bar a_2;\bar p_2,\bar p_1}
  T_{\bar a_1 \bar a_0;\bar p_00}    \bar \rho_{\rm }^{(n+1)}(\Delta x+2;t-1;\bar a_{n+1},\bar a_n,\ldots,\bar a_0) 
\end{align}
if $n$ and $\Delta x$ are both even. The summation variables $\bar a_j \in \{-1,0,1\}$ for $j = 1,\ldots,n$; $\bar a_0$ takes the values $-1$ and $1$ only. In the same manner, we find
\begin{align}
  \label{eq:rhoevol2}
  \bar \rho_{\rm }^{(n)}(\Delta x;t;\bar p_n,\ldots,\bar p_0) =&\,
  \sum_{\bar a_0,\ldots,\bar a_{n}} \sum_{\bar p_{n+1}} T_{\bar a_{n}\bar a_{n-1};\bar p_{n+1}\bar p_{n}} \ldots T_{\bar a_2,\bar a_1;\bar p_3,\bar p_2} T_{\bar a_00;\bar p_1 \bar p_0}  \bar \rho_{\rm }^{(n)}(\Delta x;t-1;\bar a_{n},\ldots,\bar a_{0})
   \\ \nonumber &\, \mbox{}
  +
  \sum_{\bar a_0,\ldots,\bar a_{n+1}} \sum_{\bar p_{n+1}} T_{\bar a_{n+1}\bar a_{n};\bar p_{n+1}\bar p_{n}} \ldots T_{\bar a_3,\bar a_2;\bar p_3,\bar p_2} T_{\bar a_1 \bar a_0;\bar p_1 \bar p_0}
   \bar \rho_{\rm }^{(n+1)}(\Delta x+1;t-1;\bar a_{n+1},\bar a_n,\ldots,\bar a_0)
  , 
\end{align}
\end{widetext}
if $n$ is even and $\Delta x$ is odd. The summation variables $\bar a_0,\ldots,\bar a_n$ take the same values as for the case that $n$ and $\Delta x$ are both even; $\bar p_{n+1}$ takes the values $-1$, $0$, and $1$.
To avoid spurious even-odd effects, we will not consider the evolution equations for the right-propagating $n$-point density with $n$ odd.

\subsection{Mapping to a drift-diffusion process}
\label{sec:dd}

The evolution equations (\ref{eq:rhoevol1}) and (\ref{eq:rhoevol2}) for the right-propagating $n$-point density both involve the $(n+1)$-point density. To obtain a closed set of equations, we follow Ref.\ \cite{tan2025operatorspreadingrandomunitary} and truncate the evolution equations (\ref{eq:rhoevol1}) and (\ref{eq:rhoevol2}) at sufficiently high order $n$ by replacing the $(n+1)$-point density by its maximally random approximation,
\begin{align}
  \label{eq:rhonn}
  \lefteqn{ \bar \rho^{(n+1)}_{\rm }(\Delta x;t;\bar p_{n+1},\bar p_{n},\ldots,\bar p_0)}
  ~~~~~~~~~~~~ \nonumber \\ =&\,
  \bar \rho^{(n)}_{\rm }(\Delta x;t;\bar p_n,\ldots,\bar p_0) \bar r^{\infty}_{\bar p_{n+1}},
\end{align}
where $\bar r^{\infty}_{\bar p_x}$ was defined in Eqs.\ (\ref{eq:rinftyU})--(\ref{eq:rinftyS}). The long-time solution of the truncated evolution equation is of the form \cite{tan2025operatorspreadingrandomunitary}
\begin{align}
  \label{eq:rhoeqsol}
  \bar \rho^{(n)}(\Delta x;t;\bar p_{n},\ldots,\bar p_{0}) =&\,
    R_1^{(n)}(\Delta x;t) 
  \\ \nonumber &\, \mbox{} \times
  V_1^{(n)}((-1)^{\Delta x};\bar p_{n},\ldots,\bar p_{0}),
\end{align}
where $V_1^{(n)}(\sigma;\bar p_{n},\ldots,\bar p_{0})$ is independent of $\Delta x$ and $t$ (up to the parity of $\Delta x$) and $R^{(n)}(\Delta x;t)$ satisfies a drift-diffusion equation
\begin{align}
  \label{eq:driftdiffusion}
  \partial_t R_{{\rm }1}^{(n)}(\Delta x,t) =&\,
  (1-v_{\rm B}^{(n)} )
  \partial_{\Delta x} R_{{\rm }1}^{(n)}(\Delta x,t)
  \nonumber \\ &\, \mbox{}
  + \frac{{\cal D}^{(n)}}{2} \partial_{\Delta x}^2 
  R_{{\rm }1}^{(n)}(\Delta x,t).
\end{align}
Explicit expressions for the drift velocity $v_{\rm B}^{(n)}$ and the diffusion constant ${\cal D}^{(n)}$ in terms of the ternary-string transition matrices $T_{\sigma,\sigma'}$ of Eq.\ (\ref{eq:Tmatrix}) are given in App.\ \ref{app:5}. We verify that the results for $v_{\rm B}$ and ${\cal D}$ converge with respect to the truncation order $n$.

\subsection{Approach to a binary or ternary distribution}

In Subsec.\ \ref{subsec:3D} we showed that a closed evolution equation can be obtained if instead of the full $\rho_{pp}$ we consider a ``projected ternary'' weight $\bar \rho_{\bar p}$ that is obtained from $\rho_{pp}$ by summing over all Pauli strings $p$ with the same binary or ternary string $\bar p$. We say that the Pauli-string density matrix $\rho$ is of ``binary form'' or ``ternary form'' if it is diagonal and all Pauli strings $p$ that map to the same binary or ternary string $\bar p$ have the same probability $\rho_{pp}$. In Ref.\ \cite{tan2025operatorspreadingrandomunitary}, we showed that $\rho_{pp}$ exponentially relaxes towards such a binary form for the unitary-invariant case and calculated the corresponding relaxation time. In fact, the exponential relaxation to the binary form applies not only to the diagonal elements $\rho_{pp}$, but to the full Pauli-string density matrix $\rho$, including its off-diagonal
elements.

In analogy to the unitary-invariant case, the Pauli-string density matrix $\rho$ of an orthogonal-invariant or symplectic-invariant
random circuit relaxes towards a diagonal distribution of ternary form. In such a distribution, all Pauli strings $p$ that have the same ternary string $\bar p$ have the same probability. Explicitly, we define the operator $\BB$, which maps $\rho_{pq}$ to its associated ternary distribution, as
\begin{equation}
  \BB \rho_{pq} = \delta_{pq} \left( \prod_{x} B_{\bar p_x} \right) \bar \rho_{\bar p},
\end{equation}
where $\bar p$ is the ternary string corresponding to $p$, $\bar \rho_{\bar p}$ the projected ternary weight corresponding to $\rho_{pp}$. The factor $B_{{\bar p}_x}$ gives the uniform probability to find the generalized Pauli operator $p_x$ given the ternary label $\bar p_x$. \\
In the orthogonal-invariant case, as well as at even sites in the
symplectic-invariant case, the ternary projector is obtained by taking
the upper signs in
\begin{equation}
\label{eq:B2}
  B_{\bar p_x}
  =
  \delta_{\bar p_x,0}
  + \frac{2}{(q\mp1)(q\pm2)} \delta_{\bar p_x,1}
  + \frac{2}{q(q\mp1)} \delta_{\bar p_x,-1}.
\end{equation}
At odd sites in the symplectic-invariant case, the lower signs apply.
For $q=2$, however, the lower-sign expression in
Eq.~\eqref{eq:B2} is singular and therefore does not define the
ternary projector. In this special case, the projector at odd sites is
instead
\begin{equation}
\label{eq:projsympq2}
  B_{\bar p_x}
  =
  \delta_{\bar p_x,0}
  + \frac{1}{3}\delta_{\bar p_x,-1}.
\end{equation}
In App.\ \ref{app:6} we show that the two-norm $|| \rho_{pq} -  \BB \rho_{pq}||_2 \to 0$ in the long-time limit.

\section{General qudit dimension $q$}
\label{sec:qgeneral}

For an arbitrary qudit dimension $q$ it is not possible to find a set of generalized Pauli matrices that satisfy the orthonormality condition (\ref{eq:orthonormality}) and that have a well-defined parity under transposition or duality. As a consequence, it is not possible to construct a stochastic growth process involving only the diagonal elements $\rho_{pp}$ of the Pauli-string density matrix.
In both the orthogonal-invariant and the symplectic-invariant case, however, the Weyl-Heisenberg choice for the generalized Pauli matrices (see App.\ \ref{app:1}) has the property that $Z_x \sigma_p^{\rm T} Z_x^{\dagger}$ is again a generalized Pauli matrix, up to a phase factor. For each generalized Pauli matrix $\sigma_{p_x}$ we therefore define the ``transpose index'' $p_x^{\rm T}$ and phase $\phi_{p_x}$ via the relation
\begin{equation}
\label{eq:Ztranspose}
  Z_x \sigma_{p_x}^{\rm T} Z_x^{\dagger} = 
  \sigma_{p_x^{\rm T}} e^{-i \phi_{p_x}},
\end{equation}
if $Z_x$ is symmetric.
For the Weyl-Heisenberg generalized Pauli matrices $\sigma_{p_x}$, the index $p_x$ is represented by a pair of integers $p_x = (p_x',p_x'')$ with $p_x'$, $p_x'' \in \ZZ_q$, see App.\ \ref{app:1}, and the phase shift $\phi_{p_x}$ is given by
\begin{equation}
  \phi_{p_x}=\frac{2\pi}{q}p'_xp''_x.
\end{equation}
Analogously, for anti-symmetric $Z_x$, we define the ``dual index'' $p_x^{\rm R}$ and corresponding phase via the relation
\begin{equation}
 Z_x \sigma_{p_x}^{\rm T} Z_x^{\dagger} = 
 \sigma_{p_x^{\rm R}} e^{-i \phi_{p_x}},
\end{equation}
where $\phi_{p_x}=2 \pi p^{\prime}_x p^{\prime \prime}_x / 2+\pi(1-$ $\left.\delta_{p^{\prime \prime \prime}_x, 0}\right)$. Here the index $p_x$ is represented by a triple of integers $p_x = (p'_x, p''_x, p'''_x)$ with $p'''_x \in \{0, 1, 2, 3\}$, see
App.~\ref{app:1}.

For generalized Pauli matrices with this property, we generalize the
definition~\eqref{eq:binarydef} of the projected ternary weight as
\begin{equation}
\label{eq:ternarygen}
  \bar \rho_{\bar p}
  =
  \sum_{a,b} \rho_{ab}
  \prod_x \chi_{\bar p_x;a_x,b_x},
\end{equation}
where $\bar p_x\in\{-1,0,1\}$ and
\begin{align}
\label{eq:chioriginal}
  \chi_{\bar p_x;a_x,b_x} =&\,
  \left\{ \begin{array}{ll} \delta_{a_x,0} \delta_{b_x,0},
    & \mbox{if $\bar p_x = 0$}, \\
    \frac{1}{2} (1 - \delta_{a_x,0})
    \\ \mbox{} \times (\delta_{a_x,b_x} +\bar p_x
  \delta_{a_x,b_x^{\mathrm{T,R}}} e^{-i \phi_{a_x}}) & \mbox{else} \end{array} \right.
\end{align}
Here, $b_x^{\mathrm T}$ is used for orthogonal circuits and for even
sites in symplectic circuits, whereas $b_x^{\mathrm R}$ is used for
odd sites in symplectic circuits.
For $q = 2$, $\bar p_x$ retains the same interpretation as defined in Subsec.~\ref{subsec:3D}, and the projected ternary weight function constructed in Eq.~\eqref{eq:ternarygen} reduces to Eq.~\eqref{eq:binarydef}.

In App.\ \ref{app:4} we show that there is a well-defined stochastic growth model for the projected ternary weight $\rho_{\bar p}$, which has the same structure as the stochastic growth process for the case $q = 2^m$ considered previously. 
As in the unitary-invariant case, under time evolution, the
Pauli-string density matrix $\rho$ relaxes to a ``ternary-form'' distribution, which is a distribution of the form
\begin{equation}
  \rho_{ab} = \bar \rho_{\bar p} \prod_{x} \frac{\chi_{\bar p_x;a_x,b_x}^*}{q^2 \bar r^{\infty}_{\bar p_x}},
\end{equation}
with $\bar \rho_{\bar p}$ the ternary distribution corresponding to $\rho_{ab}$. For more details, we refer to App. \ref{app:6}. 
\begin{figure}
\centering
\includegraphics[scale=0.50, trim=0 0 
0 0, clip]{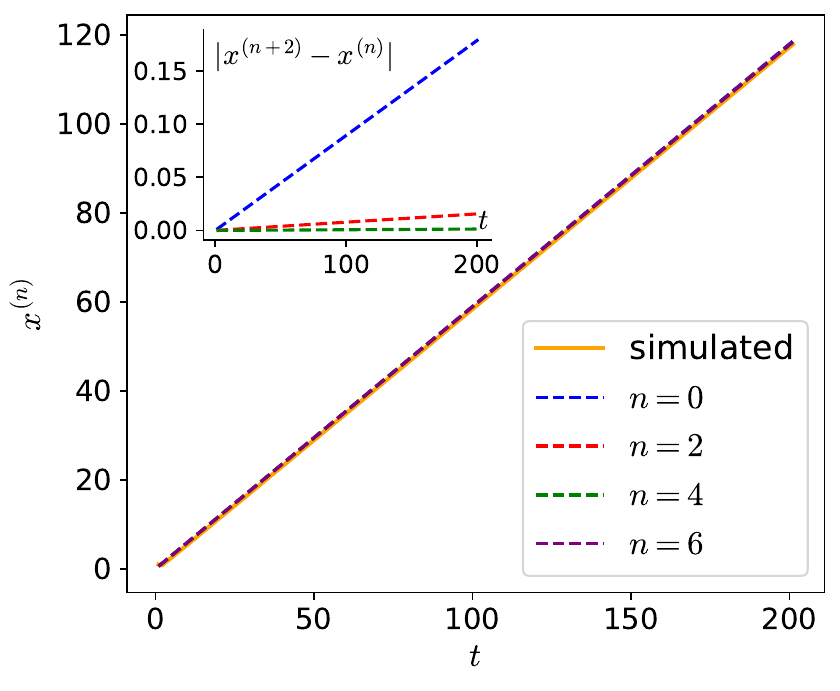} 
\includegraphics[scale=0.50, trim=0 0 
0 0]{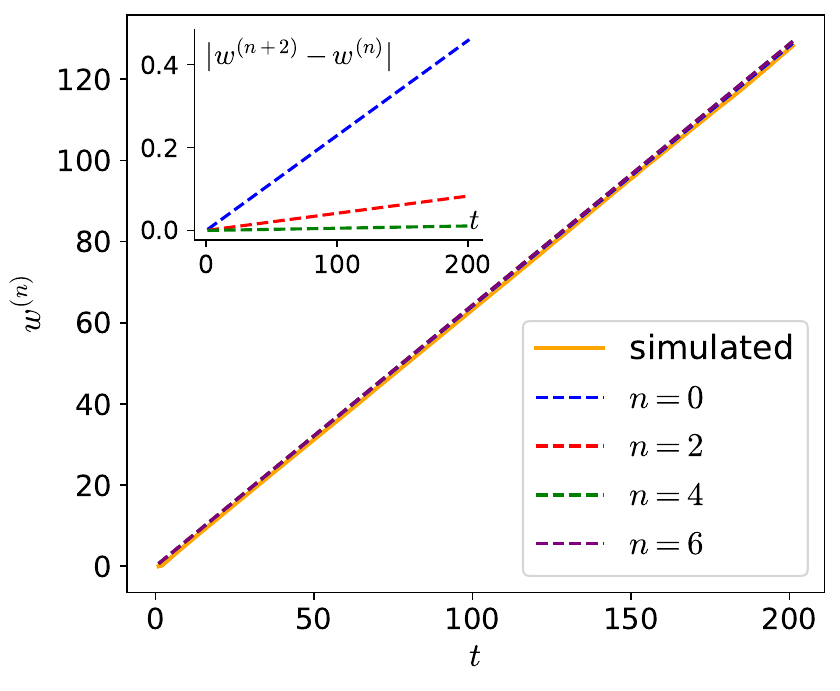} 
\vskip 0.5cm
\caption{\justifying \small
Front position (top panel) and variance of front position (bottom panel) vs. time $t$ for a random unitary circuit with Haar-orthogonal two-qudit gates and qudit dimension $q = 2$. The solid lines in both panels represent results from direct numerical simulations of the classical stochastic growth process with system size $L = 200$, based on $N = 2 \times 10^5$ independent realizations. In the top panel, the slopes of the lines correspond to the butterfly velocity $v_{\rm B}$, while in the bottom panel, the slopes yield the diffusion constant $\mathcal{D}$. The dashed curves indicate analytical results obtained from the truncation scheme at orders $n = 0$, 2, 4, and 6. Due to the rapid convergence of the truncation, the analytical curves in the main panels are visually indistinguishable. To better resolve these details, the insets display the absolute differences between successive analytical approximations, $|x^{(n+2)} - x^{(n)}|$ (top) and $|w^{(n+2)} - w^{(n)}|$ (bottom), vs.\ $t$. Here, $x^{(n)}$ and $w^{(n)}$ represent the $n$-th order analytical approximations of the front position and variance of front position, respectively.\label{fig:haar}}
\end{figure}
\section{Examples}
\label{sec:7}
\subsection{Haar-random ensembles}

For the Haar-random ensembles, the evolution matrices ${\cal U}_{x,x+1}$ are randomly chosen from the unitary, orthogonal, or symplectic group. For the Haar-unitary random circuit, the pair transition probabilities $\langle |{\cal W}_{ap;x,x+1}|^2\rangle$ are of the form (\ref{eq:Wgeneral}) with $\mathcal P_1 = 1/(q^4-1)$ and $\mathcal P_2 = \mathcal P_3 = 0$. The butterfly velocity and diffusion constant for this circuit were calculated in Ref.\ \cite{nahum_operator_2018},
\begin{equation}
  v_{\rm B} = \frac{q^2-1}{q^2+1},\ \
  {\cal D} = \frac{4 q^2}{(q^2+1)^2}.
\end{equation}
Haar-orthogonal random circuits with qudit dimension $q=2^m$ were
previously studied by Hunter-Jones in
Ref.~\cite{hunter-jones_operator_2018} using the standard orthogonal
Haar integral. In our formulation, the transition probability matrix for this
case takes the form of Eq.~\eqref{eq:Wgeneral}, with the coefficients
$\mathcal P_j$ given by Eqs.~\eqref{eq:P1Orthogonal} and
\eqref{eq:P23Orthogonal}. These results are algebraically equivalent
to the transition probabilities obtained in
Ref.~\cite{hunter-jones_operator_2018}, although the two derivations
use different notation and organize the resulting equations
differently. The corresponding formulation for arbitrary qudit
dimension $q$ is given in App.~\ref{app:7}.  

The $9 \times 9$ pair transition probability matrices appearing on the r.h.s.\ of Eq.\ (\ref{eq:Tpair}) have the block structure, 
\begin{equation}
  T =
    \begin{bmatrix}
      1&&\\
      &T_{++}&T_{+-}\\
      &T_{-+}&T_{--}
    \end{bmatrix},
  \label{eq:Tmatrix}
\end{equation}
where the sign $\pm$ of the indices of the $4 \times 4$ blocks $T_{s_{a},s_{p}}$ refer to the combined parity of the indices $a_{x}$ and $a_{x+1}$ or $p_{x}$ and $p_{x+1}$. 
For the Haar-orthogonal random circuit, the $4 \times 4$ transition matrices for the projected ternary weights are
\begin{align}
  T_{++} =&\, \frac{1}{2(q^2+2)(q^2-1)}
    \begin{pmatrix} f^2 & g^2 & 2 f & 2 f \\ f^2 & g^2 & 2 f & 2 f \\
  f^2 & g^2 & 2 f & 2 f \\ f^2 & g^2 & 2 f & 2 f \end{pmatrix}, 
  \label{eq:T1orthogonal} \\
  T_{--} =&\, \frac{1}{2 q^2(q^2-1)}
    \begin{pmatrix} f g & f g & 2 g & 2 g \\ f g & f g & 2 g & 2 g \\
  f g & f g & 2 g & 2 g \\ f g & f g & 2 g & 2 g \end{pmatrix}
  \label{eq:T2orthogonal}
\end{align}
and $T_{+-} = T_{-+} = 0$, with $f = (q+2)(q-1)$ and $g = q(q-1)$. 
Using the approximation scheme of Sec.\ \ref{sec:dd}, we then obtain for the zeroth order approximation for the butterfly velocity and diffusion constant
\begin{align}
  v_{\rm B}^{(0)} =&\, \frac{(q-1)(q^3+2q^2+2q + 3)}{(q+1)(q^3 + 2q + 1)}, 
  \label{eq:vorthogonalHaar} \\
  {\cal D}^{(0)} =&\, \frac{4}{(q+1)^2(q^3 + 2q + 1)^3}
  (q^9 + 2 q^8 + 6 q^7 + 10 q^6 
  \nonumber \\ &\, \ \ \ \ \mbox{} 
  + 11 q^5 + 19 q^4 + 3 q^3 + 15 q^2 - q - 2).
  \label{eq:DorthogonalHaar}
\end{align}

The zeroth-order expressions given in Eqs.~\eqref{eq:vorthogonalHaar} and \eqref{eq:DorthogonalHaar} coincide with those reported in Ref.~\cite{hunter-jones_operator_2018}. We further evaluate the butterfly velocity and diffusion constant using truncation orders $n=2$, $n=4$, and $n=6$. These results are presented in Fig.~\ref{fig:haar} for $q=2$, alongside direct numerical simulations of the stochastic process defined by Eq.~\eqref{eq:rhobinaryevolution}. The specific values for the butterfly velocity and diffusion constant at orders $n=0, 2, 4,$ and $6$ are summarized in Table~\ref{tab:vb_D_vs_n}. As shown, both quantities exhibit clear convergence as the truncation order increases. In contrast to the Haar-unitary circuit, Table~\ref{tab:vb_D_vs_n} demonstrates that the zeroth-order approximation is not exact for the Haar-orthogonal random circuit. Although the deviations between the higher-order and zeroth-order results are small, their existence implies that the domain-wall width does not strictly vanish in the Haar-orthogonal limit. 

Expanding Eqs.~\eqref{eq:vorthogonalHaar} and \eqref{eq:DorthogonalHaar} for large $q$ yields $v_{\text{B}}^{(0)} = 1 - 2/q^2 + \mathcal{O}(q^{-4})$ and $\mathcal{D}^{(0)} = 4/q^2 + \mathcal{O}(q^{-4})$. Without the terms of ${\cal O}(q^{-2})$, this is the same result as that obtained from an exact treatment of the limit $q \to \infty$, see Eq.\ \eqref{eq:vdqlarge}. From this, we conclude that the zeroth-order approximation is exact in the limit $q \to \infty$.

\begin{table}
\begin{tabular*}{\linewidth}{c@{\extracolsep{\fill}} ccc}
 \hline\hline
$n$ & $v_{\rm B}^{(n)}$ & $\mathcal{D}^{(n)}$ \\
\hline
0 & 0.5897435897 & 0.6408739190 \\
2 & 0.5888499236 & 0.6431654164 \\
4 & 0.5887719700 & 0.6435805364 \\
6 & 0.5887652662 & 0.6436349980 \\
\hline
sim. & 0.5886992069 & 0.6413786349 \\
\hline \hline
\end{tabular*}
\caption{\justifying \small
Butterfly velocity $v_{\rm B}^{(n)}$ and diffusion constant $\mathcal{D}^{(n)}$ for the Haar-orthogonal random circuit with $q = 2$, calculated analytically for orders $n=0, 2, 4, 6$, respectively, and compared with results obtained from numerical simulations of the stochastic evolution of the projected ternary-string distribution $\bar \rho_{\bar p}$.}
\label{tab:vb_D_vs_n}
\end{table}

\subsection{Brownian special orthogonal circuit}

In a Brownian circuit, the two-qudit operators ${\cal U}_{x,x+1}$ are obtained from the continuous time evolution with a random Hamiltonian with a Gaussian white-noise distribution \cite{Dyson1962Brownian,Dyson1972Brownian}. Specifically,
\begin{equation}
  {\cal U} = T_t e^{-i \int_0^1 dt {\cal H}(t)},
  \label{eq:UBrownian}
\end{equation}
where $T_t$ indicates the time-ordering prescription and ${\cal H}(t')$ is a $q^2 \times q^2$ random Hermitian matrix with zero mean and Gaussian delta-function correlations. For the Brownian special orthogonal circuit, $H(t)$ is anti-symmetric. It has the two-time correlation function \cite{Mehta1983gaussian,mehta1997random}
\begin{equation}
  \label{eq:HBrownian}
  \langle H_{ij}(t) H_{kl}(t') \rangle =
  \frac{\lambda}{q^2} (
  \delta_{il} \delta_{jk} -
  \delta_{ik} \delta_{jl}) \delta(t-t'),
\end{equation}
where $\lambda$ is a parameter that describes the interpolation between the trivial and Haar-random-special-orthogonal two-qudit gate operators. This ensemble satisfies the orthogonal invariance property (\ref{eq:unitary_invariance_xy}).

\begin{figure}[tb]
\centering
\includegraphics[scale=0.48, trim=0.4 0 
0 0, clip]{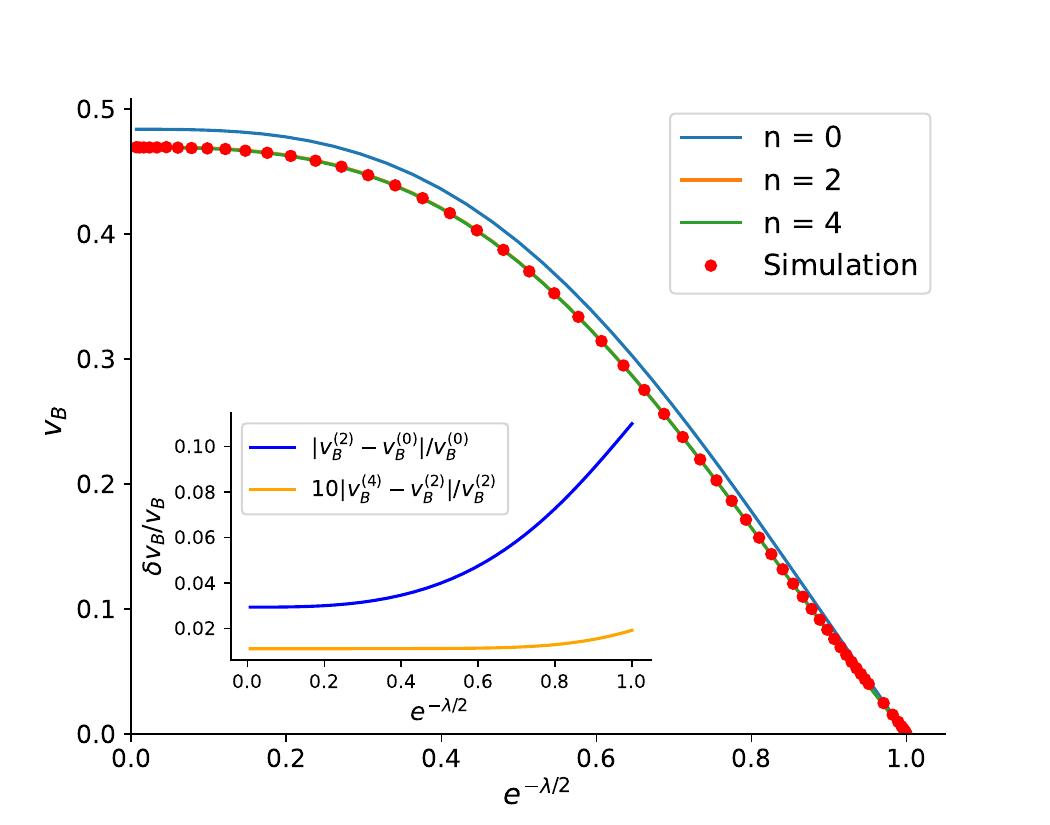} 
\includegraphics[scale=0.6, trim=0 0
0 0]{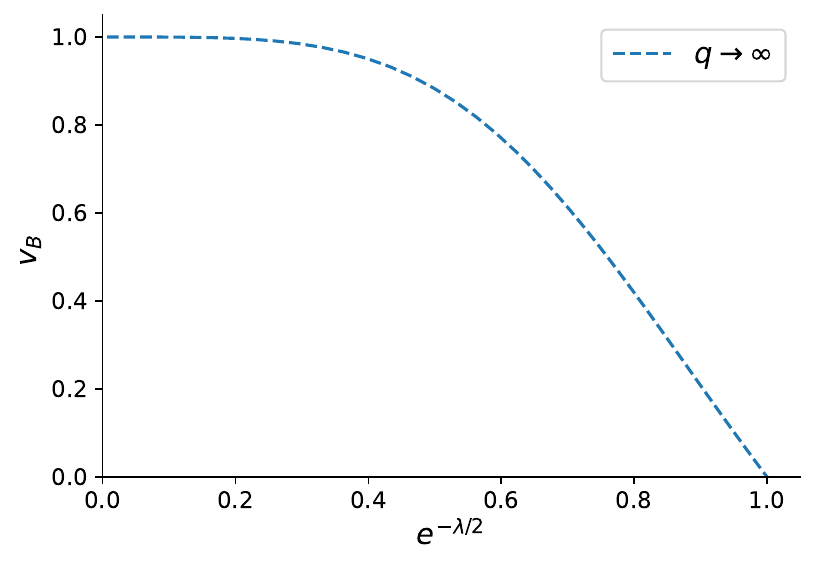} 
\vskip 0.5cm
\caption{\justifying \small
Butterfly velocity for Brownian special orthogonal circuits with local dimension $q=2$ and in the limit $q\to \infty$. The parameter $\lambda$ interpolates between a trivial circuit and the Haar-random limit. Top panel: Solid curves represent the analytical approximations $v_{\rm B}^{(n)}$ for $q=2$ at orders $n=0, 2,$ and $4$. Symbols denote data from direct numerical simulations (system size $L = 320$, evolved for 200 time steps, averaged over $N = 2 \times 10^5$ realizations). Due to the rapid convergence of the truncation scheme, the analytical results for all orders are virtually indistinguishable from the numerical data. Inset: Relative differences between the approximations of order $n=4$ and $n=2$, and between $n=2$ and $n=0$. Lower panel: The dashed line shows the result for $q\to \infty$, where the approximation scheme is exact at $n=0$.}
\label{fig:BrownianV}
\end{figure}

\begin{figure}[tb]
\centering
\includegraphics[scale=0.48, trim=0 0 
0 0, clip]{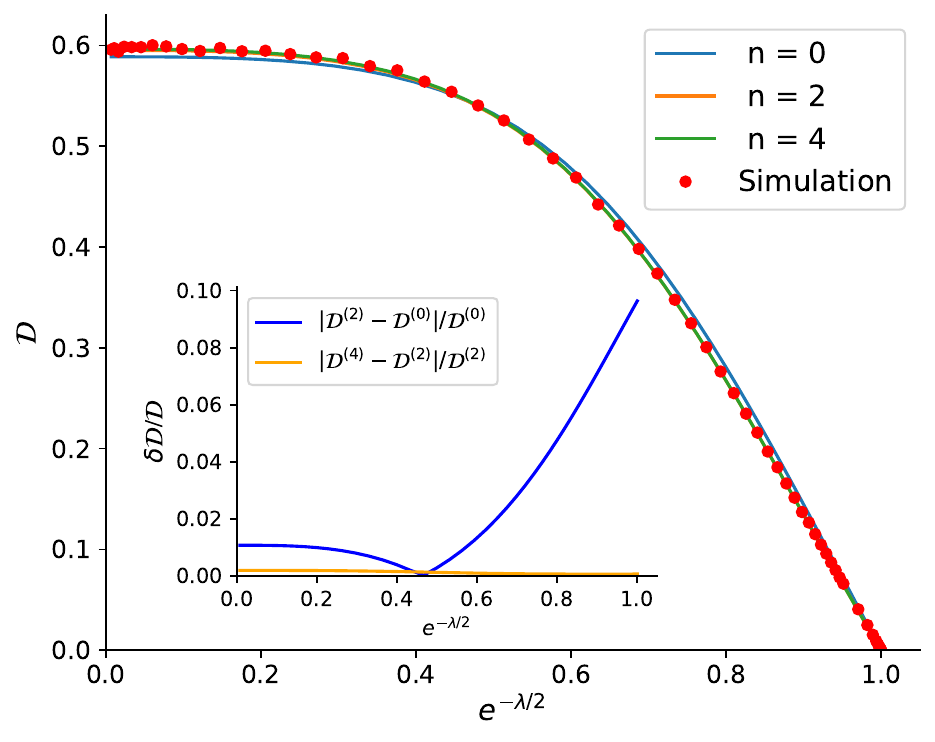} 
\includegraphics[scale=0.6, trim=0 0.5 
0 0cm]{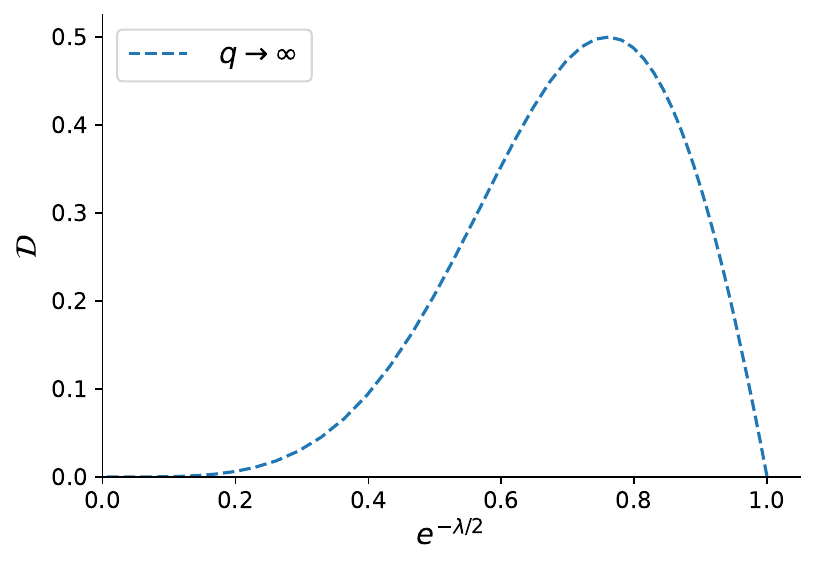} 
\vskip 0.5cm
\caption{\justifying \small
 Same as Fig.~\ref{fig:BrownianV}, but for the diffusion constant ${\cal D}$ for a Brownian special orthogonal circuit. The small fluctuations of the data points around the theoretical curve are attributed to statistical noise from the large, but finite, number of realizations $N$ in the numerical simulations. Due to the rapid convergence of the truncation scheme, the analytical results for orders $n=2$ and $4$ are indistinguishable in the upper panel. }
\label{fig:BrownianD}
\end{figure}

The moments of the distribution function for the Brownian special orthogonal circuit are given in App.\ \ref{app:8}. The butterfly velocity and the diffusion constant for the Brownian special orthogonal circuit are shown in Figs.\ \ref{fig:BrownianV} and \ref{fig:BrownianD} for $q=2$ and in the limit $q \to \infty$. In the large-$q$ limit, the zeroth-order approximation is exact, and the butterfly velocity and the diffusion constant are 
\begin{align}
  v_{\rm B}^{(0)} =&\, \frac{1 - e^{-2\lambda}}{1 + e^{-2\lambda}}, \\
  {\cal D}^{(0)} =&\, \frac{4 e^{-2\lambda} (1 - e^{-2\lambda})}{(1 + e^{-2\lambda})^2},
\end{align}
which is the same as for the unitary case. 

The evolution coefficients for Brownian symplectic circuits can be obtained from those for orthogonal circuits by making the substitution $q^2 \to -q^2$ for all averages involving two-qudit operators. The resulting butterfly velocity and diffusion constant for $q=2$ are shown in Figs.~\ref{fig:symplecticvb} and~\ref{fig:symplecticD}. (For $q \to \infty$, there is no difference between special orthogonal and symplectic circuits.) 

\subsection{$\mbox{SO}^{-}$ ensembles}

\begin{figure}
\includegraphics[scale=0.6, trim=0 0 
0 0cm]{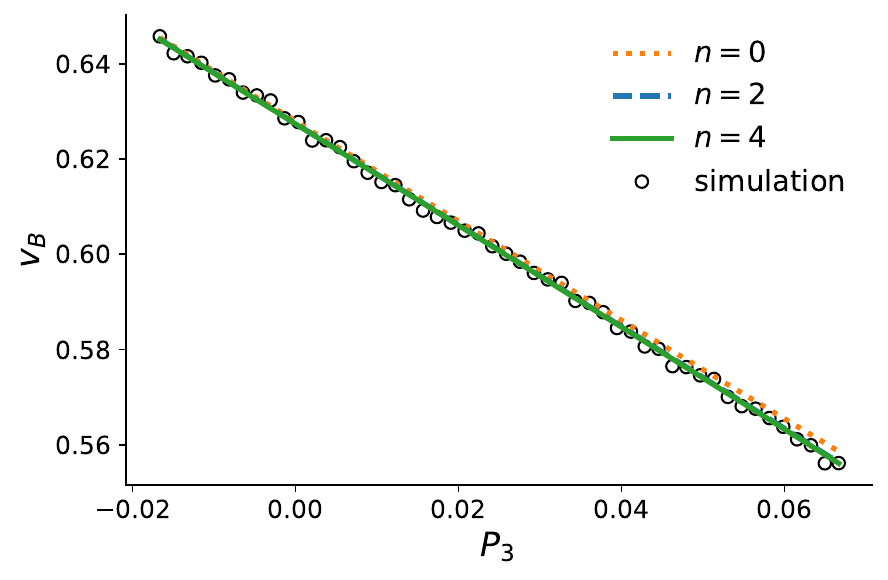} 
\vskip 0.5cm
\caption{\justifying \small
Butterfly velocity for a ${\rm SO}^{-}$ circuit at $q=2$. The parameter space of the circuit is one-dimensional and can be parameterized by $P_3$, which in turn determines other parameters $P_1$ and $P_2$. The data points represent the results of a direct numerical simulation conducted on a system of size $L=320$ for 200 time steps, with averaging over $N=5000$ independent realizations. The curves indicate analytical predictions derived from a truncation scheme at successive orders $n=0$, $2$, and $4$. Due to the rapid convergence of the truncation scheme, the analytical results for orders $n=2$ and $4$ are indistinguishable.
 }
\label{fig:SO-}
\end{figure}
An instructive example is provided by ${\rm SO}^{-}$ ensembles, which consist of orthogonal matrices with determinant $-1$. This set is topologically disconnected from the special orthogonal group ${\rm SO}(q^2)$ \cite{horn2012matrix}.
For odd qudit dimension $q$, $-{\cal U}_{x,x+1}$ has determinant $-1$ if ${\cal U}_{x,x+1}$ is an orthogonal matrix with determinant one. Since the operator spreading is unaffected by the simultaneous change ${\cal U}_{x,x+1} \to -{\cal U}_{x,x+1}$ for all $x$, the butterfly velocity and the diffusion constant in the ${\rm SO}^{-}$ circuit are the same as in the special orthogonal circuit obtained by replacing ${\cal U}_{x,x+1}$ by $-{\cal U}_{x,x+1}$ for all $x$. Such a relationship does not exist for even qudit dimension $q$. This can already be seen by noting that ${\rm SO}^{-}(q^2)$ does not contain the identity matrix or a matrix proportional to it, so that orthogonal invariance rules out the possibility of purely diagonal evolutions, thereby ensuring a nonzero minimal spreading velocity. This qualitative distinction separates the two disconnected sectors of the orthogonal ensembles.

For $q=2$, the pair transition probability $\langle |{\cal W}_{ap;x,x+1}|^2 \rangle$ takes the form given in Eq.~\eqref{eq:Wgeneral}. The parameters $\mathcal{P}_i$ for the ${\rm SO}^{-}$ ensemble with $q=2$ are
\begin{align}
\mathcal P_1 &= \begin{cases}
\frac{1}{9}-\frac{P_3}{6} & \text{if } s_{p_{x}} s_{p_{x+1}} = s_{a_{x}} s_{a_{x+1}} = 1, \\
\frac{1}{6} & \text{if } s_{p_{x}} s_{p_{x+1}} = s_{a_{x}} s_{a_{x+1}} = -1, \\
0 & \text{otherwise},
\end{cases} \nonumber\\
\mathcal P_2 &= \begin{cases}
\frac{P_3}{2} & \text{if } s_{p_{x}} s_{p_{x+1}} = s_{a_{x}} s_{a_{x+1}} = 1, \\
\frac{1}{6} & \text{if } s_{p_{x}} s_{p_{x+1}} = s_{a_{x}} s_{a_{x+1}} = -1, \\
0 & \text{otherwise},
\end{cases} \nonumber\\
\mathcal P_3 &= \begin{cases}
P_3 & \text{if } s_{p_{x}} s_{p_{x+1}} = s_{a_{x}} s_{a_{x+1}} = 1, \\
-\frac{1}{3} & \text{if } s_{p_{x}} s_{p_{x+1}} = s_{a_{x}} s_{a_{x+1}} = -1, \\
0 & \text{otherwise}.
\end{cases}
\end{align}
Here, the parameter $P_3$ is
\begin{equation}
P_3=\frac{1+ 16 \mathcal{R}_{1;3}}{75},
\end{equation}
where $\mathcal{R}_{1;3}$ is a fourth moment of the two-qudit evolution matrix ${\cal U}_{x,x+1}$,
\begin{equation}
\mathcal{R}_{1,3}= \frac{1}{q^4} \langle \operatorname{tr} {\cal U}_{x,x+1}\, \operatorname{tr} {\cal U}^3_{x,x+1} \rangle.
\end{equation}
From the expressions for ${\cal P}_1$, ${\cal P}_2$, and ${\cal P}_3$, it follows that the parameter space of the ${\rm SO}^{-}(4)$ ensemble is effectively one-dimensional: The distribution is characterized by the single real parameter $P_3$. To find the range that $P_3$ can take, we consider the structure of $q^2 \times q^2$ orthogonal matrices with $q=2$ and determinant $-1$: Every matrix ${\cal U}_{x,x+1}$ in ${\rm SO}^{-}(4)$ possesses the eigenvalues $\{-1,1,e^{i\theta},e^{-i\theta}\}$ for some $\theta \in [0,\pi]$. The bounds of $P_3$ are therefore determined by the minimum and maximum values of $\operatorname{tr} {\cal U} \operatorname{tr} {\cal U}^3 = 4 \cos \theta\, \cos 3 \theta$, which are $-\frac{9}{4}$ and $4$, respectively. This constrains $P_3$ to the interval
\begin{equation}
P_3 \in \left[-\frac{1}{60},\frac{1}{15}\right].
\end{equation}
Due to the restricted range of $P_3$, the accessible butterfly
velocities also lie within a finite interval. The maximum is attained
for
$\theta_\ast=(\pi\pm\arctan\sqrt{15})/2$,
whereas the minimum occurs at $\theta=0$ or $\theta=\pi$,
corresponding to gates with eigenvalues $(1,1,1,-1)$ or
$(-1,-1,-1,1)$. Since these two spectra differ only by an overall
factor of $-\openone$, which acts trivially under conjugation, they
generate the same operator dynamics. These involutions are the
elements of $\mathrm{SO}^{-}(4)$ closest to the trivial gate and play,
within this disconnected sector, a role analogous to that of the
identity in $\mathrm{SO}(4)$.

For a fixed value of $\theta$, orthogonal invariance requires uniform
sampling over the corresponding conjugacy class, while the probability
distribution of $\theta$ itself is not fixed by
Eq.~\eqref{eq:unitary_invariance}. The ensemble saturating the upper
bound is
$V=U\,\operatorname{diag}
\bigl(-1,1,e^{i\theta_\ast},e^{-i\theta_\ast}\bigr)U^{\mathrm T}$,
where $U$ is Haar distributed over the orthogonal group, while the
ensemble saturating the lower bound is
$V=U\,\operatorname{diag}(-1,1,1,1)U^{\mathrm T}$.
Neither extremal ensemble is Haar distributed over the
$\mathrm{SO}^{-}(4)$ sector.

Remarkably, the upper bound of the butterfly velocity in the
$\mathrm{SO}^{-}$ ensemble for $q=2$ exceeds the corresponding
Haar-orthogonal value. Intuitively, the Haar-orthogonal measure
samples the entire orthogonal group, including gates close to the
trivial one, whereas an ensemble concentrated on the most strongly
scrambling conjugacy classes of $\mathrm{SO}^{-}$ avoids this
contribution. More concretely, the condition
$\mathcal P_3=-1/3$ for
$s_{p_x}s_{p_{x+1}}=s_{a_x}s_{a_{x+1}}=-1$
suppresses the probability that odd-parity operator strings retain
their original form. Their resulting tendency to delocalize enhances
the propagation speed beyond the Haar-orthogonal value. Numerical and
analytical results for the $\mathrm{SO}^{-}$ ensembles are shown in
Fig.~\ref{fig:SO-}.

\section{Discussion and Conclusion}
\label{sec:8}
In this article, we extend the results of Ref.~\cite{tan2025operatorspreadingrandomunitary} from circuits composed of unitary‐invariant two‐qudit gates to those composed of orthogonal‐ or symplectic‐invariant gates.As in the unitary case, the operator-spreading dynamics map at late
times onto a classical stochastic growth model governed by a
drift--diffusion equation, characterized by a butterfly velocity
$v_{\rm B}$ and a diffusion constant $\mathcal D$. We derive explicit expressions for \(v_{\rm B}\) and \(\mathcal{D}\) in both the orthogonal‐ and symplectic‐invariant ensembles.

There are four key differences from the random circuits with unitary invariance:
i) 
In the unitary‐invariant case, averaging directly yields a closed Markov process on the space of Pauli strings; with orthogonal or symplectic symmetry, one must first project onto the invariant ternary subspace (in which a summation over degrees of freedom with the same symmetry has taken place) before obtaining a description in terms of a classical Markovian stochastic process for a generic dimension $q$.
ii) Instead of approaching a binary (trivial vs.\ nontrivial) distribution, the system relaxes to a three‐valued (“trivial,” “even,” “odd”) distribution. As in the unitary-invariant case, the relaxation time only depends on the parameters of the local two-qudit gate distribution.
iii) The domain wall between the trivial and maximally mixed regions is perfectly sharp in Haar-unitary circuits but has a finite width in orthogonal- or symplectic-invariant circuits, even when these circuits are composed of Haar-distributed two-qudit gates.
iv) The bounds on the butterfly velocity depend qualitatively on the
determinant sector of the orthogonal gate ensemble. In particular,
the determinant-$-1$ sector, denoted by $\mathrm{SO}^{-}$, has a
strictly positive lower bound for $v_{\rm B}$ and, for $q=2$, can
support a butterfly velocity exceeding the Haar-orthogonal value. Remarkably, neither extremal ensemble in the SO$^-$ sector is Haar-distributed, emphasizing the relevance of studying random gate ensembles that go beyond the Haar-distributed case.

Using the criterion introduced above, these differences can be
separated into effects of symmetry and of gate randomness. The
ternary structure of the Pauli-string weights and the need, for
generic qudit dimension $q$, to project onto the invariant ternary
subspace before a closed Markovian description is obtained are
shared by all orthogonal- or symplectic-invariant gate distributions
and thus originate from the symmetry alone; the relaxation time
toward the ternary form, by contrast, is set by the gate
distribution. The finite domain-wall width has both origins: reduced
gate randomness widens the domain wall already in the
unitary-invariant class \cite{tan2025operatorspreadingrandomunitary}, while the orthogonal or symplectic
constraint renders the width finite even for Haar-distributed gates.
The bounds of the butterfly velocity for even $q$, finally, fall
outside this dichotomy: the strictly positive lower bound in the
$\mathrm{SO}^-$ sector holds for any gate distribution, yet the
$\mathrm{SO}^-$ ensembles share the same orthogonal invariance as
the special orthogonal ones. This bound is instead set by the
topology of the gate manifold --- the connected component of the
orthogonal group on which the distribution is supported --- which
thus constitutes a third factor controlling operator spreading,
independent of both symmetry and randomness.

The physical significance of this enhanced butterfly velocity is
clarified by comparison with dual-unitary circuits. Local operators propagate in these circuits at the maximal light-cone velocity, $v_{\rm B}=1$ \cite{Bertini2026Exactly}, but this propagation need not be correlated with entanglement generation. The SWAP circuit, for example, transports a local operator along the light cone without growth of its support, while mapping every product state to another product state and hence having zero entangling power \cite{Zanardi2000Entangling,Wang2003Entangling}. By contrast, the conjugation-invariant random circuits studied here belong to the generic setting in which operator spreading and entanglement growth are correlated \cite{nahum2017quantum,nahum_operator_2018,zhou2019emergent}. Within this setting, the enhanced butterfly velocity found here
corresponds to faster entanglement spreading and a greater effective
entangling power of the local dynamics. It further suggests a larger
spectral gap of the second-moment operator, consistent with the
relation between local entangling power and the second-moment gap
established for solvable unitary brickwork circuits in
Ref.~\cite{Suzuki2024More}. Such an enhancement would imply faster
convergence toward the corresponding unitary, orthogonal, or
symplectic $2$-design. 

We illustrated these results using two representative examples: the Haar–orthogonal circuit and the Brownian special orthogonal circuit, which interpolates between the trivial and Haar limits. In both cases, the domain-wall width remains finite. From the fast convergence of our calculations of the butterfly velocity with the approximation order $n$, we estimate a domain-wall width of $n_{\rm DW}\sim4$. Numerical simulations of the corresponding classical stochastic growth model—without assuming a maximally random distribution far from the end of the Pauli string—show excellent agreement with our analytical predictions. Furthermore, we find that the results in the large-$q$ limit coincide with those of the unitary case.

We identify three distinct temporal regimes for random quantum circuits with orthogonal- or symplectic-invariant gate distributions:  
\emph{Short‐time binary/ternary relaxation:} At early times, independent of the system size, the weight \(\rho_{pp}(t)\) relaxes to a form that depends solely on whether each local generalized Pauli operator is trivial, even, or odd.  
\emph{Intermediate thermalization:} A localized Pauli string spreads, with a finite-width domain wall separating high‐entropy and trivial regions, propagating at velocity \(v_{\rm B}\). The crossover time for a site at position \(x\) scales as \(n_{\rm DW}/v_{\rm B}\).  
\emph{Long‐time saturation:} At times of order \(L/v_{\rm B}\) (for system size \(L\)), the distribution has extended across the entire system, completing the scrambling process~\cite{xu_scrambling_2022,xu_locality_2019,styliaris_information_2021}.

The binary or ternary relaxation time is a circuit-level measure of
Haar-like scrambling, rather than a measure of how closely the
distribution of the local gates approximates the corresponding Haar
distribution. Rather, it quantifies how rapidly the many-gate
dynamics erases distinctions between Pauli strings within the same
binary or ternary class. The Haar-special-orthogonal ensemble for
$q=2$ illustrates this distinction: a single local transition matrix
has an additional eigenvalue-$1$ state outside the ternary subspace
and therefore cannot produce ternary relaxation when repeatedly
applied. This state is not preserved by the alternating even--odd
brickwork structure, however, so the full circuit can still relax to
ternary form.

More generally, convergence to a binary or ternary form from
arbitrary initial conditions implies that no additional steady states
retain information beyond the corresponding binary or ternary
variables. Conversely, if this relaxation fails at the level of the
full circuit, additional steady-state sectors are present. Continuous
symmetries provide a natural setting in which such additional
structure arises. In a ${\rm U}(1)$-symmetric circuit, for example,
the conserved charge distinguishes charge-density operators, such as
local Pauli-$Z$ operators, from generic nonidentity operators
\cite{rakovszky_diffusive_2018,khemani_operator_2018}. The resulting
deviations from the usual binary form are particularly apparent near
the operator front, even when the bulk behind it already appears
locally random-matrix-like. We leave the incorporation of continuous
symmetries, including ${\rm U}(1)$ and ${\rm SU}(2)$, into the present
framework for future investigation
\cite{mitsuhashi2025unitary}.


\begin{acknowledgements}
We thank Adam Chaou, Vatsal Dwivedi, Jens Eisert, Ming Huang, Adam Nahum, Ryotaro Suzuki, and Guoyi Zhu for valuable discussions. Special thanks go to Yiping Deng for suggestions on figure editing. Financial support was provided by the Einstein Stiftung Berlin (Einstein Research Unit on Quantum Devices) and by the Deutsche Forschungsgemeinschaft (DFG, German Research Foundation) - Project Number 277101999 - CRC TR 183 (project A03).
\end{acknowledgements}
\appendix
\renewcommand\appendixname{APPENDIX}
\makeatletter
\def\@hangfrom@section#1#2#3{%
 #1%
 \@if@empty{#2}{\MakeTextUppercase{#3}}{%
  #2\@if@empty{#3}{}{:\ \MakeTextUppercase{#3}}}%
}%
\def\@hangfroms@section#1#2{#1\MakeTextUppercase{#2}}%
\makeatother
\section{generalized Pauli matrices}
\label{app:1}
\subsection{Orthogonal case}
The generalized Pauli matrices $\sigma_a$, $a =0,1,\ldots,q^2-1$, form a basis for operators acting on a qudit, a quantum system with $q$ degrees of freedom $|m\rangle$, $m=0,1,\ldots,q-1$. We represent the index $a = (a',a'')$ as a pair of two integers $a'$, $a'' \in \mathbb{Z}_q$. The Weyl-Heisenberg choice for the complete family of $q^2$ independent generalized Pauli matrices is \cite{patera_pauli_1988}
\begin{equation}
\label{eq:pauliortho}
\sigma_{a}=\sum_{m=0}^{q-1}\omega^{ma''}\left|m+a'\right>\left<m\right|,
\end{equation}
where $\omega=\mathrm{e}^{\frac{2\pi i}{q}}$ and $m + a'$ is taken $\mod q$.

Without loss of generality, we may choose the qudit basis, such that the involution of Eq. \eqref{eq:involution} has $Z_x=\openone$. In that case, $Z_x \sigma_a^{\mathrm{T}} Z_x^{\dagger}$, see Eq. \eqref{eq:sigmaparity}, is simply the transpose, $\sigma_a^{\mathrm{T}}$. The transpose of the generalized Pauli matrices (\ref{eq:pauliortho}) is
\begin{equation}
  \left(\sigma_a\right)^{\mathrm{T}}=\sigma_{a ^\mathrm{T}} e^{-i \phi_a},  
  \label{eq:aaphi}
\end{equation}
with $a^{\mathrm{T}}=\left(-a^{\prime}, a^{\prime \prime}\right)$ and $\phi_a=2 \pi a^{\prime} a^{\prime \prime} / q$.

The generalized Pauli matrices satisfy the orthonormality relation
\begin{align}
  \tr\sigma_{a}\sigma_{b}^\dagger= q \delta_{a,b}
\end{align}
and the commutation relation
\begin{align}
  \sigma_{a}\sigma_{b}=\omega^{b'a''-a'b''}\sigma_{b}\sigma_{a}.
\end{align}
They also satisfy the completeness relation
\begin{equation}
  \sum_{a} \mbox{tr}\, A \sigma^\dagger_{a}\, \mbox{tr}\, B^\dagger\sigma_{a}= q \mbox{tr}\, A B^\dagger  \label{eq:complete}
\end{equation}
for arbitrary $q \times q$ matrices $A$ and $B$.

Since the evolution operators act on pairs of qudits, we will often consider the tensor product $\sigma_{a_x} \otimes \sigma_{a_y}$ of generalized Pauli matrices for two qudits at positions $x$ and $y$. With the shorthand notations $a = (a_x,a_y) = (a_x',a_x'',a_y',a_y'')$ and $\Sigma_{a} = \sigma_{a_x} \otimes \sigma_{a_y}$, we observe that the $q^2 \times q^2$ matrices $\Sigma_{a}$ satisfy the orthonormality relation
\begin{equation}
  \tr \Sigma_a \Sigma_b^{\dagger} = q^2 \delta_{a,b},
  \label{eq:SigmaOrthogonal}
\end{equation}
whereas their commutation relation is
\begin{equation}
  \Sigma_a \Sigma_b = \varphi(a,b) \Sigma_b \Sigma_a,
  \label{eq:SigmaCommutation}
\end{equation}
with 
\begin{equation}
  \varphi(a,b) = \omega^{b_x'a_x''-a_x'b_x'' + b_y'a_y''-a_y'b_y''}.
\end{equation}
The function $\varphi(a,b)$ satisfies
\begin{equation}
  \varphi(0,0,b_x,b_y) = 1
    \label{eq:varphi0}
\end{equation}
and
\begin{equation}
  \varphi(a,b) = \varphi(b,a)^*,
\end{equation}
\begin{equation}
  \varphi(-a,b) = \varphi(a,b)^*.
\end{equation}
The sum rules for a generic dimension $q^2$ are
\begin{align}
  \label{eq:phiidentities}
  \sum_{a_x \neq 0} \varphi(a_x,0,b_x,b_y) =&\,
  q^2 \delta_{b_x,0}-1, \nonumber \\
  \sum_{a_y \neq 0} \varphi(0,a_y,b_x,b_y) =&\,
  q^2 \delta_{b_y,0}-1, \\
  \nonumber
  \sum_{a_x \neq 0} \sum_{a_y \neq 0} \varphi(a_x,a_y,b_x,b_y) =&\,
   1 - q^2 (\delta_{b_x,0} + \delta_{b_y,0})
  \nonumber \\ &\, \mbox{} 
  +q^4 \delta_{b_x,0} \delta_{b_y,0}.
  \nonumber
\end{align}

In addition, there are sum rules that also involve the phase $\phi_a$ of Eq.\ (\ref{eq:aaphi}). For arbitrary $\bar a_x$, $\bar a_y \in \{-1,1\}$ one has
\begin{align}
\label{eq:orthphiidentities}
& \sum_{a_x \neq 0} \frac{1+\bar a_xe^{-i\phi_{a_x}}}{2}\varphi(a_x,0,b_x,b_y) = F(b_x, \bar a_x), \nonumber\\
& \sum_{a_y \neq 0} \frac{1+\bar a_y e^{-i\phi_{a_y}}}{2}\varphi(0,a_y,b_x,b_y) = F(b_y, \bar a_y), \nonumber \\
& \sum_{a_x \neq 0} \sum_{a_y \neq 0} \frac{1+\bar a_x e^{-i\phi_{a_x}}}{2} \frac{1+\bar a_y e^{-i\phi_{a_y}}}{2}\varphi(a_x,a_y,b_x,b_y) \nonumber \\
& \qquad = F(b_x, \bar a_x) F(b_y, \bar a_y).
\end{align}
where
\begin{equation}
\label{eq:F_definition}
F(b,\bar a) = \frac{q^2 \delta_{b,0}+q\bar a e^{-i\phi_{b}}-1-\bar a } {2}.
\end{equation}
The three relations in Eq.~\eqref{eq:orthphiidentities} are essential for deriving the stochastic equations governing the projected ternary weight $\rho_{\bar p}$, as defined in Sec.~\ref{sec:qgeneral}. For $q = 2$, these relations can be interpreted as sum rules for Pauli matrices with even and odd parities under transposition. 

For dimensions $q = 2^{m}$, it is convenient to adopt an alternative operator basis, which is constructed from tensor products of the standard $2 \times 2$ Pauli matrices. 
On this basis, all generalized Pauli operators have a well-defined parity under transposition. The projections onto the ternary weights depend on the numbers of basis operators of each parity that commute or anticommute with a fixed basis operator. These numbers are listed in Table~\ref{tab:operator_counts}.

\begin{table}
\begin{tabular*}{\linewidth}{c@{\extracolsep{\fill}} cccc}
 \hline\hline
 parity $\sigma$ & parity $\sigma'$ & count $\sigma: [\sigma,\sigma'] = 0$ & count $\sigma: \{ \sigma,\sigma'\} = 0$ \\ \hline
  e & e$'$ & $q(q+2)/4$ & $q^2/4$ \\
  e & o$'$ & $q^2/4$ & $q(q+2)/4$ \\
  o & e$'$ & $q(q-2)/4$ & $q^2/4$ \\
  o & o$'$ & $q^2/4$ & $q(q-2)/4$ \\ \hline\hline
\end{tabular*}
\caption{
\justifying 
\small Numbers of generalized Pauli operators $\sigma$ for $q=2^m$ ($q>2$) that commute or anticommute with a fixed Pauli matrix $\sigma' \neq \openone$, for fixed parities of $\sigma$ and $\sigma'$. A generalized Pauli operator $\sigma$ is called even (e) or odd (o) if $\sigma = \sigma^{\rm T}$ or $\sigma = -\sigma^{\rm T}$, respectively.}
\label{tab:operator_counts}
\end{table}

\subsection{Symplectic case}

Symplectic invariance applies to qudits of half-integer spin. In this case, the qudit dimension $q$ must be even. Without loss of generality, we may choose the qudit basis such that
the involution matrix $Z_x$ of Eq. \eqref{eq:involution} takes the form
\begin{equation}
\label{eq:antisymmetric}
  Z_x=\openone_{q / 2} \otimes \sigma_2  
\end{equation}
where $\sigma_2$ is a standard $2 \times 2$ Pauli matrix. In this case, $Z_x \sigma_{a_x}^{\mathrm{T}} Z_x^{\dagger}=\sigma_{a_x}^{\mathrm{R}}$ is the dual of the generalized Pauli matrix $\sigma_{a_x}$ \cite{mehta2004random}.

Consistent with Eq. \eqref{eq:antisymmetric}, for the generalized Pauli matrices for a qudit of half-integer spin, we choose tensor products of the Weyl-Heisenberg generalized Pauli matrices and standard $2 \times 2$ Pauli matrices. Hence, we use generalized Pauli matrices of the form
\begin{equation}
\label{eq:paulisymplectic}
\sigma_a =\sum_{m=0}^{q-1} \omega^{m a^{\prime \prime}}\left|m+a^{\prime}\right\rangle\langle m| \otimes \sigma_{a^{\prime \prime \prime}},
\end{equation}
where $a=\left(a^{\prime}, a^{\prime \prime}, a^{\prime \prime \prime}\right)$ and $\sigma_{a^{\prime \prime \prime}}$ is a standard $2 \times 2$ Pauli matrix with $a^{\prime \prime \prime} \in\{0,1,2,3\}$. The dual matrix of such a generalized Pauli matrix is given by
\begin{equation}
 \left(\sigma_a\right)^{\mathrm{R}}=\sigma_{a^{\mathrm{R}}} e^{-i \phi_a},   
\end{equation}
where $a^{\mathrm{R}}=\left(-a^{\prime}, a^{\prime \prime}, a^{\prime \prime \prime}\right)$ and $\phi_a=2 \pi a^{\prime} a^{\prime \prime} / 2+\pi(1-$ $\left.\delta_{\alpha^{\prime \prime \prime}, 0}\right)$.

The matrix $\Omega_{x, y}=Z_x \otimes Z_{y}$ that describes the involution 
operation for a two-qudit gate $\mathcal{U}_{x, y}$ is antisymmetric only if $Z_x$ is symmetric and $Z_{y}$ is antisymmetric or vice versa.
For definiteness, we here focus on the case that $Z_x$ is symmetric (i.e., the qudit $x$ has integer spin) and $Z_{y}$ is antisymmetric (i.e., the qudit $y$ has half-integer spin), so that $Z_x=\openone_q$ and $Z_{y}=\openone_{q / 2} \otimes \sigma_2$ and, hence, $\Omega_{x, y}=\openone_{q^2 / 2} \otimes \sigma_2$. The tensor products $\Sigma_a$ with $a=\left(a_x^{\prime}, a_x^{\prime \prime} ; a_{y}^{\prime}, a_{y}^{\prime \prime}, a_{y}^{\prime \prime \prime}\right)$ that span the two-qudit operators are of the form $\Sigma_a=\sigma_{a_x} \otimes \sigma_{a_{y}}$, with $\sigma_{a_x}$ a generalized Pauli matrix of the form \eqref{eq:pauliortho} and $\sigma_{a_{y}}$ a generalized Pauli matrix of the form \eqref{eq:paulisymplectic}. The two-qudit matrices $\Sigma_a$ again satisfy the orthonormality relation \eqref{eq:SigmaOrthogonal}. They also satisfy the commutation relation
\begin{equation}
  \Sigma_a \Sigma_b=\varphi(a, b) \Sigma_b \Sigma_a  
\end{equation}

with
\begin{align}
\label{eq:phisym}
\varphi(a, b)= & \omega^{b_x^{\prime} a_x^{\prime \prime}-a_x^{\prime} b_x^{\prime \prime}+b_{y}^{\prime} a_{y}^{\prime \prime}-a_{y}^{\prime} b_{y}^{\prime \prime}} \nonumber\\
& \times e^{i \pi\left(1-\delta_{a_{y}^{\prime \prime \prime},0}\right)\left(1-\delta_{b_{y}^{\prime \prime \prime}, 0}\right)\left(1-\delta_{a_{y}^{\prime \prime \prime}, b_{y}^{\prime \prime \prime}}\right)}
\end{align}

The combined phase factors of both the generalized Pauli and standard Pauli matrix components determine the commutation relation in the symplectic case. 
For an arbitrary local dimension $q$, the phases defined in \eqref{eq:phisym} satisfy sum rules identical to those presented in Eqs.~\eqref{eq:phiidentities}. Furthermore, they obey additional sum rules analogous to the relations in Eq.~\eqref{eq:orthphiidentities}:
\begin{align}
\label{eq:symphiidentities}
& \sum_{a_x \neq 0} \frac{1+\bar a_xe^{-i\phi_{a_x}}}{2}\varphi(a_x,0,b_x,b_{y}) = F(b_x, \bar a_x), \nonumber\\
& \sum_{a_{y} \neq 0} \frac{1+\bar a_ye^{-i\phi_{a_{y}}}}{2}\varphi(0,a_{y},b_x,b_{y})
= F'(b_{y}, \bar a_{y}), \nonumber \\
& \sum_{a_x \neq 0} \sum_{a_{y} \neq 0} \frac{1+\bar a_x e^{-i\phi_{a_x}}}{2} \frac{1+\bar a_{y}e^{-i\phi_{a_{y}}}}{2}
  \varphi(a_x,a_{y},b_x,b_{y}) \nonumber\\
& \qquad = F(b_x, \bar a_x) F'(b_{y}, \bar a_{y}).
\end{align}
Here, the function $F(b,\bar a)$ is given in \eqref{eq:F_definition}, while
\begin{equation}
\label{eq:Fq_definition}
F'(b,\bar a) = \frac{q^2 \delta_{b,0}-q\bar ae^{-i\phi_{b}}-1-\bar a } {2}.
\end{equation}
For $q = 2$, the relations in Eq.~\eqref{eq:symphiidentities} can be interpreted as sum rules for Pauli matrices of even and odd parities within the context of the duality operation. 

For qudit dimension $q=2^m$, it is again advantageous to employ an operator basis consisting of tensor products of Pauli matrices, so that all generalized Pauli operators have a well-defined parity under the duality operation. The numbers of basis matrices that commute or anticommute with a given basis matrix of even or odd parity are summarized in Table~\ref{tab:symplectic_counts}.

\begin{table}
\begin{tabular*}{\linewidth}{c@{\extracolsep{\fill}} cccc}
 \hline\hline
 parity $\sigma$ & parity $\sigma'$ & count $\sigma: [\sigma,\sigma'] = 0$ & count $\sigma: \{ \sigma,\sigma'\} = 0$ \\ \hline
  e & e$'$ & $q(q-2)/4$ & $q^2/4$ \\
  e & o$'$ & $q^2/4$ & $q(q-2)/4$ \\
  o & e$'$ & $q(q+2)/4$ & $q^2/4$ \\
  o & o$'$ & $q^2/4$ & $q(q+2)/4$ \\ \hline\hline
\end{tabular*}
\caption{
\justifying 
\small Numbers of generalized Pauli operators $\sigma$ for $q=2^m$ ($q>2$) that commute or anticommute with a fixed Pauli matrix $\sigma' \neq \openone$, for fixed parities of $\sigma$ and $\sigma'$. A generalized Pauli operator $\sigma$ is called even (e) or odd (o) if $\sigma = \sigma^{\rm R}$ or $\sigma = -\sigma^{\rm R}$, respectively.}
\label{tab:symplectic_counts}
\end{table}


\section{calculation of $\langle W_{ap} W_{bq}^* \rangle$}
\label{app:3}

The time evolution of  $\rho_{pq}(t) = \langle \gamma_p(t) \gamma^*_q(t)\rangle$ involves the average $\langle W_{ap} W_{bq}^* \rangle$, see Eq.\ (\ref{eq:gammasqavg0}). The ensemble average $\langle W_{ap}W^{*}_{bq} \rangle$ factorizes as a product of two-qudit averages,
\begin{equation}
\label{eq:layerW}
\left<W_{ap}W^{*}_{bq}\right>=\prod_{\text{x even}}
\left\{\begin{array}{l}
\left<\mathcal{W}_{a p; x,x+1} \mathcal{W}^*_{b q; x,x+1}\right>\text { $t$ even, } \\
\left<\mathcal{W}_{a p; x-1, x }\mathcal{W}^*_{b q; x-1, x } \right> \text { $t$ odd, }
\end{array}\right.
\end{equation}
where the matrices ${\cal W}_{ap;x,y}$ and ${\cal W}_{bq;x,y}$ (with $y = x+1$ or $y = x - 1$) are expressed in terms of the pair-qudit evolution matrices ${\cal U}_{x,y}$,
\begin{align}
  \label{eq:WUapp}
  {\cal W}_{ap;x,y} =&\,
  \frac{1}{q^2}
  \mbox{tr}\, {\cal U}^\dagger_{x,y}
  \Sigma_a
  {\cal U}_{x,y}
  \Sigma_p^{\dagger}.
\end{align}
Here $\Sigma_a = \sigma_{a_x} \otimes \sigma_{a_y}$ and $\Sigma_p = \sigma_{p_x} \otimes \sigma_{p_y}$, see App.\ \ref{app:1}.

In Sec.~\ref{sec:5a}, we presented the two-qudit transition matrices
for the CUE, COE, CSE, and Haar orthogonal and symplectic ensembles
without giving their derivations. Since these calculations follow
the same general procedure, we use the Haar orthogonal ensemble as a
representative example to illustrate the diagrammatic evaluation.
This example also provides a reference point for the more involved
calculation of orthogonally or symplectically invariant gate
distributions presented below.

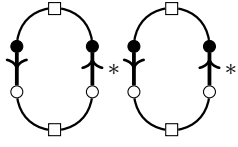
\begin{figure}[t]
\vspace{0cm}
    \centering
    \begin{tikzpicture}[
        every node/.style={
            draw,
            circle,
            minimum size=4.5pt,
            inner sep=0pt
        }
    ]


        \node[fill=black] (n1) at (0,0) {};
        \node[fill=black] (n3) at (1,0) {};

        \node[fill=white] (n2) at (0,-0.6) {};
        \node[fill=white] (n4) at (1,-0.6) {};

        \node[
            draw,
            rectangle,
            minimum size=4.5pt,
            fill=white
        ] (n6) at (0.5,0.5) {};

        \node[
            draw,
            rectangle,
            minimum size=4.5pt,
            fill=white
        ] (n5) at (0.5,-1.1) {};

        \draw[
            middlearrow={>[length=1.4mm,line width=1pt]},
            black,
            line width=.5mm
        ] (n1) -- (n2);

        \draw[
            middlearrow={<[length=1.4mm,line width=1pt]},
            black,
            line width=.5mm
        ] (n3) -- node[
            draw=none,
            right=1.5mm,
            inner sep=0pt
        ] {$\ast$} (n4);

        \draw[line width=.3mm]
            (n1) to[bend left=35] (n6.west);

        \draw[line width=.3mm]
            (n6.east) to[bend left=35] (n3);

        \draw[line width=.3mm]
            (n2) to[bend right=35] (n5.west);

        \draw[line width=.3mm]
            (n5.east) to[bend right=35] (n4);


        \node[fill=black] (m1) at (1.55,0) {};
        \node[fill=black] (m3) at (2.55,0) {};

        \node[fill=white] (m2) at (1.55,-0.6) {};
        \node[fill=white] (m4) at (2.55,-0.6) {};

        \node[
            draw,
            rectangle,
            minimum size=4.5pt,
            fill=white
        ] (m6) at (2.05,0.5) {};

        \node[
            draw,
            rectangle,
            minimum size=4.5pt,
            fill=white
        ] (m5) at (2.05,-1.1) {};

        \draw[
            middlearrow={>[length=1.4mm,line width=1pt]},
            black,
            line width=.5mm
        ] (m1) -- (m2);

        \draw[
            middlearrow={<[length=1.4mm,line width=1pt]},
            black,
            line width=.5mm
        ] (m3) -- node[
            draw=none,
            right=1.5mm,
            inner sep=0pt
        ] {$\ast$} (m4);

        \draw[line width=.3mm]
            (m1) to[bend left=35] (m6.west);

        \draw[line width=.3mm]
            (m6.east) to[bend left=35] (m3);

        \draw[line width=.3mm]
            (m2) to[bend right=35] (m5.west);

        \draw[line width=.3mm]
            (m5.east) to[bend right=35] (m4);

    \end{tikzpicture}
\caption{ \justifying \small
Diagrammatic representation of the product $W_{ap} W_{bq}^*$ before taking the ensemble average.}
\label{fig:Wdiagram}
\end{figure}

Both the Haar-orthogonal example and the subsequent calculation for
orthogonally or symplectically invariant gate distributions use a
generalization of the diagrammatic method developed by Beenakker and
one of the authors for averages over the unitary group
\cite{brouwer_diagrammatic_1996} (following a structure originally
developed by Weingarten \cite{weingarten1978} and Samuel
\cite{samuel1980} for large $q$ and finite $q$, respectively) to
integrations over the orthogonal and symplectic groups
\cite{collins2006integration,MATSUMOTO_2013}.
In this diagrammatic approach, elements of the two-qudit evolution matrix $\mathcal{U}_{x,y}$ and
its complex conjugate are represented by single directed lines,
\begin{equation}
\label{eq:Udiagr}
\begin{tikzpicture}[node distance=8mm]
    \node(n1) at (0,0.1) {};
    \node[below=of n1](n2) at (0,0) {};
    \draw[
    middlearrow={>[length=1.4mm,line width=1pt]},
    black,
    line width=.5mm
    ]
    (n1) to node[left,draw=none]{$\mathcal{U}=\ $} (n2);

    \node(n3) at (2,0.1) {};
    \node[below=of n1](n4) at (2,0) {};
    \draw[
    middlearrow={<[length=1.4mm,line width=1pt]},
    black,
    line width=.5mm
    ]
    (n3) to
    node[left,draw=none]{$\mathcal{U}^*=\ $}
    node[auto,draw=none]{*}
    (n4);
\end{tikzpicture}.
\end{equation}
The generalized Pauli matrices
$\Sigma_a$, $\Sigma_b$, $\Sigma_p$, and $\Sigma_q$ are represented
by solid lines terminating in open squares. Spatial indices are
omitted from the diagrams for simplicity. The product $W_{ap} W_{bq}^*$ is thus represented by diagrams with four directed lines and four lines with open squares, whereas the joining of lines indicates matrix multiplication, see Fig.\ \ref{fig:Wdiagram}. 
\subsection{Special case: Haar-orthogonal and Haar-symplectic ensembles}

We first consider the Haar orthogonal ensemble, for which
$\mathcal{U}_{x,y}$ itself is a $q^2\times q^2$
Haar-distributed orthogonal matrix. The average over the orthogonal group is evaluated by
pairing the first and second indices of the matrix elements of
$\mathcal{U}$ and/or $\mathcal{U}^*$. Diagrammatically, these pairings
are represented by dashed contraction lines joining the corresponding
filled dots, which denote first indices, and open dots, which denote
second indices. When evaluating a diagram, each closed loop consisting of solid lines with squares and dashed contraction lines contributes a trace of the corresponding generalized Pauli matrices or a product of them. The loop structure consisting of solid lines with arrows an dashed contraction lines determines the Weingarten coefficient: $V_2$ if there is one loop involving all four dashed contraction lines, $V_{1,1}$ if there are two loops involving two dashed lines each \cite{brouwer_diagrammatic_1996}.

\begin{figure}
\vspace{-2cm}
\centering
\includegraphics[
scale=0.4,
trim={8bp 20bp 8bp 10bp},
clip
]{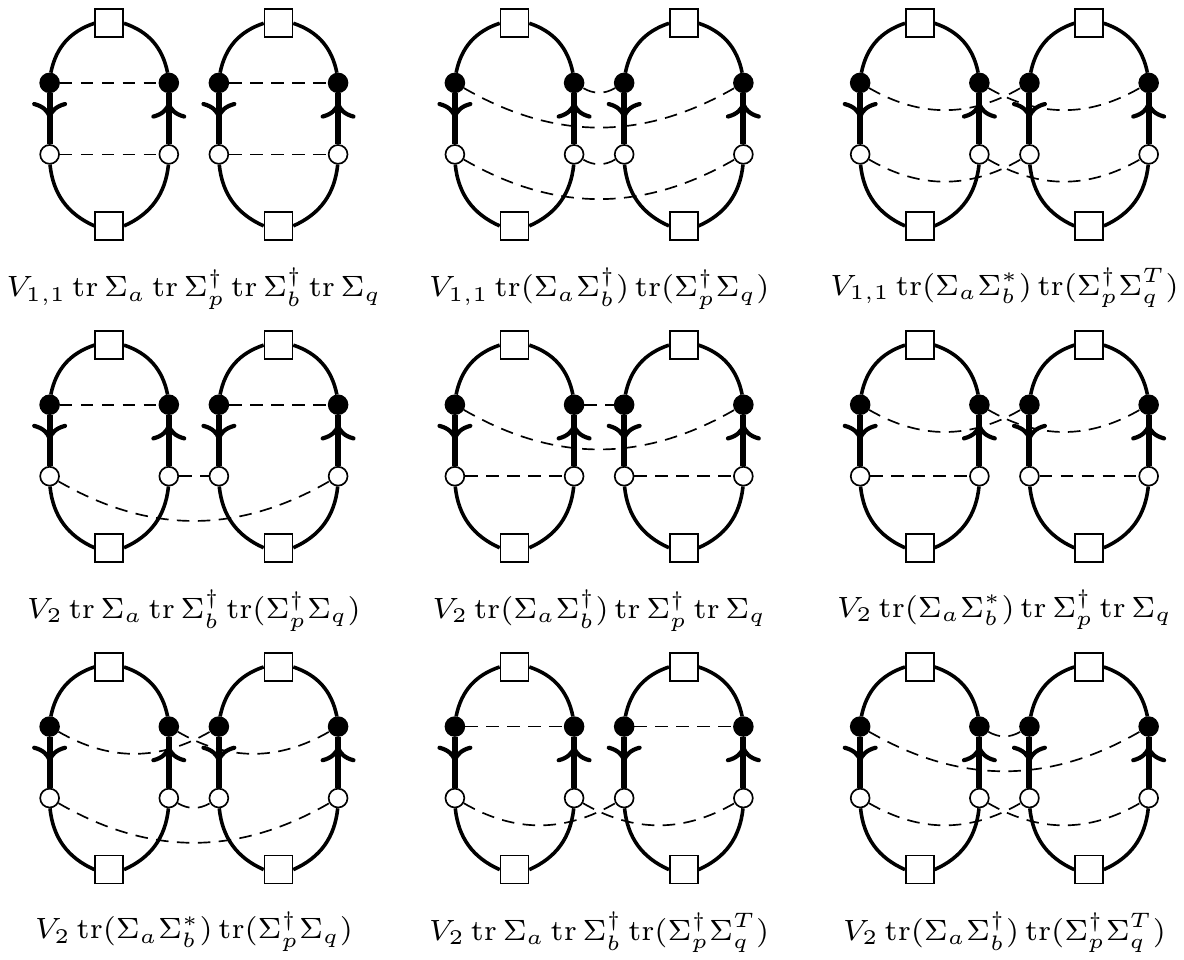}
\vspace{-2cm}
\caption{ \justifying \small
Diagrams contributing to
$\langle\mathcal{W}_{ap;x,x+1}
\mathcal{W}_{bq;x,x+1}^*\rangle$
for Haar-distributed orthogonal gates.
The Weingarten numbers $V_{1,1}$ and $V_2$ are given in Eq.\ (\ref{eq:V11V2}).
}
\label{fig:Haarcontraction}
\end{figure}

The diagrams contributing to
$\langle\mathcal{W}_{ap;x,x+1}
\mathcal{W}_{bq;x,x+1}^*\rangle$
are shown in Fig.~\ref{fig:Haarcontraction}, as well as the evaluation of each diagram following the rules outlined above. Since an orthogonal matrix is real,
$\mathcal{U}_{x,y}=\mathcal{U}_{x,y}^*$, and the asterisks may be
omitted in these diagrams.
The two Weingarten numbers entering this calculation are
\begin{align}
\label{eq:V11V2}
V_{1,1}
&=
\frac{q^2+1}
{q^2(q^2-1)(q^2+2)},
\nonumber\\
V_2
&=
-\frac{1}
{q^2(q^2-1)(q^2+2)}.
\end{align}
Summing the contributions of all diagrams and evaluating the traces of products of generalized Pauli matrices, we verify that the average $\langle\mathcal{W}_{ap;x,x+1}
\mathcal{W}_{bq;x,x+1}^*\rangle$ for
qudit dimensions $q=2^m$ has the property of
Eq.~\eqref{eq:WW1}. The resulting transition probability matrix
$\langle|\mathcal{W}_{ap;x,y}|^2\rangle$ is then given by
Eqs.~\eqref{eq:P1Orthogonal} and \eqref{eq:P23Orthogonal}.
For the symplectic case the diagrams are the same as those shown in Fig.\ \ref{fig:Haarcontraction} and one arrives at the same result as for the orthogonal case up to the substitution $q^2 \to -q^2$.
The transition matrices for the COE and CSE ensembles, see Eq.\ (\ref{eq:PCOECSE}), can be obtained by applying the corresponding
integration rules in the same manner \cite{hunter-jones_operator_2018}.

\subsection{General case: orthogonal- and symplectic-invariant gate distribution}

We now turn to the general case of a two-qudit evolution matrix
$\mathcal{U}_{x,y}$ whose distribution is invariant under
orthogonal or symplectic changes of basis. In contrast to the Haar
case, $\mathcal{U}_{x,y}$ is not itself averaged over a Haar
ensemble. Instead, its basis invariance allows the replacement
\begin{equation}
\mathcal{U}_{x,y}
\to
\mathcal{V}_{x,y}
\mathcal{U}_{x,y}
\mathcal{V}_{x,y}^{\dagger},
\label{eq:Usubst}
\end{equation}
where $\mathcal{V}_{x,y}$ is a statistically independent
$q^2\times q^2$ Haar-distributed orthogonal or symplectic matrix,
followed by averaging over $\mathcal{V}_{x,y}$. The resulting second moment contains more Haar-distributed matrix
factors than in the preceding Haar-orthogonal example, leading to a
substantially larger number of possible contraction patterns.
  
We retain the single-line convention introduced above for
$\mathcal{U}$ and $\mathcal{U}^*$ and represent
$\mathcal{V}$ and $\mathcal{V}^*$ by double lines,
\begin{equation}
\label{eq:Vdiagr}
\begin{tikzpicture}[every node/.style={draw, circle, minimum size=4.5pt},node distance=6mm]
    \node[fill, inner sep=0pt, blue](n1) at (0,0.1) {};
    \node[inner sep=0pt,below=of n1,blue] (n2) at (0,0) {};
     \draw[double distance between line centers=0.9mm, middlearrow={>[length=1.4mm,line width=1pt,open]}, blue, line width=.2mm] (n1) to node [left, draw=none]{${\cal V}=$} (n2);
     
    \node[fill, inner sep=0pt, blue](n3) at (2,0.1) {};
    \node[inner sep=0pt,below=of n1,blue] (n4) at (2,0) {};
    \draw[double distance between line centers=0.9mm, middlearrow={<[length=1.4mm,line width=1pt,open]}, blue, line width=.2mm]  (n3) to node [left, draw=none]{${\cal V}^*=$} node [auto, draw=none]{*.} (n4);
\end{tikzpicture}
\end{equation}
where the spatial indices $x$ and $y$ are omitted for simplicity.  

Only terms in which the first and second indices of all factors
$\mathcal{V}$ and $\mathcal{V}^*$ coincide pairwise contribute to
$\langle\mathcal{W}_{ap;x,y}
\mathcal{W}_{bq;x,y}^*\rangle$. As in the Haar-orthogonal
calculation, these pairings are represented diagrammatically by
dashed contraction lines joining the corresponding filled dots,
which denote first indices, and open dots, which denote second
indices.

Following the approach in Ref.~\cite{tan2025operatorspreadingrandomunitary}, we consider the two loop structures in the diagrams separately:  
(i) loops containing only dashed and doubled lines, and  
(ii) loops composed of dashed and solid lines.  
Loops of type (i) determine the Weingarten function. These are closed loops formed by alternating dashed lines (contractions) and double lines (representing the Haar matrices ${\cal V}$ and ${\cal V}^*$). Defining the ``length'' $c_i$ of each loop as the number of double lines associated with ${\cal V}$, with $i=1,\ldots,k$ and $k$ the total number of loops, the corresponding Weingarten number is denoted by $V_{c_1,\ldots,c_k}$.  
The five Weingarten numbers that appear in our calculation for the orthogonal-invariant case are
\begin{align}
\label{eq:unified_weingarten}
V_{1,1,2} =&\, \frac{- q^6 - 6q^4 - 3 q^2 + 6}{D(q)}, \nonumber \\
V_{2,2} =&\, \frac{q^4 + 5 q^2 + 18}{D(q)}, \nonumber \\
V_{1,3} =&\, \frac{2q^4 + 8 q^2}{D(q)}, \\
V_{1,1,1,1} =&\, \frac{q^8 + 7 q^6 + q^4 - 35 q^2 - 6}{D(q)}, \nonumber \\
V_4 =&\, \frac{-5 q^2 - 6}{D(q)}, \nonumber
\end{align}
where
\begin{equation}
  \label{eq:Dqdef}
D(q) = \begin{multlined}[t]
    ( q^2-3)( q^2-2)( q^2-1) q^2( q^2+1) \\
    \times ( q^2+2)( q^2+4)( q^2+6).
    \end{multlined}
\end{equation}
Weingarten numbers for the symplectic-invariant case are obtained from Eqs.\ (\ref{eq:unified_weingarten}) and (\ref{eq:Dqdef}) by the substitution $q^2 \to - q^2$ \cite{collins2006integration,MATSUMOTO_2013}. (Recall that the integration is over orthogonal or symplectic matrices ${\cal V}_{x,y}$ of size $q^2 \times q^2$.)

Loops of type (ii) are built from alternating single directed lines (for ${\cal U}$ and ${\cal U}^*$) and dashed lines connecting open dots, as well as loops consisting of alternating single lines ending in open squares (for $\Sigma_a$, $\Sigma_b$, $\Sigma_p$, and $\Sigma_q$) and dashed lines connecting open dots. Each such loop contributes a trace over the product of matrices ${\cal U}$ and ${\cal U}^*$ or over generalized Pauli matrices $\Sigma_a$, $\Sigma_b$, $\Sigma_p$, and $\Sigma_q$, following their order within the loop.

\begin{figure*}[t!]
    \centering
\includegraphics[trim=0.5cm 8cm 0.5cm 8cm, clip, scale=0.8]{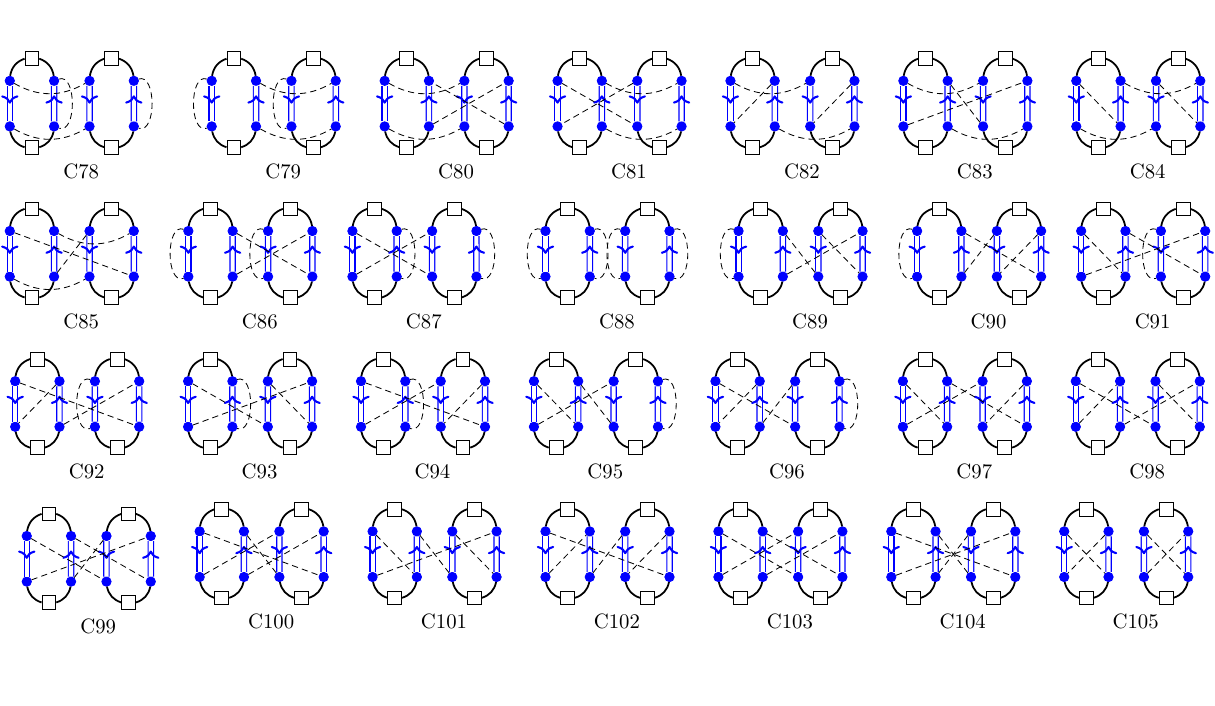}
\caption{
    \justifying \small
Contractions $C_j$ of closed dots (corresponding to the generalized Pauli matrices) and their contributions to $\langle {\cal W}_{ap;x,y} {\cal W}_{bq;x,y} \rangle$. The coefficients $B_j$ ($j = 1, 2, \ldots, 105$) represent the contributions arising from the contractions of open dots, each of which involves a sum over 105 distinct terms. In the figure, each double line with two closed dots represents the product ${\cal V}_{x,y} {\cal U}_{x,y} {\cal V}_{x,y}^{\dagger}$, cf. Eqs.\ (\ref{eq:Udiagr}) and (\ref{eq:Vdiagr}).
}
    \label{fig:diagrams}
\end{figure*}

In the diagrammatic evaluation of the second moment $\langle \mathcal{W}_{ap;x,y} \mathcal{W}^*_{bq;x,y} \rangle$, there are 105 possible contractions for the open dots and 105 for the closed dots, yielding a total of $105^2 = 11\,025$ diagrams. The diagrams are the same for the orthogonal and symplectic cases, but the outcome of the calculation of the diagrams is not because of the different Weingarten numbers. Each set of 105 contraction patterns corresponds to the diagrams of the Brauer algebra \cite{brauer1937algebras}. We organize the diagrams according to the contractions of the closed dots, which determines the loops of type (ii) and, hence, the appearance of Kronecker deltas involving the indices $a$, $b$, $p$, and $q$. The 105 possible contraction patterns and their corresponding contributions are shown in Fig.\ \ref{fig:diagrams}. Each contribution $C_j$ contains a factor $B_j$ ($j=1,2,\ldots,105$), representing the contractions of the open dots, along with a product of traces over the generalized Pauli matrices $\Sigma_a$, $\Sigma_b^{\dagger}$, $\Sigma_p^{\dagger}$, and $\Sigma_q$.
\begin{figure*}
    \centering \includegraphics[scale=0.8]{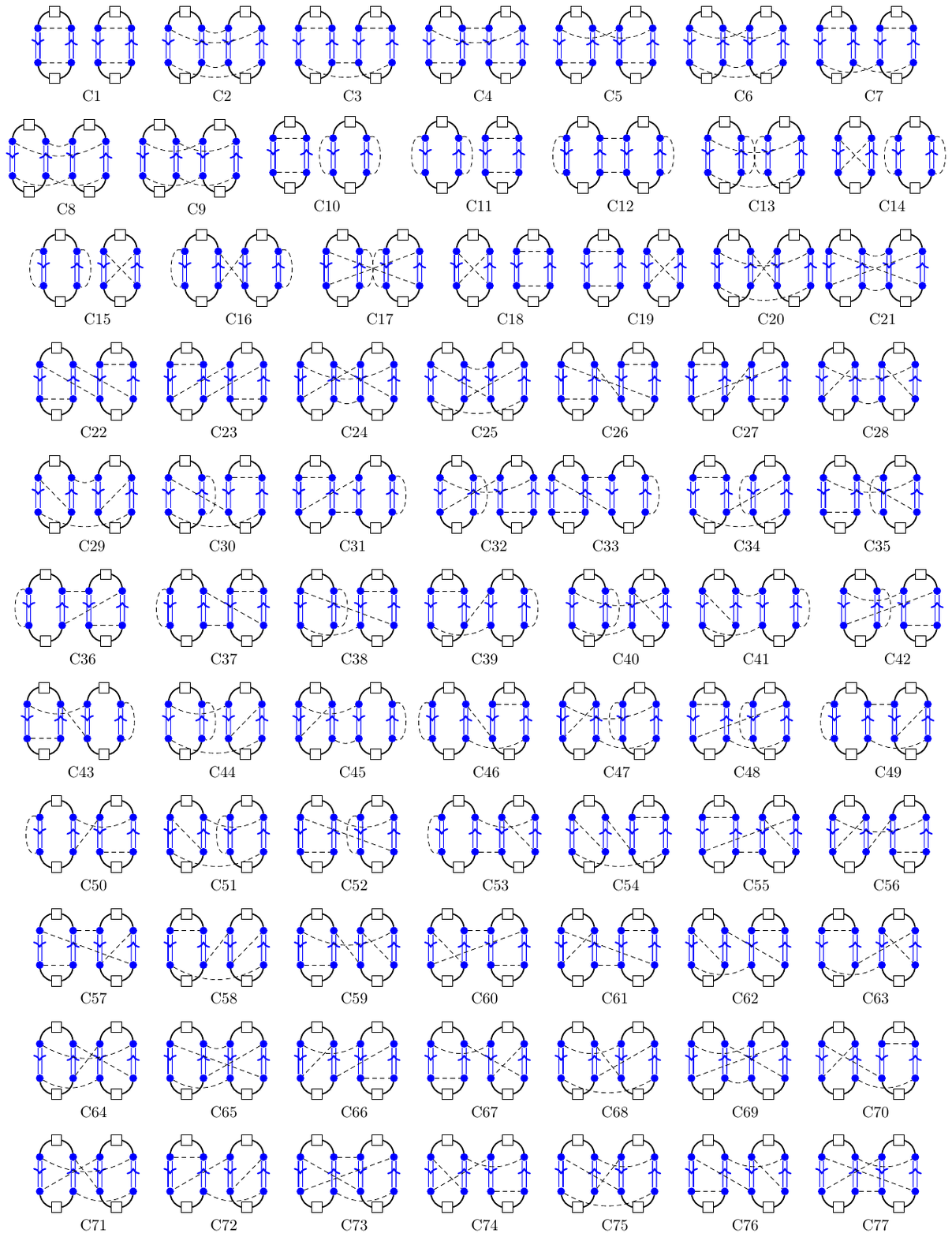}
\end{figure*}
The expressions $C_j$ of those diagrams are 
\begin{widetext}
\begin{align*}
C_1 &= B_1\tr\Sigma_a\tr\Sigma_p^\dagger\tr\Sigma_b^\dagger\tr\Sigma_q & C_2 &= B_2\tr\Sigma_a\Sigma_b^\dagger\tr\Sigma_p^\dagger\Sigma_q & C_3 &= B_3\tr\Sigma_a\tr\Sigma_b^\dagger\tr\Sigma_p^\dagger\Sigma_q \\
C_4 &= B_4\tr\Sigma_a\Sigma_b^\dagger\tr\Sigma_p^\dagger\tr\Sigma_q & C_5 &= B_5\tr\Sigma_a\Sigma_b^*\tr\Sigma_p^\dagger\tr\Sigma_q & C_6 &= B_6\tr\Sigma_a\Sigma_b^*\tr\Sigma_p^\dagger\Sigma_q \\
C_7 &= B_7\tr\Sigma_a\tr\Sigma_b^\dagger\tr\Sigma_p^\dagger\Sigma_q^T & C_8 &= B_8\tr\Sigma_a\Sigma_b^\dagger\tr\Sigma_p^\dagger\Sigma_q^T & C_9 &= B_9\tr\Sigma_a\Sigma_b^*\tr\Sigma_p^\dagger\Sigma_q^T \\
C_{10} &= B_{10}\tr\Sigma_a\tr\Sigma_p^\dagger\tr\Sigma_b^\dagger\Sigma_q & C_{11} &= B_{11}\tr\Sigma_a\Sigma_p^\dagger\tr\Sigma_b^\dagger\tr\Sigma_q & C_{12} &= B_{12}\tr\Sigma_a\Sigma_p^\dagger\Sigma_q\Sigma_b^\dagger \\
C_{13} &= B_{13}\tr\Sigma_a\Sigma_b^\dagger\Sigma_q\Sigma_p^\dagger & C_{14} &= B_{14}\tr\Sigma_a\Sigma_p^*\tr\Sigma_b^\dagger\Sigma_q & C_{15} &= B_{15}\tr\Sigma_a\Sigma_p^\dagger\tr\Sigma_b^\dagger\Sigma_q^T \\
C_{16} &= B_{16}\tr\Sigma_a\Sigma_p^\dagger\Sigma_b^*\Sigma_q^T & C_{17} &= B_{17}\tr\Sigma_a\Sigma_q^T\Sigma_b^*\Sigma_p^\dagger & C_{18} &= B_{18}\tr\Sigma_a\Sigma_p^*\tr\Sigma_b^\dagger\tr\Sigma_q \\
C_{19} &= B_{19}\tr\Sigma_a\tr\Sigma_p^\dagger\tr\Sigma_b^\dagger\Sigma_q^T & C_{20} &= B_{20}\tr\Sigma_a\Sigma_b^\dagger\Sigma_p^*\Sigma_q^T & C_{21} &= B_{21}\tr\Sigma_a\Sigma_q^T\Sigma_p^*\Sigma_b^\dagger \\
C_{22} &= B_{22}\tr\Sigma_a\Sigma_q\tr\Sigma_b^\dagger\tr\Sigma_p^\dagger & C_{23} &= B_{23}\tr\Sigma_a\tr\Sigma_p^\dagger\Sigma_b^\dagger\tr\Sigma_q & C_{24} &= B_{24}\tr\Sigma_a\Sigma_b^\dagger\Sigma_p^\dagger\Sigma_q \\
C_{25} &= B_{25}\tr\Sigma_a\Sigma_q\Sigma_p^\dagger\Sigma_b^\dagger & C_{26} &= B_{26}\tr\Sigma_a\Sigma_q^T\tr\Sigma_p^\dagger\tr\Sigma_b^\dagger & C_{27} &= B_{27}\tr\Sigma_a\tr\Sigma_p^\dagger\Sigma_b^*\tr\Sigma_q \\
C_{28} &= B_{28}\tr\Sigma_a\Sigma_b^\dagger\Sigma_q^T\Sigma_p^* & C_{29} &= B_{29}\tr\Sigma_a\Sigma_p^*\Sigma_q^T\Sigma_b^\dagger & C_{30} &= B_{30}\tr\Sigma_a\Sigma_q\Sigma_p^\dagger\tr\Sigma_b^\dagger \\
C_{31} &= B_{31}\tr\Sigma_a\tr\Sigma_p^\dagger\Sigma_q\Sigma_b^\dagger & C_{32} &= B_{32}\tr\Sigma_a\Sigma_b^\dagger\Sigma_p^\dagger\tr\Sigma_q & C_{33} &= B_{33}\tr\Sigma_a\Sigma_q\Sigma_b^\dagger\tr\Sigma_p^\dagger \\
C_{34} &= B_{34}\tr\Sigma_a\tr\Sigma_p^\dagger\Sigma_b^\dagger\Sigma_q & C_{35} &= B_{35}\tr\Sigma_a\Sigma_b^\dagger\Sigma_q\tr\Sigma_p^\dagger & C_{36} &= B_{36}\tr\Sigma_a\Sigma_p^\dagger\Sigma_b^\dagger\tr\Sigma_q \\
C_{37} &= B_{37}\tr\Sigma_a\Sigma_p^\dagger\Sigma_q\tr\Sigma_b^\dagger & C_{38} &= B_{38}\tr\Sigma_a\Sigma_q^T\Sigma_p^\dagger\tr\Sigma_b^\dagger & C_{39} &= B_{39}\tr\Sigma_a\tr\Sigma_p^\dagger\Sigma_b^*\Sigma_q^T \\
C_{40} &= B_{40}\tr\Sigma_a\Sigma_b^\dagger\Sigma_q^T\Sigma_p^\dagger & C_{41} &= B_{41}\tr\Sigma_a\Sigma_p^*\Sigma_q\Sigma_b^\dagger & C_{42} &= B_{42}\tr\Sigma_a\Sigma_b^*\Sigma_p^\dagger\tr\Sigma_q \\
C_{43} &= B_{43}\tr\Sigma_a\Sigma_b^*\Sigma_q^T\tr\Sigma_p^\dagger & C_{44} &= B_{44}\tr\Sigma_a\Sigma_b^*\Sigma_q\Sigma_p^\dagger & C_{45} &= B_{45}\tr\Sigma_a\Sigma_b^*\Sigma_q^T\Sigma_p^* \\
C_{46} &= B_{46}\tr\Sigma_a\Sigma_p^\dagger\Sigma_q^T\tr\Sigma_b^\dagger & C_{47} &= B_{47}\tr\Sigma_a\Sigma_b^\dagger\Sigma_q\Sigma_p^* & C_{48} &= B_{48}\tr\Sigma_a\tr\Sigma_p^\dagger\Sigma_q^T\Sigma_b^* \\
C_{49} &= B_{49}\tr\Sigma_a\Sigma_p^\dagger\Sigma_q^T\Sigma_b^\dagger & C_{50} &= B_{50}\tr\Sigma_a\Sigma_p^\dagger\Sigma_b^*\tr\Sigma_q & C_{51} &= B_{51}\tr\Sigma_a\Sigma_p^*\Sigma_q^T\Sigma_b^* \\
C_{52} &= B_{52}\tr\Sigma_a\Sigma_q^T\Sigma_b^*\tr\Sigma_p^\dagger & C_{53} &= B_{53}\tr\Sigma_a\Sigma_p^\dagger\Sigma_q\Sigma_b^* & C_{54} &= B_{54}\tr\Sigma_a\Sigma_p^*\Sigma_q^T\tr\Sigma_b^\dagger \\
C_{55} &= B_{55}\tr\Sigma_a\tr\Sigma_p^\dagger\Sigma_q\Sigma_b^* & C_{56} &= B_{56}\tr\Sigma_a\Sigma_b^\dagger\Sigma_p^*\tr\Sigma_q & C_{57} &= B_{57}\tr\Sigma_a\Sigma_q^T\Sigma_b^\dagger\tr\Sigma_p^\dagger \\
C_{58} &= B_{58}\tr\Sigma_a\tr\Sigma_p^\dagger\Sigma_b^*\Sigma_q & C_{59} &= B_{59}\tr\Sigma_a\Sigma_b^\dagger\Sigma_q^T\tr\Sigma_p^\dagger & C_{60} &= B_{60}\tr\Sigma_a\Sigma_p^*\Sigma_b^\dagger\tr\Sigma_q \\
C_{61} &= B_{61}\tr\Sigma_a\Sigma_q^T\Sigma_p^*\tr\Sigma_b^\dagger & C_{62} &= B_{62}\tr\Sigma_a\Sigma_p^*\Sigma_q\tr\Sigma_b^\dagger & C_{63} &= B_{63}\tr\Sigma_a\tr\Sigma_p^\dagger\Sigma_b^\dagger\Sigma_q^T \\
C_{64} &= B_{64}\tr\Sigma_a\Sigma_b^\dagger\Sigma_p^*\Sigma_q & C_{65} &= B_{65}\tr\Sigma_a\Sigma_q^T\Sigma_p^\dagger\Sigma_b^\dagger & C_{66} &= B_{66}\tr\Sigma_a\Sigma_b^*\Sigma_p^*\tr\Sigma_q \\
C_{67} &= B_{67}\tr\Sigma_a\Sigma_b^*\Sigma_q\tr\Sigma_p^\dagger & C_{68} &= B_{68}\tr\Sigma_a\Sigma_b^*\Sigma_p^*\Sigma_q^T & C_{69} &= B_{69}\tr\Sigma_a\Sigma_b^*\Sigma_p^\dagger\Sigma_q \\
C_{70} &= B_{70}\tr\Sigma_a\Sigma_q\Sigma_p^*\tr\Sigma_b^\dagger & C_{71} &= B_{71}\tr\Sigma_a\Sigma_b^\dagger\Sigma_p^\dagger\Sigma_q^T & C_{72} &= B_{72}\tr\Sigma_a\tr\Sigma_p^\dagger\Sigma_q^T\Sigma_b^\dagger \\
C_{73} &= B_{73}\tr\Sigma_a\Sigma_q\Sigma_p^*\Sigma_b^\dagger & C_{74} &= B_{74}\tr\Sigma_a\Sigma_p^*\Sigma_b^*\tr\Sigma_q & C_{75} &= B_{75}\tr\Sigma_a\Sigma_q\Sigma_p^\dagger\Sigma_b^* \\
C_{76} &= B_{76}\tr\Sigma_a\Sigma_q\Sigma_b^*\tr\Sigma_p^\dagger & C_{77} &= B_{77}\tr\Sigma_a\Sigma_q^T\Sigma_p^*\Sigma_b^* & C_{78} &= B_{78}\tr\Sigma_a\Sigma_b^*\Sigma_q^T\Sigma_p^\dagger \\
C_{79} &= B_{79}\tr\Sigma_a\Sigma_p^\dagger\Sigma_q^T\Sigma_b^* & C_{80} &= B_{80}\tr\Sigma_a\Sigma_b^*\Sigma_p^*\Sigma_q & C_{81} &= B_{81}\tr\Sigma_a\Sigma_q\Sigma_p^*\Sigma_b^* \\
C_{82} &= B_{82}\tr\Sigma_a\Sigma_b^*\Sigma_q\Sigma_p^* & C_{83} &= B_{83}\tr\Sigma_a\Sigma_b^*\Sigma_p^\dagger\Sigma_q^T & C_{84} &= B_{84}\tr\Sigma_a\Sigma_p^*\Sigma_q\Sigma_b^* \\
C_{85} &= B_{85}\tr\Sigma_a\Sigma_q^T\Sigma_p^\dagger\Sigma_b^* & C_{86} &= B_{86}\tr\Sigma_a\Sigma_p^\dagger\Sigma_b^\dagger\Sigma_q & C_{87} &= B_{87}\tr\Sigma_a\Sigma_q\Sigma_b^\dagger\Sigma_p^\dagger \\
C_{88} &= B_{88}\tr\Sigma_a\Sigma_p^\dagger\tr\Sigma_b^\dagger\Sigma_q & C_{89} &= B_{89}\tr\Sigma_a\Sigma_p^\dagger\Sigma_b^\dagger\Sigma_q^T & C_{90} &= B_{90}\tr\Sigma_a\Sigma_p^\dagger\Sigma_b^*\Sigma_q \\
C_{91} &= B_{91}\tr\Sigma_a\Sigma_p^*\Sigma_b^\dagger\Sigma_q & C_{92} &= B_{92}\tr\Sigma_a\Sigma_q^T\Sigma_b^*\Sigma_p^* & C_{93} &= B_{93}\tr\Sigma_a\Sigma_q\Sigma_b^*\Sigma_p^\dagger \\
C_{94} &= B_{94}\tr\Sigma_a\Sigma_q^T\Sigma_b^\dagger\Sigma_p^\dagger & C_{95} &= B_{95}\tr\Sigma_a\Sigma_p^*\Sigma_b^*\Sigma_q^T & C_{96} &= B_{96}\tr\Sigma_a\Sigma_q\Sigma_b^\dagger\Sigma_p^* \\
C_{97} &= B_{97}\tr\Sigma_a\Sigma_p^*\Sigma_b^*\Sigma_q & C_{98} &= B_{98}\tr\Sigma_a\Sigma_q\Sigma_b^*\Sigma_p^* & C_{99} &= B_{99}\tr\Sigma_a\Sigma_q\tr\Sigma_p^\dagger\Sigma_b^* \\
C_{100} &= B_{100}\tr\Sigma_a\Sigma_q^T\tr\Sigma_p^\dagger\Sigma_b^\dagger & C_{101} &= B_{101}\tr\Sigma_a\Sigma_p^*\Sigma_b^\dagger\Sigma_q^T & C_{102} &= B_{102}\tr\Sigma_a\Sigma_q^T\Sigma_b^\dagger\Sigma_p^* \\
C_{103} &= B_{103}\tr\Sigma_a\Sigma_q\tr\Sigma_p^\dagger\Sigma_b^\dagger & C_{104} &= B_{104}\tr\Sigma_a\Sigma_q^T\tr\Sigma_p^\dagger\Sigma_b^* & C_{105} &= B_{105}\tr\Sigma_a\Sigma_p^*\tr\Sigma_b^\dagger\Sigma_q^T
\end{align*}
(For the symplectic case, the transpose $T$ should be replaced by the dual $R$ , complex conjugation is considered in the quaternion sense, and $\tr$ should be replaced by $-\tr$, see Refs.\ \cite{mehta1997random,brouwer_diagrammatic_1996}.)

It remains to calculate the coefficients $B_j$ and the products of traces of generalized Pauli matrices. In principle, the calculation of each coefficient $B_j$ again involves the evaluation of 105 separate diagrams. To avoid having to perform this calculation for every coefficient $B_j$ separately, we make use of the fact that many diagrams in Fig.\ \ref{fig:diagrams} have the same form, implying that their coefficients $B_j$ are identical. This gives the relations
\begin{equation} \label{eq:B_equalities}
\begin{alignedat}{2}
B_1 &= B_2; & B_3 &= B_4; \\
B_5 &= B_6=B_7=B_8; & B_{10} &= B_{11}=B_{12}=B_{13}; \\
B_{14} &= B_{15}=B_{16}=B_{17}; & B_{18} &= B_{19}=B_{20}=B_{21}; \\
B_{22} &= B_{23}=B_{24}=B_{25}; & B_{26} &= B_{27}=B_{28}=B_{29}; \\
B_{30} &= \dots = B_{33} = B^*_{34} = \dots = B^*_{37};~~~~~ & B_{54} &= \dots = B_{57} = B^*_{58} = \dots = B^*_{61}; \\
B_{38} &= \dots = B_{45} = B^*_{46} = \dots = B^*_{53}; & B_{62} &= \dots = B_{69} = B^*_{70} = \dots = B^*_{77}; \\
B_{78} &= B^*_{79}; & B_{80} &= B^*_{81}; \\
B_{82} &= B_{83}=B_{84}=B_{85}; & B_{86} &= B^*_{87}; \\
B_{89} &= \dots = B_{92} = B^*_{93} = \dots = B^*_{96}; & B_{97} &= B_{98}=B_{99}=B_{100}; \\
B_{101} &= B_{102}; & B_{104} &= B_{105}.
\end{alignedat}
\end{equation}
\end{widetext}
In addition, we use that $\langle \mathcal{W}_{a p ; x, y}\mathcal{W}^*_{b q; x,y}\rangle = 0$ if $\Sigma_a = \openone$ and $\Sigma_p \neq \openone$. For the orthogonal case, this gives the constraints
\begin{align}
\label{eq:Bequality}
B_{1}+ q^2B_{3}+B_5+2\Re B_{30}+2\Re B_{54} &=0; \nonumber\\
(1+ q^2)B_{5}+B_9+2\Re B_{40}+2\Re B_{70} &=0; \nonumber\\
B_{14}+(1+ q^2)B^*_{40}+B_{70}+B^*_{78}+B_{84}+B_{90} &=0; \nonumber\\
B_{18}+B_{28}+B_{40}+ q^2B^*_{54}+B_{70}+B^*_{90}+B_{101} &=0;\nonumber \\
B_{10}+B_{24}+ q^2B^*_{30}+B^*_{40}+B^*_{70}+B_{86}+B_{90} &=0;\nonumber \\
B_{40}+(1+ q^2)B^*_{70}+B_{80}+B_{84}+B_{90}+B_{100} &=0; \nonumber\\
 q^2B_{24}+2\Re B_{30}+2\Re B_{70}+B_{100}+B_{103} &=0; \nonumber\\
 q^2B_{28}+2\Re B_{40}+2\Re B_{54}+B_{100}+B_{105} &=0.
\end{align}
Similarly, from $\langle \mathcal{W}_{a p ; x, y}\mathcal{W}^*_{b p'; x,y}\rangle = 1$ if $a = p = b = q = 0$ we find
\begin{multline}
\label{eq:Bnormality}
q^4 B_1 + q^2(B_3+B_5+2B_{10}+2B_{18}+B_{24}+B_{28}) \\
+ 2(B_{14}+\Re B_{30}-\Re B_{40}+\Re B_{54}+\Re B_{70}) \\
+ B_{88}+B_{105}=1.
\end{multline}
for the orthogonal case. To obtain the corresponding constraints for the symplectic case, one has to replace $q^2$ by $-q^2$ in Eqs.\ (\ref{eq:Bequality}) and (\ref{eq:Bnormality}).

For qudit size $q=2^m$, the pair-transition probability retains the form given in Eq.~\eqref{eq:Wgeneral}, with the coefficients $\mathcal{P}_i$ defined piecewise:
\begin{align}
   \mathcal P_1= &\,  \begin{cases}
  P_1 & \mbox{if $s_{p_{x}} s_{p_{x+1}} = s_{a_{x}} s_{a_{x+1}} = 1$}, \\
  P'_1 & \mbox{if $s_{p_{x}} s_{p_{x+1}} = s_{a_{x}} s_{a_{x+1}} = -1$}, \\
  P''_1 & \mbox{if $s_{p_{x}} s_{p_{x+1}} \neq s_{a_{x}} s_{a_{x+1}}$}
  \end{cases} \nonumber\\
\mathcal P_2=&\,
\begin{cases}
  P_2 & \mbox{if $s_{p_{x}} s_{p_{x+1}} = s_{a_{x}} s_{a_{x+1}} = 1$}, \\
  P'_2 & \mbox{if $s_{p_{x}} s_{p_{x+1}} = s_{a_{x}} s_{a_{x+1}} = -1$}, \\
  P''_2 & \mbox{if $s_{p_{x}} s_{p_{x+1}} \neq s_{a_{x}} s_{a_{x+1}}$}
  \end{cases} \nonumber\\
 \mathcal P_3=&\,
\begin{cases}
  P_3 & \mbox{if $s_{p_{x}} s_{p_{x+1}} = s_{a_{x}} s_{a_{x+1}} = 1$}, \\
  P'_3 & \mbox{if $s_{p_{x}} s_{p_{x+1}} = s_{a_{x}} s_{a_{x+1}} = -1$}, \\
  0 & \mbox{if $s_{p_{x}} s_{p_{x+1}} \neq s_{a_{x}} s_{a_{x+1}}$}
  \end{cases}.
\end{align}
The parameters $P_i, P'_i, P''_i$ can be expressed in terms of the coefficients $B_j$ as
\begin{align}
\label{eq:Par}
P_1 =&\,
    B_1+B_9+2B_5 +  q^{-2}(2B_{10}+2B_{18}+2B_{24} \nonumber \\ & \mbox{}
    +2B_{28} +8\Re B_{40}+8\Re B_{70}+2\Re B_{78} \nonumber \\ & \mbox{}
    +2\Re B_{80}+4B_{84}),
    \nonumber\\
P_2 =&\, 2 q^{-2}(B_{14}+\Re B_{86}+B_{100}+B_{101}+4\Re B_{90}),\nonumber \\
P_3 =&\, 2B_{14}+2B_{100}+B_{88}+B_{103}+2B_{105},\nonumber \\
P'_1 =&\,   B_1+B_9-2B_5 +  q^{-2}(2B_{10}+2B_{18}+2B_{24} \nonumber\\ & \mbox{}
    + 2B_{28}-8\Re B_{40}-8\Re B_{70}+2\Re B_{78}\nonumber \\ & \mbox{}
  +2\Re B_{80}+4B_{84}), \nonumber\\
P'_2 =&\, 2 q^{-2}(B_{14}+\Re B_{86}+B_{100}+B_{101}-4\Re B_{90}), \nonumber \\
P'_3 =&\, -2B_{14}-2B_{100}+B_{88}+B_{103}+2B_{105}, \nonumber\\
P''_1 =&\,
    B_1-B_9+  q^{-2}(2B_{10}+2B_{18}+2B_{24}+2B_{28} \nonumber \\ & \ \ \ \ \mbox{}
    -2\Re B_{78}-2\Re B_{80}-4B_{84}), \nonumber \\
P''_2 =&\, 2 q^{-2}(-B_{14}+\Re B_{86}-B_{100}+B_{101}).
\end{align}
For these coefficients, the normalization condition $  \sum_{p_x,p_y} \left| \mathcal{W}_{ap;x,y}\right|^2 =
  1 $ implies the constraints 
\begin{align}
\label{eq:pnorm}
\begin{multlined}
\frac{q^4+ q^2-2}{2}P_1+\frac{ q^2-2}{2}P_2+P_3 \\
+\frac{q^4- q^2}{2}P''_1-\frac{ q^2}{2}P''_2 = 1,
\end{multlined}\nonumber \\
\begin{multlined}
\frac{q^4- q^2}{2}P'_1+\frac{ q^2}{2}P'_2+P'_3 \\
+\frac{q^4+ q^2-2}{2}P''_1-\frac{ q^2+2}{2}P''_2 = 1.
\end{multlined}
\end{align}
Equations (\ref{eq:Par}) and (\ref{eq:pnorm}) are for the orthogonal case. For the symplectic case, one again has to replace $q^2 \to -q^2$.

Although these relations significantly reduce the number of coefficients $B_j$ that have to be calculated separately, the number of remaining coefficients is still very large. A further simplification can be obtained if the two-qudit gate matrices ${\cal U}$ satisfy a further symmetry. Below, we consider the cases that ${\cal U}$ is orthogonal, symplectic, symmetric, and self-dual in detail.

\subsection{Orthogonal and symplectic ensembles}
\label{subsec:orthsympensembles}

When the two-qudit gate operators ${\cal U}$ satisfy $\mathcal{U}^{\dagger} =\mathcal{U}^{\rm T}$ (orthogonal case) or $\mathcal{U}^\dagger=\mathcal{U}^R$ (symplectic case), {\em i.e.}, ${\cal U}$ is an orthogonal or symplectic $q^2 \times q^2$ matrix, respectively, the symmetry constraints on the coefficients $B_j$ simplify significantly. In both cases, we find that additional equalities hold for the coefficients $B_j$,
\begin{align}
\label{eq:UnifiedEqualities}
B_1 =&\, B_2 = B_9, \nonumber\\
B_3 =&\, B_4 = B_5 = B_6 = B_7 = B_8, \nonumber\\
B_{10} =&\, B_{11} = B_{12} = B_{13} = B_{78} = B_{79}, \nonumber\\
B_{14} =&\, B_{15} = B_{16} = B_{17} = B_{86} = B_{87}, \nonumber\\
B_{18} =&\, B_{19} = B_{20} = B_{21} = B_{80} = B_{81},\nonumber \\
B_{22} =&\, \dots = B_{29} = B_{82} = \dots = B_{85}, \nonumber\\
B_{30} =&\, \dots = B_{53},\nonumber \\
B_{54} =&\, \dots = B_{77}, \nonumber\\
B_{89} =&\, \dots = B_{96}, \nonumber\\
B_{97} =&\, \dots = B_{102}, \nonumber\\
B_{103} =&\, B_{104} = B_{105},
\end{align}
These relations reduce the constraints of Eqs.\ (\ref{eq:Bequality}) to
\begin{equation}
\label{eq:BBB}
\begin{aligned}
2B_{30}+2B_{54}+B_1+(1+ q^2)B_3 &= 0 \\
B_{24}+B_{10}+B_{14}+(1+ q^2)B_{30}+B_{54}+B_{90} &= 0 \\
B_{24}+B_{18}+B_{30}+(1+ q^2)B_{54}+B_{90}+B_{101} &= 0 \\
 q^2B_{24}+2B_{30}+2B_{54}+B_{101}+B_{103} &= 0.
\end{aligned}
\end{equation}
The normality condition \eqref{eq:Bnormality} simplifies to
\begin{multline}
q^4 B_1+2q^2(B_3+B_{10}+B_{18}+B_{24}) \\
+2B_{14}+4(B_{30}+B_{54})+B_{88}+B_{103}=1.
\end{multline}
Consequently, the parameters $P_i$, $P'_i$ and $P''_i$ defined in Eq.~\eqref{eq:Par} become
\begin{align}
\label{eq:UnifiedSpecificPara}
P_1 = &\,
    2B_1+2B_3 \nonumber \\ &\, \mbox{} + q^{-2}(4B_{10}+4B_{18}+8B_{24} 
    +8B_{30}+8B_{54}), \nonumber \\
P_2 =&\,  q^{-2}(4B_{14}+4B_{101}+8B_{90}), \nonumber \\
P_3 =&\, 2B_{14}+2B_{101}+B_{88}+3B_{103},\nonumber \\
P'_1 =&\, 
    2B_1-2B_3 \nonumber \\ &\, \mbox{} + q^{-2}(4B_{10}+4B_{18}+8B_{24} 
    -8B_{30}-8B_{54}), \nonumber \\
P'_2 =&\,  q^{-2}(4B_{14}+4B_{101}-8B_{90}), \nonumber\\
P'_3 =&\, B_{88}+3B_{103}-2B_{14}-2B_{101},\nonumber \\
P''_1 =&\, 0, \nonumber\\
P''_2 =&\, 0
\end{align}
and instead of the normalization constraints of Eq.~\eqref{eq:pnorm} one has
\begin{align}
  \label{eq:normcond}
-\frac{ q^2+2}{2}P_1+\frac{ q^2}{2}P'_1-P_2 &= 0, \nonumber\\
\frac{q^4+ q^2-2}{2}P_1+\frac{ q^2-2}{2}P_2+P_3 &= 1, \nonumber\\
\frac{q^4- q^2}{2}P'_1+\frac{ q^2}{2}P'_2+P'_3 &= 1.
\end{align}
Equations (\ref{eq:BBB})--(\ref{eq:normcond}) are for the orthogonal case; for the symplectic case, one has to replace $q^2$ by $-q^2$. 

This system leaves three free parameters: $P_3$, $P'_3$, and $P'_2$. The remaining coefficients $B_j$ can be expressed in terms of the moments $\mathcal{R}$ of the two-qudit gate operator by straightforward evaluation of the corresponding diagrams,
\begin{widetext}
\begin{align}
    B_{14}=&\,  q^2\left[( q^2+2)V_{2,2}+4V_{1,3}+(12+2 q^2)V_{4}\right]+ q^4 \mathcal{R}_{1;1}[(4+ q^2)V_{1,1,2}+ q^2V_{2,2}+(8+4 q^2)V_{1,3}+12V_{4}]\nonumber\\
    &\, \mbox{} + q^2  \mathcal{R}_{2}[ q^2V_{1,1,2}+(4+ q^2)V_{2,2}+8V_{1,3}+(12+4 q^2)V_4]+ q^2  \mathcal{R}_4 (2V_{2,2}+4V_{1,3})
  \nonumber \\ &\, \mbox{}
  + q^6  \mathcal{R}_{1,1;2}(V_{1,1,1,1}+V_{2,2}+4V_{1,3})
    + q^4 \mathcal{R}_{1,3}(4V_{1,1,2}+4V_4)+ q^4 \mathcal{R}_{2;2}(V_{1,1,2}+2V_4)+ q^8 \mathcal{R}_{1,1;1,1}V_{1,1,2},\nonumber\\
    B_{88}=&\,  q^2\left[3 q^2V_{2,2}+18V_{4}\right]+ q^4 \mathcal{R}_{1;1} (6 q^2V_{1,1,2}+24V_{1,3}) + q^2  \mathcal{R}_{2} (6 q^2V_{2,2}+24V_4)\nonumber+6 q^2  \mathcal{R}_4 V_4 \nonumber \\ &\, \mbox{} +6 q^6  \mathcal{R}_{1,1;2}V_{1,1,2}
    +8 q^4 \mathcal{R}_{1,3} V_{1,3}+3 q^4 \mathcal{R}_{2;2} V_{2,2}+ q^8 \mathcal{R}_{1,1;1,1}V_{1,1,1,1},\nonumber\\
    B_{90}=&\,  q^2\left[3V_{2,2}+6V_{1,3}+(9+3 q^2)V_{4}\right]+ q^4 \mathcal{R}_{1;1}[3V_{1,1,2}+3V_{2,2}+(6+3 q^2)V_{1,3}+(12+3 q^2)V_{4}]\nonumber\\
    &\, \mbox{} + q^2  \mathcal{R}_{2}[3V_{1,1,2}+3V_{2,2}+(6+3 q^2)V_{1,3}+(12+3 q^2)V_4]+ q^2  \mathcal{R}_4 (3V_{1,1,2}+3V_{4})
  \nonumber \\ &\, \mbox{}
+ q^6  \mathcal{R}_{1,1;2}(3V_{1,1,2}+ 3V_{4})
  +q^4 \mathcal{R}_{1,3}(V_{1,1,1,1}+3V_{2,2}+4V_{1,3})+3 q^4 \mathcal{R}_{2;2} V_{1,3}+ q^8 \mathcal{R}_{1,1;1,1}V_{1,3},\nonumber\\
    B_{101}=&\,  q^2\left[2V_{1,1,2}+( q^2+4)V_{2,2}+4V_{1,3}+(8+2 q^2)V_{4}\right]+ q^4 \mathcal{R}_{1;1}[4V_{2,2}+8V_{1,3}+(12+6 q^2)V_{4}]\nonumber\\
    &\, \mbox{} + q^2  \mathcal{R}_{2}[4V_{1,1,2}+(8+4 q^2)V_{1,3}+(12+2 q^2)V_4]+ q^2  \mathcal{R}_4 (V_{1,1,1,1}+V_{2,2}+4V_{1,3})
  \nonumber \\ &\, \mbox{}
+ q^6  \mathcal{R}_{1,1;2}(2V_{2,2}+4V_{1,3}) + q^4 \mathcal{R}_{1,3}(4V_{1,1,2}+4V_4)+ q^4 \mathcal{R}_{2;2}(2V_{1,1,2}+V_4)+ q^8 \mathcal{R}_{1,1;1,1}V_{4},\nonumber\\
    B_{103}=&\,  q^2\left[4V_{1,1,2}+(3 q^2+4)V_{2,2}+10V_{4}\right]+ q^4 \mathcal{R}_{1;1}[2 q^2V_{2,2}+8V_{1,3}+(16+4 q^2)V_{4}]\nonumber\\
    &\, \mbox{} + q^2  \mathcal{R}_{2}[2 q^2V_{1,1,2}+16V_{1,3}+(8+4 q^2)V_4]+ q^2  \mathcal{R}_4 (4V_{1,1,2}+2V_{4})+ q^6  \mathcal{R}_{1,1;2}(2V_{1,1,2}+4V_{4})\nonumber\\
    &\, \mbox{} +8 q^4 \mathcal{R}_{1,3}V_{1,3}+ q^4 \mathcal{R}_{2;2}(V_{1,1,1,1}+2V_{2,2})+ q^8 \mathcal{R}_{1,1;1,1}V_{2,2},
\end{align}
\end{widetext}
where, again, for the symplectic case, one has to replace $q^2$ by $-q^2$. The moments $\mathcal{R}$ of the two-qudit gate operators are defined as
\begin{align}
  {\cal R}_{1;1} =&\, q^{-4} \langle (\mbox{tr}\, {\cal U}_{x,x+1})^4\, \rangle, \nonumber \\
  \mathcal{R}_{2}=& q^{-2} \langle \mbox{tr}\, {\cal U}_{x,x+1}^2\, \rangle , \nonumber \\
  \mathcal{R}_4=&
  q^{-2} \langle \mbox{tr}\, {\cal U}_{x,x+1}^4\, \rangle ,\nonumber\\
  \mathcal{R}_{1;3}=&
  q^{-4} \langle \mbox{tr}\, {\cal U}_{x,x+1}\, \mbox{tr}\, {\cal U}^3_{x,x+1} \rangle,
 \nonumber \\
  {\cal R}_{2;2} =&\, q^{-4} \langle (\mbox{tr}\, {\cal U}_{x,x+1}^2)^2\, \rangle,  \nonumber \\
  {\cal R}_{1,1;1,1} =&\, q^{-8} \langle (\mbox{tr}\, {\cal U}_{x,x+1})^4\,  \rangle, \nonumber \\
  {\cal R}_{1,1;2} =&\, q^{-6} \langle (\mbox{tr}\, {\cal U}_{x,x+1})^2\, \mbox{tr}\, {\cal U}_{x,x+1}^{2} \rangle. \label{eq:R112}
\end{align}
The definitions (\ref{eq:R112}) apply both to the orthogonal and the symplectic case.

In the large-$q$ limit, the expressions for the average $\langle {\cal W}_{ap;x,y} {\cal W}^*_{bq;x,y} \rangle$ can be further simplified. We find that $B_{88} \sim {\cal O}(1)$, $B_{14} \sim \mathcal{O}(q^{-2})$, $B_{90} \sim B_{103} \sim\mathcal{O}(q^{-4})$, and $B_{101} \sim \mathcal{O}(q^{-6})$. As a consequence, of the eight coefficients $P_j$ and $P_j'$, one has $P_3 = P_3' = B_{88} +{\cal O}(q^{-2})$, whereas $P_j \sim {\cal O}(q^{-4})$ for $j=1,2$, and $(P_j-P_j')/P_j \sim {\cal O}(q^{-2})$ for all $j=1,2,3$. This results in a simple closed form for the zeroth-order truncated evolution equations in the limit of large qudit size $q$, which contains the coefficient $P_3$ only,
\begin{align}
  \label{eq:rhoqlarge}
 \bar \rho_{\rm }^{(0)}(\Delta x;t)= &P_3\bar \rho_{\rm }^{(0)}(\Delta x+1;t-1)    \nonumber\\
  \bar \rho_{\rm }^{(0)}(\Delta x+1;t)= &\bar \rho_{\rm }^{(0)}(\Delta x+2;t-1)\nonumber\\
  &+(1-P_3)\bar \rho_{\rm }^{(0)}(\Delta x+1;t-1).
\end{align}
(Corrections to these equations appear to order $\mathcal{O}(q^{-2})$.) The butterfly velocity and diffusion constant are easily found from Eqs.\ (\ref{eq:rhoqlarge}),
\begin{align}
\label{eq:vdqlarge}
\lim_{q\to\infty}v_B =&\, \frac{1-P_3}{1+P_3},\nonumber \\ \lim_{q\to\infty}\mathcal{D} =&\, \frac{4P_3(1-P_3)}{(1+P_3)^2}.
\end{align}
The fact that for the limit of large $q$ the equations (\ref{eq:rhoqlarge}) are closed without having to make further approximations implies that the truncation procedure of Sec.\ \ref{sec:dd} is exact in the limit $q \to \infty$.
In the large-$q$ limit, we obtain that 
\begin{equation}
  \lim_{q \to \infty}
  P_3 =  \lim_{q \to \infty} \mathcal{R}_{1,1;1,1}.
\label{eq:P3infty}
\end{equation}

\subsection{Symmetric and self-dual ensembles}
\label{subsec:unified_symm_dual}

The symmetry constraints on the coefficients $B_j$ also simplify significantly if the two-qudit gate operators ${\cal U}$ are symmetric or self-dual matrices, ${\cal U} = {\cal U}^{\rm T}$ or ${\cal U} = {\cal U}^{\rm R}$ for the orthogonal or symplectic case, respectively. The circular orthogonal and symplectic ensembles (COE and CSE, respectively) from random matrix theory are examples of ensembles with such symmetry \cite{mehta1997random}. For symmetric or self-dual ${\cal U}$, the relations between the coefficients $B_j$ are
\begin{align}
\label{eq:identitysymmetric}
B_1 =&B_2 = B_{18} = \dots = B_{21} = B_{104} = B_{105}, \nonumber \\
B_3 =&\, B_4 = B_{26} = \dots = B_{29} \nonumber \\ =&\, B_{54} = \dots = B_{61} = B_{101} = B_{102}, \nonumber \\
B_5 =&\, B_6 = \dots = B_8 = B_{22} = \dots = B_{25} \nonumber \\ =&\, B_{62} = \dots = B_{77} = B_{82} = \dots = B_{85} \nonumber \\ =&\, B_{97} = \dots = B_{100}, \nonumber \\
B_9 =&\, B_{80} = B_{81} = B_{103}, \nonumber \\ B_{10} =&\, B_{11} = \dots = B_{17}, \nonumber \\ B_{78} =&\, B^*_{79} = B^*_{86} = B_{87}, \nonumber \\
B_{30} =&\, B_{31} = \dots = B_{33} = B^*_{34} = \dots = B^*_{37} \nonumber \\ =&\, B_{38} = \dots = B_{45} = B^*_{46} = \dots = B^*_{53} \nonumber \\ =&\, B^*_{89} = \dots = B^*_{92} = B_{93} = \dots = B_{96}.
\end{align}
These relations simplify the expressions in \eqref{eq:Bequality} to a set of three linear constraints,
\begin{align}
\label{eq:Bsymm}
2\Re B_{30} +  B_5 + B_1 + (q^2+2)B_3 &= 0, \nonumber \\
(q^2+3)B_5 + B_9 + 2\Re B_{30} &= 0, \nonumber \\
B_{10} + (q^2+2)B^*_{30} + 2B_5 + B_{86} &= 0.
\end{align}
Similarly, the normality condition (\ref{eq:Bnormality}) simplifies to
\begin{multline}
\label{eq:normsymm}
(q^2+1)^2 B_1 + 2(q^2+1)(B_3+B_5+B_{10}) \\
+ 4\Re B_{30} + B_{88} = 1.
\end{multline}
Equations (\ref{eq:Bsymm})--(\ref{eq:normsymm}) are for the orthogonal case; for the symplectic case, one has to replace $q^2$ by $-q^2$. 
Because of the constraints of Eqs. \eqref{eq:identitysymmetric}, (\ref{eq:Bsymm}) and (\ref{eq:normsymm}), Eq.~\eqref{eq:Par} becomes
\begin{align}
P_1 =&\, \frac{B_9 (q^4 + 4 q^2 - 2) - 2 \mathrm{Re}B_{30} (2 q^4 + 7 q^2 + 8)}{q^2 (q^2 + 3)}\nonumber\\
&\, \mbox{} -\frac { B_3 (q^6 + 7 q^4 + 14 q^2 + 6)}{q^2 (q^2 + 3)}, \nonumber\\
P_2 =&\, \frac{2 \left[ B_9 + B_3 (q^2 + 3) - \mathrm{Re}B_{30} (q^4 + q^2 - 8) \right]}{q^2 (q^2 + 3)}, \nonumber\\
P_3 =&\, B_3 (q^6+4q^4+q^2-4)  +B_9 \left(4-q^2-\frac{8}{q^2+3} \right) \nonumber\\
&\, \mbox{}  +2\mathrm{Re}B_{30}  \left(q^4+q^2-\frac{16}{q^2+3} \right)- 2 q^2B_{10}+1 , \nonumber\\
P'_{1} =&\, \frac{B_9 (q^4 + 8 q^2 + 14) - 2 \mathrm{Re}B_{30} (2 q^4 + 11 q^2 + 16)}{q^2 (q^2 + 3)}\nonumber \\
&\, \mbox{} -\frac{ B_3 (q^6 + 7 q^4 + 14 q^2 + 6)}{q^2 (q^2 + 3)},\nonumber \\
P'_{2} =&\, \frac{2 \left[ B_9 + B_3 (q^2 + 3) -\mathrm{Re} B_{30} (q^4 + 9 q^2 + 16) \right]}{q^2 (q^2 + 3)},\nonumber \\
P'_{3} =&\,1+ B_3 (q^6+4q^4+q^2-4) + B_9 \left(4-q^2-\frac{4 }{q^2 + 3} \right) \nonumber\\
&\, \mbox{} + 2  \mathrm{Re}B_{30} \left(q^4+2q^2-\frac{8}{q^2+3} \right) - 2 B_{10}(q^2+2), \nonumber\\
P''_{1} =&\, -\frac{B_9 q^4 + (4 B_9 - 2 \mathrm{Re}B_{30}) q^2 + 6 B_9 - 4 B_{10} (q^2 + 3)}{q^2 (q^2 + 3)}\nonumber\\
&\, \mbox{} -\frac{ B_3 (q^6 + 7 q^4 + 14 q^2 + 6)}{q^2 (q^2 + 3)},\nonumber \\
P''_{2} =&\, \frac{2B_3}{q^2}-\frac{2 \left[ \mathrm{Re}B_{30} (q^4 + 5q^2)  - 3 B_9 + 2 B_{10} (q^2 + 3)  \right]}{q^2 (q^2 + 3)}.
\end{align}
The remaining coefficients $B_j$ can be expressed in terms of the moments $\mathcal{R}$ of the two-qudit gate operator,
\begin{widetext}
\begin{align}
B_{3}=&\,  q^2\left[V_{1,1,1,1}+( 2q^2+6)V_{1,1,2}+5V_{2,2}+(4q^2+16)V_{1,3}+(2 q^2+20)V_{4}\right]+ q^4 \mathcal{R}_{2;2}(V_{2,2}+3V_4) \nonumber\\
&+q^4 \mathcal{R}_{1;1}[4V_{1,1,2}+4V_{2,2}+(4q^2+8)V_{1,3}+(4 q^2+16)V_{4}]+ q^6 \mathrm{Re} \mathcal{R}_{1,1;2}(2V_{2,2}+2V_{4}) + q^8 \mathcal{R}_{1,1;1,1}V_{4},\nonumber\\
 B_{9}=&\,  q^2\left[8V_{1,1,2}+4(q^2+1)V_{2,2}+16V_{1,3}+4(q^2+5)V_{4}\right]+ q^4 \mathcal{R}_{1;1}[16V_{1,3}+(16+8 q^2)V_{4}]\nonumber\\
    &\, \mbox{} + q^4 \mathcal{R}_{2;2}(V_{1,1,1,1}+2V_{1,1,2}+V_{2,2})+ q^6  \mathrm{Re}\mathcal{R}_{1,1;2}(2V_{1,1,2}+2V_{2,2})+ q^8 \mathcal{R}_{1,1;1,1}V_{2,2},\nonumber\\
B_{10}=&\,  q^2\left[2q^2V_{1,1,2}+( 2q^2+8)V_{2,2}+16V_{1,3}+(4 q^2+24)V_{4}\right]  + 4q^4 \mathcal{R}_{2;2}V_4+ 4q^6  \mathrm{Re}\mathcal{R}_{1,1;2}V_{1,3}
 + q^8 \mathcal{R}_{1,1;1,1}V_{1,1,2}\nonumber\\
   &\, \mbox{}
 + q^4 \mathcal{R}_{1;1}[q^2V_{1,1,1,1}+(q^2+8)V_{1,1,2}+ 2q^2V_{2,2}+(4 q^2+8)V_{1,3}+16V_{4}] ,\nonumber\\
\mathrm{Re} B_{30}=&\,  q^2\left[6V_{1,1,2}+6V_{2,2}+(4q^2+12)V_{1,3}+(4 q^2+24)V_{4}\right]+q^4 \mathcal{R}_{2;2} (2V_{1,3}+2V_4)+ q^6  \mathrm{Re}\mathcal{R}_{1,1;2}(V_{1,1,2}+V_{1,3}+ 2V_{4}) \nonumber\\
&\, \mbox{}
+ q^4 \mathcal{R}_{1;1}[V_{1,1,1,1}+(2q^2+3)V_{1,1,2}+6V_{2,2}+(2q^2+10)V_{1,3}+(4q^2+12)V_{4}]+ q^8 \mathcal{R}_{1,1;1,1}V_{1,3}
.
\end{align}
\end{widetext}
The relevant moments $\mathcal{R}$ required to calculate these coefficients are defined as:
\begin{align}
\label{eq:moments}
  \mathcal{R}_{1;1} &= q^{-4} \langle \operatorname{tr} \mathcal{U}_{x,x+1}\, \operatorname{tr} \mathcal{U}^\dagger_{x,x+1} \rangle, \nonumber \\
  \mathcal{R}_{2;2} &= q^{-4} \langle \operatorname{tr} \mathcal{U}_{x,x+1}^2\, \operatorname{tr} \mathcal{U}_{x,x+1}^{\dagger 2} \rangle, \nonumber \\
  \mathcal{R}_{1,1;1,1} &= q^{-8} \langle (\operatorname{tr} \mathcal{U}_{x,x+1})^2\, (\operatorname{tr} \mathcal{U}_{x,x+1}^{\dagger})^2 \rangle, \nonumber \\
  \mathcal{R}_{1,1;2} &= q^{-6} \langle (\operatorname{tr} \mathcal{U}_{x,x+1})^2\, \operatorname{tr} \mathcal{U}_{x,x+1}^{\dagger 2} \rangle.
\end{align}
Following an argument analogous to that in App.~\ref{subsec:orthsympensembles}, we find that,
in the limit $q\to\infty$, the butterfly velocity and diffusion
constant are again given by Eq.~\eqref{eq:vdqlarge}, with $P_3$ evaluated
for the symmetric or self-dual ensemble. In the large-$q$ limit, we obtain the same form of $P_3$ as shown in Eq. \eqref{eq:P3infty}.  

\subsection{Combined symmetries}
\label{subsec:combined_symmetries}

Finally, we consider the most constrained ensembles, where the matrix $\mathcal{U}$ possesses both symmetries. Specifically, we examine the Symmetric Orthogonal case ($\mathcal{U}=\mathcal{U}^{*}=\mathcal{U}^{T}$) and the Self-dual Symplectic case ($\mathcal{U}=\mathcal{U}^R=\mathcal{U}^\dagger$).

The combined symmetries impose constraints on the coefficients $B_j$,
\begin{align}
\label{eq:combined_equalities}
B_1 =&\, B_2 = B_9 = B_{18} = \dots = B_{21} = B_{80} = B_{81}  \nonumber \\
    =&\, B_{103} = B_{104} = B_{105}; \nonumber \\
B_3 =&\, B_4 = \dots = B_8 = B_{22} = \dots = B_{29} \nonumber \\ =&\, B_{54} = \dots = B_{77} = B_{82}\dots =B_{85} 
\nonumber \\ =&\, B_{97} = \dots = B_{102}; \nonumber \\
B_{10} =&\, B_{11} = \dots = B_{17} = B_{78} = B_{79} = B_{86} = B_{87}; \nonumber \\
B_{30} =&\, B_{31} = \dots = B_{53} = B_{89} = \dots = B_{96}.
\end{align}
The constraints of Eqs.\ (\ref{eq:Bequality}) and (\ref{eq:Bnormality}) simplify to
\begin{align}
\label{eq:Bcombined1}
2 B_{30} + B_1 + (3 +  q^2)B_3 =&\, 0, \nonumber\\
2B_{10} + (2 +  q^2)B_{30} + 2B_{3} =&\, 0
\end{align}
and
\begin{align}
\label{eq:Bcombined2}
& (1+ q^2)^2 B_1 + 4(1+ q^2)B_3 \nonumber \\ & ~~~~  \mbox{} + 2(1+ q^2)B_{10} + 4 B_{30} + B_{88} = 1.
\end{align}
These relations are for the orthogonal case; for the symplectic case, one has to replace $q^2 \to -q^2$. Due to the constraints of Eqs. \eqref{eq:combined_equalities}, \eqref{eq:Bcombined1} and \eqref{eq:Bcombined2},  the parameters defined in Eq.~\eqref{eq:Par} become
\begin{align*}
P_1 &= B_{30} \left( -6 - \frac{4}{q^2} \right) - 2 B_3 (q^2 + 4),\nonumber \\
P_2 &= B_{30} \left( -2 + \frac{4}{q^2} \right),\nonumber \\
P_3 &= B_3 (q^6 + 5 q^4 + 2 q^2 - 8) + B_{30} (3 q^4 + 6 q^2 - 8) + 1,\nonumber \\
P'_1 &= -\frac{2 \left[ B_{30} (3 q^2 + 10) + B_3 (q^4 + 6 q^2 + 8) \right]}{q^2},\nonumber \\
P'_2 &= -\frac{2 B_{30} (q^2 + 6)}{q^2}, \nonumber\\
P'_3 &= B_3 (q^6 + 5 q^4 + 2 q^2 - 8) + B_{30} (3 q^4 + 8 q^2 - 4) + 1,\nonumber\\
P''_1&=0,\nonumber\\
P''_2&=0.
\end{align*}
Due to the high degree of symmetry, only two independent moments are required to determine the coefficients $B_j$ :
\begin{align}
B_{3}=&\,  q^2\left[V_{1,1,1,1}+( 2q^2+6)V_{1,1,2}+(4q^2+16)V_{1,3}\right.\nonumber\\&\left.+6V_{2,2}+(5 q^2+20)V_{4}\right] +q^4 \mathcal{R}_{1;1}[4V_{1,1,2}\nonumber\\ &+(2q^2+4)V_{2,2}+(4q^2+8)V_{1,3}+(6 q^2+16)V_{4}] \nonumber\\
&+ q^8 \mathcal{R}_{1,1;1,1}V_{4},\nonumber\\
B_{30}=&\,  q^2[6V_{1,1,2}+6V_{2,2}+(6q^2+12)V_{1,3}\nonumber\\
&\, \mbox{}+(6 q^2+24)V_{4}]
+ q^4 \mathcal{R}_{1;1}[V_{1,1,1,1}+6V_{2,2}\nonumber\\ &+(3q^2+3)V_{1,1,2}+(3q^2+10)V_{1,3}\nonumber\\ &+(6q^2+12)V_{4}]+ q^8 \mathcal{R}_{1,1;1,1}V_{1,3}
,
\end{align}
with
\begin{align}
\label{eq:combined_moments}
  \mathcal{R}_{1;1} =&\,q^{-4} \langle \operatorname{tr} \mathcal{U}_{x,x+1}\, \operatorname{tr} \mathcal{U}_{x,x+1} \rangle, \nonumber \\
  \mathcal{R}_{1,1;1,1} =&\, q^{-8}\langle (\operatorname{tr} \mathcal{U}_{x,x+1})^2\, (\operatorname{tr} \mathcal{U}_{x,x+1})^2 \rangle.
\end{align}
An analogous large-$q$ argument shows that the butterfly velocity and
diffusion constant retain the form given in Eq.~\eqref{eq:vdqlarge}, with
$P_3$ evaluated for the ensembles considered in this subsection. As a concrete example, Ref.~\cite{TanBrownian} considers the choice
$V=O\,\mathrm{diag}(-1,1,\ldots,1)\,O^{\mathrm T}$, where $O$ is Haar
distributed over the orthogonal group.  In the large-$q$ limit, we obtain the same form of $P_3$ as shown in Eq. \eqref{eq:P3infty}. 

\section{projected transition matrix }
\label{app:4}
\subsection{The case $q=2^m$}
For $q=2^m$, to show that the projected binary-string distribution $\bar \rho_{\bar p}$ satisfies a closed evolution equation, it is sufficient to consider the adjacent qudits $x$ and $x+1$, which are subject to a two-qudit gate. From Eqs.\ (\ref{eq:gammasqavg}) and (\ref{eq:binarydef}) we then have
\begin{align}
  \label{eq:barrhoevol}
  \bar \rho_{\bar p_{x},\bar p_{x+1}}(t) =&\,
  \sum_{a_x,a_{x+1}}
  \rho_{a_{x},a_{x+1};a_{x},a_{x+1}}(t-1)
  \nonumber \\ &\, \mbox{} \times
  \sum_{p_x \to \bar p_x} \sum_{p_{x+1} \to \bar p_{x+1}}
  \langle |W_{a_x,a_{x+1};p_x,p_{x+1}}|^2 \rangle.
\end{align}
In this expression, depending on $\bar p_{x}$ and $\bar p_{x+1}$, the summations over $p_x$ and $p_{x+1}$ are either restricted to $p_{x} = 0$ or $\bar p_{x+1} = 0$ or extend over all $p_{x} \neq 0$ or $p_{x+1} \neq 0$. We may then use the general form (\ref{eq:Wgeneral}) of $\langle |W_{a_x,a_{x+1};p_x,p_{x+1}}|^2 \rangle$, Eq.\ (\ref{eq:varphi0}), and the sum rules (\ref{eq:phiidentities}) to see that $\sum_{p_x \to \bar p_x} \sum_{p_{x+1} \to \bar p_{x+1}} \langle |W_{a_x,a_{x+1};p_x,p_{x+1}}|^2 \rangle$ depends only on whether or not $a_{x}$ and $a_{x+1}$ are equal to $0$, {\em i.e.}, they only depend on the projected binary indices $\bar a_x$ and $\bar a_{x+1}$ associated with $a_{x}$ and $a_{x+1}$. This means that the right-hand side of Eq.\ (\ref{eq:barrhoevol}) can be expressed in terms of the projected binary string distribution $\bar \rho_{\bar a_x,\bar a_{x+1}}(t-1)$, which is what we set out to prove. Therefore, we can obtain a closed Markovian evolution equation of the form of Eq.\ (\ref{eq:rhobinaryevolution}).
Since the transition matrix $T_{\bar{a}\bar{p}}$ is constructed from a sum of the non-negative pair probabilities $\langle |W_{a_x,a_{x+1};p_x,p_{x+1}}|^2 \rangle$, its elements are also non-negative.
 Similarly, for both the orthogonal- and symplectic-invariant cases, one can employ the sum rules in Eqs.~\eqref{eq:orthphiidentities} and \eqref{eq:symphiidentities} to obtain the projected pair transition probability matrix $T_{\bar a \bar p}$, which exhibits the block structure shown in Eq.~\eqref{eq:Tmatrix}.

For the orthogonal and symplectic invariant circuit, the blocks of the transition matrix have the structure
\begin{align}
\label{eq:T_matrices}
T_{++} &= \begin{pmatrix}
    T_{{\rm e}{\rm e};{\rm e}{\rm e}} & T_{{\rm e}{\rm e};{\rm o}{\rm o}} & T_{{\rm e}{\rm e};{\rm e}\mathbb{I}} & T_{{\rm e}{\rm e};\mathbb{I}{\rm e}} \\
    T_{{\rm o}{\rm o};{\rm e}{\rm e}} & T_{{\rm o}{\rm o};{\rm o}{\rm o}} & T_{{\rm o}{\rm o};{\rm e}\mathbb{I}} & T_{{\rm o}{\rm o};\mathbb{I}{\rm e}} \\
    T_{{\rm e}\mathbb{I};{\rm e}{\rm e}} & T_{{\rm e}\mathbb{I};{\rm o}{\rm o}} & T_{{\rm e}\mathbb{I};{\rm e}\mathbb{I}} & T_{{\rm e}\mathbb{I};\mathbb{I}{\rm e}} \\
    T_{\mathbb{I}{\rm e};{\rm e}{\rm e}} & T_{\mathbb{I}{\rm e};{\rm o}{\rm o}} & T_{\mathbb{I}{\rm e};{\rm e}\mathbb{I}} & T_{\mathbb{I}{\rm e};\mathbb{I}{\rm e}}
\end{pmatrix}, \displaybreak[0]\nonumber \\[1em]
T_{+-} &= \begin{pmatrix}
    T_{{\rm e}{\rm e};{\rm e}{\rm o}} & T_{{\rm e}{\rm e};{\rm o}{\rm e}} & T_{{\rm e}{\rm e};{\rm o}\mathbb{I}} & T_{{\rm e}{\rm e};\mathbb{I}{\rm o}} \\
    T_{{\rm o}{\rm o};{\rm e}{\rm o}} & T_{{\rm o}{\rm o};{\rm o}{\rm e}} & T_{{\rm o}{\rm o};{\rm o}\mathbb{I}} & T_{{\rm o}{\rm o};\mathbb{I}{\rm o}} \\
    T_{{\rm e}\mathbb{I};{\rm e}{\rm o}} & T_{{\rm e}\mathbb{I};{\rm o}{\rm e}} & T_{{\rm e}\mathbb{I};{\rm o}\mathbb{I}} & T_{{\rm e}\mathbb{I};\mathbb{I}{\rm o}} \\
    T_{\mathbb{I}{\rm e};{\rm e}{\rm o}} & T_{\mathbb{I}{\rm e};{\rm o}{\rm e}} & T_{\mathbb{I}{\rm e};{\rm o}\mathbb{I}} & T_{\mathbb{I}{\rm e};\mathbb{I}{\rm o}}
\end{pmatrix}, \displaybreak[0] \nonumber \\[1em]
T_{-+} &= \begin{pmatrix}
    T_{{\rm e}{\rm o};{\rm e}{\rm e}} & T_{{\rm e}{\rm o};{\rm o}{\rm o}} & T_{{\rm e}{\rm o};{\rm e}\mathbb{I}} & T_{{\rm e}{\rm o};\mathbb{I}{\rm e}} \\
    T_{{\rm o}{\rm e};{\rm e}{\rm e}} & T_{{\rm o}{\rm e};{\rm o}{\rm o}} & T_{{\rm o}{\rm e};{\rm e}\mathbb{I}} & T_{{\rm o}{\rm e};\mathbb{I}{\rm e}} \\
    T_{{\rm o}\mathbb{I};{\rm e}{\rm e}} & T_{{\rm o}\mathbb{I};{\rm o}{\rm o}} & T_{{\rm o}\mathbb{I};{\rm e}\mathbb{I}} & T_{{\rm o}\mathbb{I};\mathbb{I}{\rm e}} \\
    T_{\mathbb{I}{\rm o};{\rm e}{\rm e}} & T_{\mathbb{I}{\rm o};{\rm o}{\rm o}} & T_{\mathbb{I}{\rm o};{\rm e}\mathbb{I}} & T_{\mathbb{I}{\rm o};\mathbb{I}{\rm e}}
\end{pmatrix}, \displaybreak[0]\nonumber \\[1em]
T_{--} &= \begin{pmatrix}
    T_{{\rm e}{\rm o};{\rm e}{\rm o}} & T_{{\rm e}{\rm o};{\rm o}{\rm e}} & T_{{\rm e}{\rm o};{\rm o}\mathbb{I}} & T_{{\rm e}{\rm o};\mathbb{I}{\rm o}} \\
    T_{{\rm o}{\rm e};{\rm e}{\rm o}} & T_{{\rm o}{\rm e};{\rm o}{\rm e}} & T_{{\rm o}{\rm e};{\rm o}\mathbb{I}} & T_{{\rm o}{\rm e};\mathbb{I}{\rm o}} \\
    T_{{\rm o}\mathbb{I};{\rm e}{\rm o}} & T_{{\rm o}\mathbb{I};{\rm o}{\rm e}} & T_{{\rm o}\mathbb{I};{\rm o}\mathbb{I}} & T_{{\rm o}\mathbb{I};\mathbb{I}{\rm o}} \\
    T_{\mathbb{I}{\rm o};{\rm e}{\rm o}} & T_{\mathbb{I}{\rm o};{\rm o}{\rm e}} & T_{\mathbb{I}{\rm o};{\rm o}\mathbb{I}} & T_{\mathbb{I}{\rm o};\mathbb{I}{\rm o}}
\end{pmatrix}.
\end{align}
Here the indices $\mathbb{I}$, ${\rm e}$, and ${\rm o}$ refer to the ternary indices $\bar{p} = 0$, $1$, and $-1$, corresponding to the identity, symmetric, and antisymmetric non-identity channels, respectively.
The non-vanishing elements of the projected transition matrix $T_{\bar a\bar p}$ are 
\begin{widetext}
\begin{align}
\label{eq:Telement}
T_{{\rm e}{\rm e};{\rm e}{\rm e}}&=P_1\left(\frac{q^2+q-2}{2}\right)\frac{q^2+\sigma q-2}{2}+P_2\left(\frac{q-2}{2}\right)\frac{\sigma q-2}{2}+P_3,\nonumber\\
T_{{\rm o}{\rm o};{\rm e}{\rm e}}&=\frac{(q-2)(\sigma q-2)}{\sigma q^2}T_{{\rm e}{\rm e};{\rm o}{\rm o}}=P_1\left(\frac{q^2+q-2}{2}\right)\frac{q^2+\sigma q-2}{2}+P_2\left(\frac{q+2}{2}\right)\frac{\sigma q +2}{2},\nonumber\\
T_{{\rm e}\mathbb{I};{\rm e}{\rm e}}&=\frac{q^2+\sigma q-2}{2}T_{{\rm e}{\rm e};{\rm e}\mathbb{I}}=P_1\left(\frac{q^2+q-2}{2}\right)\frac{q^2+\sigma q-2}{2}+P_2\left(\frac{q-2}{2}\right)\left(\frac{q^2+\sigma q-2}{2}\right),\nonumber\\
T_{\mathbb{I}{\rm e} ;{\rm e}{\rm e}}&=\frac{q^2+q-2}{2}T_{{\rm e}{\rm e};\mathbb{I}{\rm e}}=P_1\left(\frac{q^2+q-2}{2}\right)\frac{q^2+\sigma q-2}{2}+P_2\left(\frac{\sigma q-2}{2}\right)\left(\frac{q^2+q-2}{2}\right),\nonumber\\
T_{{\rm e}{\rm o};{\rm e}{\rm o}}&=P'_1\left(\frac{q^2-\sigma q}{2}\right)\left(\frac{q^2+q-2}{2}\right)+P'_2\left(\frac{q-2}{2}\right)\left(\frac{\sigma q}{2}\right)+P'_3,\nonumber\\
T_{{\rm o}{\rm e};{\rm o}{\rm e}}&=P'_1\left(\frac{q^2-q}{2}\right)\left(\frac{q^2+\sigma q-2}{2}\right)+P'_2\left(\frac{\sigma q-2}{2}\right)\left(\frac{q}{2}\right)+P'_3,\nonumber\\
T_{{\rm e}{\rm o};{\rm o}{\rm e}}&=\frac{q+2\sigma }{q+2}T_{{\rm o}{\rm e};{\rm e}{\rm o}}=P'_1\left(\frac{q^2-q}{2}\right)\left(\frac{q^2+\sigma q-2}{2}\right)+P'_2\left(\frac{\sigma q+2}{2}\right)\left(\frac{q}{2}\right),\nonumber\\
T_{{\rm o}\mathbb{I};{\rm e}{\rm o}}&=\frac{(q+2)(q-\sigma)}{2}T_{{\rm e}{\rm o};{\rm o}\mathbb{I}}=P'_1\left(\frac{q^2-\sigma q}{2}\right)\left(\frac{q^2+q-2}{2}\right)-P'_2\left(\frac{q+2}{2}\right)\left(\frac{q^2-\sigma q}{2}\right),\nonumber\\
T_{\mathbb{I}{\rm o};{\rm o}{\rm e}}&=\frac{(q+2\sigma )(q-1)}{2}T_{{\rm o}{\rm e};\mathbb{I}{\rm o}}=P'_1\left(\frac{q^2-q}{2}\right)\left(\frac{q^2+\sigma q-2}{2}\right)-P'_2\left(\frac{\sigma q+2}{2}\right)\left(\frac{q^2-q}{2}\right),\nonumber\\
T_{\mathbb{I}{\rm o};{\rm e}{\rm o}}&=\frac{q^2+q-2}{2}T_{{\rm e}{\rm o};\mathbb{I}{\rm o}}=P'_1\left(\frac{q^2-\sigma q}{2}\right)\left(\frac{q^2+q-2}{2}\right)+P'_2\left(\frac{q^2+q-2}{2}\right)\left(\frac{\sigma q}{2}\right),\nonumber\\
T_{{\rm o}\mathbb{I};{\rm o}{\rm e}}&=\frac{q^2+\sigma q-2}{2}T_{{\rm o}{\rm e};{\rm o}\mathbb{I}}=P'_1\left(\frac{q^2-q}{2}\right)\left(\frac{q^2+\sigma q-2}{2}\right)+P'_2\left(\frac{q^2+\sigma q-2}{2}\right)\left(\frac{q}{2}\right),\nonumber\\
T_{{\rm o}{\rm o};{\rm o}{\rm o}}&=P_1\left(\frac{q^2-q}{2}\right)\frac{q^2-\sigma q}{2}+P_2\left(\frac{q}{2}\right)\frac{\sigma q}{2}+P_3,\nonumber\\
T_{{\rm e}\mathbb{I};{\rm o}{\rm o}}&=\frac{q^2(q-\sigma)}{2(q+2)}T_{{\rm o}{\rm o};{\rm e}\mathbb{I}}=P_1\left(\frac{q^2-q}{2}\right)\frac{q^2-\sigma q}{2}-P_2\left(\frac{q}{2}\right)\left(\frac{q^2-\sigma q}{2}\right),\nonumber\\
T_{\mathbb{I}{\rm e};{\rm o}{\rm o}}&=\frac{q^2(q-1)}{2(q+2\sigma)}T_{{\rm o}{\rm o};\mathbb{I}{\rm e}}=P_1\left(\frac{q^2-q}{2}\right)\frac{q^2-\sigma q}{2}-P_2\left(\frac{\sigma q}{2}\right)\left(\frac{q^2-q}{2}\right),\nonumber\\
T_{{\rm e}\mathbb{I};{\rm e}\mathbb{I}}&=P_1\left(\frac{q^2+q-2}{2}\right)+P_2\left(\frac{q-2}{2}\right)+P_3 ,\nonumber \\
T_{\mathbb{I}{\rm e};\mathbb{I}{\rm e}}&=P_1\left(\frac{q^2+\sigma q-2}{2}\right)+P_2\left(\frac{\sigma q-2}{2}\right)+P_3 ,\nonumber \\
T_{{\rm e}\mathbb{I};\mathbb{I}{\rm e}}&=\frac{q^2+\sigma q-2}{q^2+q-2}T_{\mathbb{I}{\rm e};{\rm e}\mathbb{I}}=P_1\left(\frac{q^2+\sigma q-2}{2}\right)+P_2\left(\frac{q^2+\sigma q-2}{2}\right),\nonumber\\
T_{\mathbb{I}{\rm o};{\rm o}\mathbb{I}}&=\frac{q+\sigma }{q+1}T_{{\rm o}\mathbb{I};\mathbb{I}{\rm o}}=P'_1\left(\frac{q^2-q}{2}\right)+P'_2\left(\frac{q^2-q}{2}\right),\nonumber\\
T_{{\rm o}\mathbb{I};{\rm o}\mathbb{I}}&=P'_1\left(\frac{q^2-q}{2}\right)+P'_2\left(\frac{q}{2}\right)+P'_3,\nonumber\\
T_{\mathbb{I}{\rm o};\mathbb{I}{\rm o}}&=P'_1\left(\frac{q^2-\sigma q}{2}\right)+P'_2\left(\frac{\sigma q}{2}\right)+P'_3,\nonumber\\
T_{{\rm e}{\rm e};{\rm e}{\rm o}}&=\frac{q}{q+2\sigma }T_{{\rm e}{\rm o};{\rm e}{\rm e}}=P''_1\left(\frac{q^2-\sigma q}{2}\right)\frac{q^2+q-2}{2}-P''_2\left(\frac{q-2}{2}\right)\frac{\sigma q}{2},\nonumber\\
T_{{\rm e}{\rm e};{\rm o}{\rm e}}&=\frac{q}{q+2}T_{{\rm o}{\rm e};{\rm e}{\rm e}}=P''_1\left(\frac{q^2-q}{2}\right)\frac{q^2+\sigma q-2}{2}-P''_2\left(\frac{\sigma q-2}{2}\right)\frac{q}{2},\nonumber\\
T_{{\rm o}{\rm o};{\rm e}{\rm o}}&=\frac{q-2\sigma }{q-2}T_{{\rm o}{\rm o};{\rm o}{\rm e}}=\frac{q+2}{q}T_{{\rm e}{\rm o};{\rm o}{\rm o}}=\frac{q+2}{q}T_{{\rm o}{\rm e};{\rm o}{\rm o}}=P''_1\left(\frac{q^2-\sigma q}{2}\right)\frac{q^2+q-2}{2}-P''_2\left(\frac{q+2}{2}\right)\frac{\sigma q}{2},\nonumber\\
T_{{\rm e}\II;{\rm e}{\rm o}}&=\frac{q^2-\sigma q}{2}T_{{\rm e}{\rm o};{\rm e}\II}=P''_1\left(\frac{q^2-\sigma q}{2}\right)\frac{q^2+q-2}{2}+P''_2\left(\frac{q-2}{2}\right)\left(\frac{q^2-\sigma q}{2}\right),\nonumber\\
T_{\II {\rm e} ;{\rm o}{\rm e}}&=\frac{q^2-q}{2}T_{{\rm o}{\rm e};\II {\rm e} }=P''_1\left(\frac{q^2-q}{2}\right)\frac{q^2+\sigma q-2}{2}+P''_2\left(\frac{\sigma q-2}{2}\right)\left(\frac{q^2-q}{2}\right),\nonumber\\
T_{{\rm o}\II;{\rm e}{\rm e} }&=\frac{q^2+\sigma q-2}{2}T_{{\rm o}{\rm e};{\rm e}\II}=\frac{q^2+\sigma q-2}{2}T_{{\rm o}\II;{\rm e}\II  }=\frac{(q+2)(q^2+\sigma q-2)}{2q}T_{{\rm e}{\rm e};{\rm o}\II }=\frac{q+2}{q}T_{{\rm e}\II;{\rm o}{\rm e}},\nonumber\\
&=\frac{(q+2)(q^2+\sigma q-2)}{2q}T_{{\rm e}\II  ;{\rm o}\II}=P''_1\left(\frac{q^2+q-2}{2}\right)\frac{q^2+\sigma q-2}{2}-P''_2\left(\frac{q+2}{2}\right)\frac{q^2+\sigma q-2}{2},\nonumber\\
T_{\II {\rm o};{\rm e}{\rm e}}&=\frac{q^2+q-2}{2}T_{{\rm e}{\rm o};\II {\rm e} }=\frac{q^2+q-2}{2}T_{\II {\rm o};\II {\rm e}}=\frac{(q+2\sigma )(q^2+q-2)}{2q}T_{{\rm e}{\rm e};\II {\rm o}}=\frac{q+2\sigma }{q}T_{\II {\rm e} ;{\rm e}{\rm o}},\nonumber\\
&=\frac{(q+2\sigma )(q^2+q-2)}{2q}T_{\II {\rm e};\II {\rm o}}=P''_1\left(\frac{q^2+q-2}{2}\right)\frac{q^2+\sigma q-2}{2}-P''_2\left(\frac{\sigma q+2}{2}\right)\frac{q^2+q-2}{2},\nonumber\\
T_{{\rm o}\II;{\rm o}{\rm o} }&=\frac{q^2-\sigma q}{2}T_{{\rm o}{\rm o};{\rm o}\II }=P''_1\left(\frac{q^2-q}{2}\right)\frac{q^2-\sigma q}{2}+P''_2\frac{q}{2}\left(\frac{q^2-\sigma q}{2}\right),\nonumber\\
T_{\II {\rm o};{\rm o}{\rm o}}&=\frac{q^2-q}{2}T_{{\rm o}{\rm o};\II {\rm o}}=P''_1\left(\frac{q^2-q}{2}\right)\frac{q^2-\sigma q}{2}+P''_2\frac{\sigma q}{2}\left(\frac{q^2-q}{2}\right),\nonumber\\
T_{{\rm o}\II ;\II {\rm e}}&=\frac{q^2+\sigma q-2}{q^2+q-2}T_{\II {\rm o};{\rm e}\II  }=\frac{q^2+\sigma q-2}{q^2-q}T_{\II {\rm e};{\rm o}\II }=\frac{q+2\sigma }{q}T_{{\rm e}\II  ;\II {\rm o}}=P''_1\left(\frac{q^2+\sigma q-2}{2}\right)+P''_2\left(\frac{q^2+\sigma q-2}{2}\right),
\end{align}
\end{widetext}
where $\sigma=1$ is for the orthogonal-invariant ensembles, and $\sigma=-1$ is for symplectic-invariant ensembles. This general formulation \eqref{eq:Telement} holds for all $q = 2^m$ with $m > 0$ in the orthogonal-invariant case, and with $m > 1$ in the symplectic-invariant case. 

For the symplectic case, at odd time steps, the second and fourth indices have to be swapped with the first and third indices. (Recall that for the symplectic ensemble we take the qudits to have integer spin for even $x$ and half-integer spin and odd $x$.)

For a half-integer-spin qudit of size $q = 2$, the label ``e'' (corresponding to ternary index $1$) does not occur, because all single-qudit operators are either equal to the unit operator or they are antisymmetric under the dual operation. Hence, for even time steps, the second and fourth indices can take the values ``o'' (corresponding to ternary index $-1$) and ``$\openone$'' only. For the symplectic-invariant ensemble with $q=2$, the non-vanishing elements of the ternary-projected transition matrix $T_{\bar a \bar p}$ at even time steps are
\begin{align}
T_{\mathbb{I}{\rm o};\mathbb{I}{\rm o}} &= 3P'_1 - P'_2 + P'_3 ,\nonumber\\
T_{{\rm o}\mathbb{I};\mathbb{I}{\rm o}} &= 3(P'_1 + P'_2) ,\nonumber\\
T_{{\rm e}{\rm o};\mathbb{I}{\rm o}} &= 3P'_1 - P'_2 ,\nonumber\\
T_{\mathbb{I}{\rm o};{\rm o}\mathbb{I}} &= P'_1 + P'_2 ,\nonumber\\
T_{{\rm o}\mathbb{I};{\rm o}\mathbb{I}} &= P'_1 + P'_2 + P'_3 ,\nonumber\\
T_{{\rm e}{\rm o};{\rm o}\mathbb{I}} &= P'_1 - P'_2 ,\nonumber\\
T_{{\rm o}{\rm o};{\rm o}{\rm o}} &= 3P_1 - P_2 + P_3 ,\nonumber\\
T_{{\rm e}\mathbb{I};{\rm o}{\rm o}} &= 3P_1 - 3P_2 ,\nonumber\\
T_{{\rm o}{\rm o};{\rm e}\mathbb{I}} &= 2P_1 - 2P_2 ,\nonumber\\
T_{{\rm e}\mathbb{I};{\rm e}\mathbb{I}} &= 2P_1 + P_3 ,\nonumber\\
T_{\mathbb{I}{\rm o};{\rm e}{\rm o}} &= 6P'_1 - 2P'_2 ,\nonumber\\
T_{{\rm o}\mathbb{I};{\rm e}{\rm o}} &= 6P'_1 - 6P'_2 ,\nonumber\\
T_{{\rm e}{\rm o};{\rm e}{\rm o}} &= 6P'_1 + P'_3,\nonumber\\
T_{{\rm o}{\rm o};\II {\rm o}}&=3P''_1-P''_2,\nonumber\\
T_{{\rm e}\II;\II {\rm o}}&=3P''_1+3P''_2,\nonumber\\
T_{{\rm o}{\rm o}; {\rm o}\II}&=P''_1+P''_2,\nonumber\\
T_{{\rm e}\II; {\rm o}\II}&=P''_1-P''_2,\nonumber\\
T_{\II {\rm o}; {\rm o}{\rm o}}&=3P''_1-P''_2,\nonumber\\
T_{{\rm o}\II ; {\rm o}{\rm o}}&=3P''_1+3P''_2,\nonumber\\
T_{{\rm e} {\rm o}; {\rm o}{\rm o}}&=3P''_1+P''_2,\nonumber\\
T_{\II {\rm o}; {\rm e}\II}&=2P''_1+2P''_2,\nonumber\\
T_{{\rm o}\II ; {\rm e}\II}&=2P''_1-2P''_2,\nonumber\\
T_{{\rm e}{\rm o} ; {\rm e}\II}&=2P''_1,\nonumber\\
T_{{\rm o}{\rm o}; {\rm e}{\rm o}}&=6P''_1+2P''_2,\nonumber\\
T_{{\rm e}\II; {\rm e}{\rm o}}&=6P''_1.
\end{align}
\subsection{General qudit size $q$}

For a general qudit size $q$, which is not a power of $2$, we can also obtain a closed Markovian evolution equation of the form of Eq.\ (\ref{eq:rhobinaryevolution}),
where the transition matrix $T_{\bar a \bar p}$ has a block structure as in Eq.\ \eqref{eq:Tmatrix}.

This matrix is constructed by summing the transition probabilities $W_{ab;pq}$ over all Pauli strings $p$ and $q$ with the weight factors $\chi_{\bar p;p,q}$ defined in Eq.\ (\ref{eq:chioriginal}), depending on whether the qudit has integer or half-integer spin. We find that this summation depends only on the ternary structure of the input indices, denoted $\bar{a}$. Consequently, the elements of $T_{\bar{a}\bar{p}}$ are built from sums over terms of the form 
\begin{equation*}
    \frac{1}{q^4} \left\langle |\mbox{tr}\, {\cal U}_{x,y} \Sigma_{a;s} {\cal U}_{x,y}^{\dagger} \Sigma_{p;s}^{\dagger} |^2 \right\rangle,
\end{equation*}
where the matrices $\Sigma_{a;s}$ and $\Sigma_{p;s}$ are defined as
\begin{align}
 \Sigma_{a;s} &=\frac{1}{4} (\sigma_{a_x} + s_{a_x} \sigma^T_{a_x}) \otimes (\sigma_{a_{x+1}} + s_{a_{x+1}} \sigma^T_{a_{x+1}})\nonumber, \\
 \Sigma_{p;s} &= \frac{1}{4}(\sigma_{p_x} + s_{p_x} \sigma^T_{p_x}) \otimes (\sigma_{p_{x+1}} + s_{p_{x+1}} \sigma^T_{p_{x+1}})
\end{align}
and $s_{a_x}$, $s_{a_{x+1}}$, $s_{p_x}$, and $s_{p_{x+1}}$ take the value $-1$ and $1$.

For any $q > 2$, the elements of the projected transition matrix $T_{\bar{a}\bar{p}}$ retain the form given in Eq.~\eqref{eq:Telement}.
%


\section{Drift-diffusion equation}
\label{app:5}

To derive the drift-diffusion equation (\ref{eq:driftdiffusion}), we follow Ref.\ \cite{tan2025operatorspreadingrandomunitary} and combine $\bar \rho^{(n)}_{\rm }(x;t)$ and $\bar \rho^{(n)}_{\rm }(x+1,t)$ into a two-component spinor,
\begin{align}
  \label{eq:Rdef}
  R_{\rm }^{(n)}(\Delta x;t) = \begin{pmatrix} \bar \rho_{\rm }^{(n)}(\Delta x;t) \\
  \bar \rho_{\rm }^{(n)}(\Delta x + 1;t) \end{pmatrix},
\end{align}
where now $\Delta x$ is always even.
The evolution equations (\ref{eq:rhoevol1})--(\ref{eq:rhoevol2}) with the truncation prescription (\ref{eq:rhonn}), may then be represented as
\begin{align}
  R_{\rm }^{(n)}(\Delta x;t) =&\,
  D^{(n)} R_{\rm }^{(n)}(\Delta x;t-1)  \nonumber \\ &\, \mbox{}
  + D'^{(n)} R_{\rm }^{(n)}(\Delta x+2;t-1),
  \label{eq:Revol}
\end{align}
where $D^{(n)}$ and $D'^{(n)}$ are $(4\times3^{n}) \times (4\times3^{n})$ matrices.
The transition matrices $D^{(n)}$ and $D'^{(n)}$ satisfy the normalization rule
\begin{equation}
  \sum_{i} (D^{(n)}_{ij} + D'^{(n)}_{ij}) = 1
  \label{eq:Dsum}
\end{equation}
for each $j=1,\ldots,4\times3^{n}$. We denote the left-eigenvectors, right-eigenvectors, and eigenvalues of $D^{(n)} + D'^{(n)}$ by $\tilde V_j^{(n)}$, $V_j^{(n)}$, and $d_{j}^{(n)}$, respectively. Equation (\ref{eq:Dsum}) ensures that the largest eigenvalue $d_1^{(n)} = 1$, with left-eigenvector $\tilde V_1^{(n)} = (1,1,\ldots,1)^{\rm T}$. In the two-component spinor notation of Eq.\ (\ref{eq:Rdef}), the right-eigenvectors $V_1^{(n)}$ are written
\begin{equation}
    V_1^{(n)} = \begin{pmatrix} V_1^{(n)}(1) \\
    V_1^{(n)}(-1) \end{pmatrix}.
\end{equation}
The right-eigenvector $V_1^{(n)}$ also appears in Eq.\ (\ref{eq:rhoeqsol}) of the main text, where we write its dependence on the ternary labels $\bar p_0$, \ldots, $\bar p_n$ explicitly.

In the long-time limit, the solution of Eq.\ (\ref{eq:Revol}) is of the form
\begin{equation}
  R^{(n)}_{\rm }(\Delta x,t) = R_{{\rm }1}^{(n)}(\Delta x,t) V_1^{(n)},
  \label{eq:Rntlong}
\end{equation}
where $R_{{\rm }1}^{(n)}(\Delta x,t)$ satisfies the drift-diffusion equation (\ref{eq:driftdiffusion}), with

\begin{equation}
  v_{\rm B}^{(n)} = 1 - 2 d'^{(n)}_{11},
  \label{eq:vBn} 
\end{equation}
the butterfly velocity and
\begin{equation}
  {\cal D}^{(n)} = 4 d'^{(n)}_{11}[1 -  d'^{(n)}_{11}]
  + 8 \sum_{j \neq 1} \frac{d'^{(n)}_{1j} d'^{(n)}_{j1}}{1 - d_j^{(n)}},
  \label{eq:DDn}
\end{equation}
the diffusion constant governing the diffusive spreading of the front. (Recall that $\Delta x$ is measured with respect to a ballistic propagation at unit velocity so that the drift velocity measured with respect to $\Delta x$ is $v_{\rm B} - 1$.) In Eqs.\ (\ref{eq:vBn}) and (\ref{eq:DDn}) we abbreviated 
\begin{equation}
  d_{ij}'^{(n)} = \tilde V_i^{(n){\rm T}} D'^{(n)} V_j^{(n)}.
  \label{eq:dijdef}
\end{equation}
In Eq.\ (\ref{eq:Rntlong}), the function $R_1^{(n)}(\Delta x,t)$ describes the propagation of the ends of the Pauli strings, whereas the right-eigenvector $V_1^{(n)}$ contains information on the structure of the Pauli strings immediately behind the end. 

The drift-diffusion equation can also be derived for the case that a complete basis of left- and right-eigenvectors of $D^{(n)} + D'^{(n)}$ does not exist. We refer to Ref.\ \cite{tan2025operatorspreadingrandomunitary} for details.

In the case of symplectic-invariant circuits, a closed drift-diffusion equation can only be derived by considering two consecutive time steps as a single unit. We denote the truncated projected transition matrix for even time steps as $D^{(n)} + D'^{(n)}$, and for odd time steps as $\widetilde D^{(n)} + \widetilde D'^{(n)}$. This coarse-graining yields a drift-diffusion equation identical in form to Eq. \eqref{eq:driftdiffusion}, but with a modified butterfly velocity and diffusion constant given by
\begin{equation}
    v_B^{(n)}=1-\tilde d'^{(n)}_{11}
\end{equation}
where
\begin{equation}
\tilde d'^{(n)}_{11}=\tilde V^{(n){\rm T}}_1\left[ \widetilde D'^{(n)}+ D'^{(n)}(\widetilde D^{(n)}+\widetilde D'^{(n)})\right]V^{(n)}_1.
\end{equation}
The diffusion constant is defined as
\begin{align}
    \mathcal{D}^{(n)}=&2\tilde {d}'^{(n)}_{11}(1-\tilde {d}'^{(n)}_{11})\nonumber\\
    &+4\tilde V^{(n){\rm T}}_1\left[ \widetilde D'^{(n)}+ D'^{(n)}(\widetilde D^{(n)}+\widetilde D'^{(n)})\right] \widetilde G^{(n)}\nonumber \\
    &\times \left[ ( D^{(n)}+ D'^{(n)})\widetilde D'^{(n)}+ D'^{(n)}(\widetilde D^{(n)}+\widetilde D'^{(n)})\right]\nonumber\\
    &\times V^{(n)}_1+4\tilde V^{(n){\rm T}}_1  D'^{(n)}\widetilde D'^{(n)} V^{(n)}_1
\end{align}
with 
\begin{align}
  \widetilde G^{(n)} =&\, 
\left[ \openone - (D^{(n)} + D'^{(n)})(\widetilde D^{(n)} +\widetilde D'^{(n)})
  \right. \nonumber \\ &\, \mbox{} \ \ \ \ \left. +
  V_1^{(n)} \tilde V_1^{(n){\rm T}}\right]^{-1} 
  - V_1^{(n)} \tilde V_1^{(n){\rm T}}.
\end{align}
Here, $\tilde V_1^{(n)}$ and $V_1^{(n)}$ are the left and right eigenvectors, respectively, of the two-step propagator $(D^{(n)} + D'^{(n)})(\widetilde D^{(n)} + \widetilde D'^{(n)})$ corresponding to the eigenvalue $1$. One can verify that the butterfly velocity and diffusion constant recover the original forms in Eqs. \eqref{eq:vBn} and \eqref{eq:DDn} when $D^{(n)}=\widetilde D^{(n)}$ and $D'^{(n)}=\widetilde D'^{(n)}$.

\section{Convergence to the ternary form}
\label{app:6}

We first give an outline of the arguments that show that the two-norm $\| \rho -  \BB \rho\|_2 \to 0$ in the long-time limit. In this appendix, we use the notation
\begin{equation}
  \WW_{ab,pq} = W_{ap} W_{bq}^*.
\end{equation}
Both $\WW$ and its ensemble average $\langle \WW \rangle$ are
operators on the Pauli-string density matrix $\rho$. Because the
amplitudes $ W_{ap}$ factorize into pair contributions,
Eq.~\eqref{eq:layerW}, the layer operator $\WW$ is a tensor product of pair
transition matrices, one for each two-qudit gate in the layer. For
a single two-qudit gate $U$, the pair transition matrix is defined
as
\begin{equation}
  W(U)_{ab;pq} =
  {\cal W}_{ap;x,y}\, {\cal W}^*_{bq;x,y}
  \Big|_{{\cal U}_{x,y} = U},
  \label{eq:Wpairdef}
\end{equation}
with the pair amplitudes ${\cal W}_{ap;x,y}$ given in
Eq.~\eqref{eq:WUapp}.\\
For qudit size $q = 2^m$, the ternary-state projector $\BB$ is
defined in Eq.~\eqref{eq:B2}; for general qudit dimension $q$ it is
defined as
\begin{equation}
\label{eq:Bgeneric}
\begin{split}
    \BB_{p_x, q_x; \bar{p}_x} ={}& \frac{2(1 - \delta_{p_x, 0}) \chi^*_{+1}(p_x, q_x) \delta_{\bar{p}_x, +1}}{q^2 + q - 2} \\
    &+ \frac{2(1 - \delta_{p_x, 0}) \chi^*_{-1}(p_x, q_x) \delta_{\bar{p}_x, -1}}{q^2 - q} \\
    &+ \delta_{p_x, q_x} \delta_{p_x, 0} \delta_{\bar{p}_x, 0},
\end{split}
\end{equation}
with $\chi_{\bar p;a,b}$ defined in Eq.\ (\ref{eq:chioriginal}) for qudits.
If $q=2^m$, we can choose the tensor products of Pauli matrices as the operator basis, in which case Eq.~\eqref{eq:Bgeneric} simplifies to Eq.~\eqref{eq:B2}. 

Using the fact that the ternary-state projector $\BB$ and the evolution operator $\langle \WW \rangle$ for $\rho$ commute,
\begin{equation}
  \BB \rho_{pq}(t) = \sum_{a,b} \langle \WW_{ab,pq} \rangle \BB \rho_{ab}(t-1),
  \label{eq:Bcommutation}
\end{equation}
we find that for a succession of two time steps the two-norm
\begin{equation}
  \left\| \rho(t) - \BB \rho(t) \right\|_2 =
  \sqrt{\sum_{pq}\left|\rho_{pq}(t) - \BB \rho_{pq}(t)\right|^2},
  \label{eq:twotimestepbound}
\end{equation}
satisfies the bound
\begin{align}
\label{eq:ineqtern}
  \lefteqn{\left\| \rho(t+2) - \BB \rho(t+2) \right\|_2} ~~~~~~~~ \nonumber \\ \le&\,
  \|\langle \WW_2 \rangle \langle \WW_1 \rangle \|'_{\infty} \, \left\| \rho(t) - \BB \rho(t) \right\|_2,
\end{align}

where $\|\langle \WW_2 \rangle \langle \WW_1 \rangle \|'_{\infty}$
denotes the largest singular value of the product of the two
single-layer averaged transition matrices, evaluated after
excluding the subspace of density matrices with
$\rho - \BB \rho = 0$. (Since the gates of the two layers are drawn
independently, this product equals the average
$\langle\WW_2\WW_1\rangle$ of the two-layer transition matrix.) Per
qudit pair, this subspace is three-dimensional, spanned by the
trivial state and the maximally random even-parity and odd-parity
states. 
The largest singular value of the pair transition matrix $\langle W(U)\rangle$ for a pair of neighboring qudits is $1$. Its multiplicity is generically two, corresponding to the trivial state and the maximally random nontrivial state. Exceptions are the case of the trivial circuit, for which all singular values of the transition matrix are $1$, and the case where the transition matrices are orthogonal or symplectic. In the latter cases, the multiplicity is three for $q > 2$, because there are separate maximally random even-parity and odd-parity states, and four for $q = 2$ in the orthogonal case, as discussed further below. 

For the operator norm $\|\langle \WW_2 \rangle \langle \WW_1 \rangle \|'_{\infty}$ we find the bound
\begin{equation}
 \|\langle \WW_2 \rangle \langle \WW_1 \rangle\|'^2_{\infty}\le \lambda^2_W  
\end{equation}
with
\begin{align}
  \label{eq:lambdaWestimate}
  \lambda^2_W
  =& \sigma_2^2\!\left[\sigma_2^2 + (1-\sigma_2^2)\,\|\PP_1\|'^2_\infty\right]
     \nonumber \\
     &+ (1-\sigma_2^2)[\chi+(1-\chi)\sigma_2^2]\|\PP_2\|'^2_\infty.
\end{align}
Here $\PP_j$ is the projector onto the eigenvalue-1 eigenspace of
$ \langle \WW_j\rangle$,
$j = 1,2$, associated with the two consecutive layers of the
brickwork circuit, and $\sigma_2 \in [0,1)$ is the largest singular
value of the pair transition matrix $\langle W(U)\rangle$ other
than 1. Since $\langle\WW\rangle$ is a tensor product of pair transition matrices $\langle W(U)\rangle$,
$\sigma_2$ is also the largest sub-unit singular value of a full
layer. Finally,
$\chi := \|\PP_1 \PP_2\|_\infty'^2$; because the two layers are offset by
one site in the brickwork pattern, their eigenvalue-1 subspaces do
not coincide, and $\chi$ characterizes the largest singular value
of the product $\langle\WW_2\rangle\langle\WW_1\rangle$ in a
Haar-distributed circuit, evaluated after projecting out vectors
satisfying $\rho - \BB\rho = 0$.
For Haar-distributed ensembles, the second largest singular value vanishes, $\sigma_2 = 0$, which leads to
\begin{equation}
\lambda_W^2 = \chi \, \|\PP_2\|_\infty'^2 .
\end{equation}
For qudit dimension $q>2$, the quantity $\|\PP_2\|_\infty'^2$ vanishes, implying $\lambda_W = 0$, in agreement with the behavior of fully unitary circuits. 
In contrast, for $q=2$, if the local gate belongs to the ${\rm SO}(4)$ or ${\rm SO}^-(4)$ ensemble, one finds $\|\PP_2\|_\infty'^2 = 1$, because of the additional singular vector, beyond the ternary states, associated with the largest singular value of $\langle \WW \rangle$. 
Consequently, in this case,
\begin{equation}
\lambda_W^2 = \chi .
\end{equation}
The parameter $\chi$ is strictly smaller than 1 and is independent of the specific properties of the probability distribution of the gate ensemble. 
Its precise value can be obtained numerically. 
Since all contributing parameters are strictly less than one, it follows that $\lambda_W < 1$. 
Moreover, since these parameters are independent of the system size, $\lambda_W$ remains strictly smaller than $1$ as the system size increases.

We define the decay time for convergence to a ternary distribution, $\tau_{\rm t}$, as
\begin{equation}
\tau_{\rm t} = \frac{-1}{\ln \lambda_W}.
\end{equation}
For Haar-distributed ensembles other than ${\rm SO}(4)$ and ${\rm SO}^{-}(4)$, the convergence to the ternary distribution occurs already after the first time step. 

\subsection{Detailed derivation}


Before giving the full derivation, we first outline the logical structure. We denote the ensemble of two-qudit gate operators by ${\cal E}$ and the transition matrix corresponding to the two-qudit operator $U$ by $W(U)$, Eq.~\eqref{eq:Wpairdef}. We call $\rho$ a \emph{unimodular eigenvector} of $\langle W \rangle$
if $\langle W \rangle \rho = c\rho$ with $|c|=1$. This condition is
{\em a priori} weaker than the steady-state condition of
Sec.~\ref{sec:4}, which requires $c=1$; in the following we
determine all unimodular eigenvectors for unitary-, orthogonal-,
and symplectic-invariant ensembles.\\
1. We show that the largest singular value of the ensemble average $\langle W\rangle$ is $1$;\\
2. We show that every unimodular eigenvector of $\langle W\rangle$ satisfies $W(U)\rho=c\,\rho$ for all $U\in\mathcal{E}$;\\
3. We show that if the ensemble is unitary-invariant and $\rho_{i\mu}$ are unimodular eigenvectors, then $W(V)\rho_{i\mu}=\rho_{i\mu}$ for all $q^2 \times q^2$ unitary matrices $V$;\\
4. We determine the unimodular eigenvectors for unitary-, orthogonal-, and symplectic-invariant ensembles, and identify which of them are the steady states of Sec.~\ref{sec:4};\\\
5. We derive the inequality \eqref{eq:ineqtern};\\
6. We give an upper bound for
$\| \langle \WW_2 \rangle \langle \WW_1 \rangle \|'_{\infty}$.

We begin by bounding the singular values of the averaged pair
transition matrix $\langle W(U) \rangle$. The positive semi-definiteness of the variance of $W$ implies that for any normalized two-qudit state $\|\widetilde{\rho}\|=1$, one has the inequality
\begin{align}
\lefteqn{\widetilde \rho^\dagger\left<\left( W-\langle W\rangle\right)\left( W^\dagger-\langle W^\dagger\rangle\right)\right>\widetilde\rho} ~~~~~~~~ \nonumber\\
 =&\, \widetilde \rho^\dagger\left(\left< W W^\dagger\right>-\langle W\rangle\langle W^\dagger\rangle\right)\widetilde\rho\nonumber\\
 =&\, 1-\widetilde \rho^\dagger\langle W\rangle\langle W^\dagger\rangle\widetilde \rho \nonumber \\ \ge&\, 0.
\end{align}
Here $\langle\ldots\rangle$ denotes the ensemble average over $\mathcal{E}$, and we have used $\langle W\rangle^\dagger=\langle W^\dagger\rangle$, together with the unitarity of $W_{ab;pq}$, which is guaranteed by
\begin{align}
 (W W^\dagger)_{ab;a'b'} &= \sum_{pq} W_{ab;pq}
 W^*_{a'b';pq} \nonumber \\
 &= \frac{1}{q^8} \sum_{pq}
 \tr U^\dagger \Sigma_a U \Sigma_p^\dagger \,
 \tr U^\dagger \Sigma_b^\dagger U \Sigma_q \nonumber \\
 &\quad \times \tr U^\dagger \Sigma_{a'}^\dagger U \Sigma_p \,
 \tr U^\dagger \Sigma_{b'} U \Sigma_q^\dagger \nonumber \\
 &= \delta_{aa'} \delta_{bb'}.
\end{align}
It follows that
\begin{equation}
   1 \ge \max_{\|\widetilde \rho\|_2 = 1} \left( \widetilde \rho^\dagger\langle W \rangle\langle W^\dagger \rangle\widetilde \rho \right) = \|\langle W\rangle \|_\infty^2,
\end{equation}
so that the largest singular value of \( \langle W \rangle \) is bounded from above by $1$.

If $\rho$ is a unimodular eigenvector, we obtain that 
\begin{equation}
    \left< W \right>^\dagger\rho=c^*\rho.
\end{equation}
This follows from the observation that
\begin{align}
    \| \left< W \right>^\dagger\rho-c^*\rho\|_2^2 =&\, \|\left< W \right>^\dagger\rho\|_2^2-\|\rho\|_2^2\nonumber\\
    \le &\, \|\left< W \right>^\dagger\|^2_\infty\|\rho\|_2^2-\|\rho\|_2^2\nonumber \\ \le&\, 0,
\end{align} 
where we have used $\|\left< W \right>^\dagger\|^2_\infty=\|\left< W \right>\|^2_\infty\le 1 $, and non-negativity of the norm. It follows that a unimodular eigenvector also satisfies
\begin{equation}
  \left< W \right>  \left< W \right>^\dagger  \rho = \left< W \right>^\dagger  \left< W \right> \rho=\rho.
\end{equation}
Moreover, we have
\begin{align}
  \label{eq:Wrho}
    \left\langle\|W\rho - c\rho\|_2^2\right\rangle =&\, \left\langle \rho^\dagger W^\dagger W \rho + \rho^\dagger \rho - 2\,\Re\left(\rho^\dagger W \rho\right) \right\rangle \nonumber \\
    =&\, 2 \rho^\dagger \rho - 2\,\Re\left(c^*\rho^\dagger \left\langle W \right\rangle \rho\right) \nonumber \\ =&\, 0
\end{align}
if $\langle W \rangle \rho = c\rho$ with $|c|=1$. Since the squared two-norm is non-negative for any $W$, Eq.\ (\ref{eq:Wrho}) implies that $W\rho = c\rho$ must hold for all $W(U)$, without taking the ensemble average. Therefore, $\rho$ is a simultaneous eigenvector with eigenvalue $c$ for the entire ensemble of $W$. 

One may argue that the unimodular eigenvectors $\rho$ would depend on the structure of the ensemble. We now demonstrate that for a unitary-invariant ensemble, the only unimodular eigenvectors are the two steady states identified in Sec.~\ref{sec:4}.

\subsection{Unitary invariant ensembles}
Equivalently, vectorizing the two-qudit Pauli operators $\Sigma_a$,
the pair transition matrix of Eq.~\eqref{eq:Wpairdef} can be written
in terms of the Hilbert-Schmidt inner product,
\begin{equation}
  W(U)_{ab;pq} = \left<ab^*\right| U^* \otimes U \otimes U \otimes
  U^* \left|pq^*\right>,
  \label{eq:WUUUU}
\end{equation}
or, in matrix notation,
\begin{equation}
  W(U) = U^* \otimes U \otimes U \otimes U^*.
\end{equation}
Here, we have used the identities \cite{macedo2013typing}
\begin{equation}
    \tr(A^\dagger B) = \left<\mathrm{vec}(A)|\mathrm{vec}(B)\right>,
\end{equation}
\begin{equation}
    \left|\mathrm{vec}(ABC)\right> = C^T \otimes A\, \left|\mathrm{vec}(B)\right>
\end{equation}
and we defined
\begin{align}
  \left<a\right|=&(\mathrm{vec}(\Sigma^\dagger_a))^\dagger,\nonumber\\
\left<b^*\right|=&(\mathrm{vec}(\Sigma^\dagger_b))^{\rm T},\nonumber\\
\left|p\right>=&(\mathrm{vec}(\Sigma^\dagger_p)),\nonumber\\
\left|q^*\right>=&(\mathrm{vec}^*(\Sigma^\dagger_q)).
\end{align}
Due to the invariance of the ensemble-averaged operator $\langle W \rangle$ under unitary transformations $V$ from the ensemble $\mathcal{E}$, we find
\begin{align}
  \langle W \rangle =&\,  \langle W(VUV^\dagger) \rangle
  \nonumber \\ =&\, W(V) \langle W(U) \rangle W^\dagger(V)
\end{align}
for all $q^2 \times q^2$ unitary matrices $V$.

Now, suppose $\{\rho_{i\mu}\}$ is an orthonormal basis set of eigenvectors of $\langle W \rangle$ with eigenvalue $c_i$ with $|c_i| = 1$,
\begin{equation}
    \langle W \rangle \rho_{i\mu} = c_i\,\rho_{i\mu}.
\label{eq:wavg}
\end{equation}
Then
\begin{align}
  W(V)\,\langle W \rangle\,W^\dagger(V)\,W(V)\rho_{i\mu}
  &= \langle W \rangle\,W(V)\rho_{i\mu} \nonumber\\
  &= c_i\,W(V)\rho_{i\mu}
\end{align}
for all $q^2 \times q^2$ unitary matrices $V$, so that $W(V) \rho_{i \mu}$ is also an eigenvector of $\langle W \rangle$ at the same eigenvalue $c_i$. Hence, for an arbitrary unitary matrix $V$, $W(V) \rho_{i \mu}$ can be expressed as a linear combination of the basis vectors $\rho_{i \mu}$,
\begin{align}
  W(V)\rho_{i\mu} =&\, \sum_{\nu} C^i_{\mu\nu}(V)\,\rho_{i\nu},
\end{align}
where the coefficients $C^i_{\mu \nu}(V)$ satisfy the orthonormality constraint
\begin{align}
  \sum_{\nu} C^i_{\mu\nu}(V)\,C^{i*}_{\mu\nu}(V)=&\, 1.
\end{align}
The matrix $C^i_{\mu\nu}(V)$ further satisfies
\begin{equation}
    C^i_{\mu\nu}(V^\dagger)= C^{i*}_{\nu\mu}(V),
\end{equation}
\begin{equation}
  \sum_{\nu'} C^i_{\mu\nu'}(V)\,C^i_{\nu'\nu}(U)= C^i_{\mu\nu}(U V),
\end{equation}
where we have used $W(UV)=W(U)W(V)$.

We first consider the case that the ensemble $\mathcal{E}$ of
two-qudit operators $U$ forms a group, and relax this assumption
later. Because $\mathcal{E}$ is invariant under unitary conjugation,
$VUV^\dagger \in \mathcal{E}$ for all unitary matrices $V$, so the
group formed by $\mathcal{E}$ is a normal subgroup of the full
unitary group ${\rm U}(q^2)$ \cite{dummit2003abstract}.If $\mathcal{E}$ is a group, then $U^2 \in \mathcal{E}$, and
Eqs.~\eqref{eq:wavg} and \eqref{eq:Wrho} give $W(U^2)\rho_{i\mu} =
c_i\,\rho_{i\mu}$, while $W(U^2) = W^2(U)$ gives
\begin{equation}
  W(U^2)\rho_{i\mu} = W^2(U)\rho_{i\mu} = c_i^{\,2}\,\rho_{i\mu}.
\end{equation}
Comparing the two gives $c_i = c_i^{\,2}$, hence $c_i = 1$.
Any $q^2 \times q^2$ unitary matrix $V$ can be written as
\begin{equation}
    V=(\det V)^{1/q^2}\,V',
  \label{eq:VVprime}
\end{equation}
where $V'\in {\rm SU}(q^2)$. Moreover, any special unitary matrix can be expressed as a commutator \cite{fraleigh2003first,sepanski2007compact},
\begin{equation}
  \label{eq:Vcomm}
    V'
=ABA^{-1}B^{-1},
\end{equation}
with $A,B\in {\rm U}(q^2)$. Therefore,
\begin{align}
  C^{i}_{\mu\nu}(V)
  =&\, C^{i}_{\mu\nu}\!\left((\det V)^{1/q^2} V'\right) \nonumber\\
  =&\, \sum_{\nu'} C^{i}_{\mu\nu'}(V')\,C^{i}_{\nu'\nu}\!\left((\det V)^{1/q^2}\openone\right) \nonumber \\
   =&\, C^{i}_{\mu\nu}(V'),
\end{align}
where we used $C^{i}_{\nu'\nu} ((\det V)^{1/q^2}\openone)=C^{i}_{\nu'\nu}(\openone)=\delta_{\nu'\nu}$. We also note that
\begin{equation}
    W(V)=W(V'),
\end{equation}
because $V$ and $V'$ only differ by a factor, see Eq.~\eqref{eq:VVprime},
and $V$ and $V^*$ appear equally often (twice each) in the tensor
product defining $W(V)$, see Eq.~\eqref{eq:WUUUU}.\\
Likewise, any matrix $U_i\in\mathcal{E}$ can be written as $U_i=(\det U_i)^{1/q^2}U'_i$ with $U'_i\in {\rm SU}(q^2)$. If $\mathcal{E}$ is a group, then the set $\{U'_i\}$ also forms a group. Indeed, if $U_i U_j=U_k$, then
\begin{align}
  U_iU_j=&\, (\det (U_i U_j))^{1/q^2}\,U'_iU'_j \nonumber \\ =&\, (\det U_k)^{1/q^2}\,U'_k,
\end{align}
so that
\begin{equation}
   U'_iU'_j=U'_k.
\end{equation}
Since $\openone\in\mathcal{E}$, it follows that $\openone\in\{U'_i\}$. Associativity holds automatically. Hence $\{U'_i\}$ is a group whenever $\mathcal{E}$ is. It is then a normal subgroup of ${\rm SU}(q^2)$. 

We now make use of the fact that there are only two kinds of normal subgroups of ${\rm SU}(q^2)$ \cite{dummit2003abstract}: (i) subgroups of its center $\{e^{2\pi i k/q^2}\openone\,|\,k=0,1,\dots,q^2-1\}$; and (ii) the entire group ${\rm SU}(q^2)$. The first case yields $W(U)$ behaving like $W(\openone)$, corresponding to the trivial circuit and producing no convergence to the binary or ternary state; thus, only the second case is nontrivial. Therefore, we must consider the case that
\begin{equation}
  W(V)\rho_{i\mu}=\rho_{i\mu}
\end{equation}
for all $q^2 \times q^2$ unitary matrices $V$, if $\langle W \rangle \rho_{i \mu} = c_i \rho_{i \mu}$ with $|c_i| = 1$ for the ensemble average $\langle W \rangle$ over all $U \in {\cal E}$.

If the ensemble $\mathcal{E}$ does not form a group, we can reach the same conclusion as follows:
For any $U\in \mathcal{E}$ with $U=(\det U)^{1/q^2}U'$, define
\begin{equation}
  \mathcal{S}=\{\, V U'^{\,k} V^\dagger \mid k\in \mathbb{Z},\; V\in {\rm U}(q^2)\,\}.
\end{equation}
Let $G$ be the set of all finite products of elements of $\mathcal{S}$. Then $G$ is a group. Indeed, if $g_1,g_2\in G$ with $g_1=s_1s_2\cdots$ and $g_2=m_1m_2\cdots$ for $s_i,m_i\in\mathcal{S}$, then
\begin{equation}
  g_1 g_2 = s_1 s_2 \cdots m_1 m_2 \cdots,
\end{equation}
which is again a finite product of elements of $\mathcal{S}$. Moreover, for $A=s_1 s_2 \cdots s_m\in G$,
\begin{equation}
  A^{-1}=s_m^{-1}\cdots s_2^{-1}s_1^{-1}\in G.
\end{equation}
Since $\openone\in\mathcal{S}$, we also have $\openone\in G$, and associativity holds as before. Hence $G$ is a (normal) subgroup of ${\rm SU}(q^2)$. By the same reasoning as above, $G$ is either central or all of ${\rm SU}(q^2)$. For a nontrivial circuit, one must be in the noncentral case, {\em i.e.}, $G={\rm SU}(q^2)$.

Furthermore, for any $A = V U'^k V^{\dagger} \in \mathcal{S}$, one has
\begin{align}
  W(V U'^{\,k} V^\dagger)\rho_{i\mu}
   =&\, \sum_{\nu} C^{i}_{\mu\nu}(V U'^{\,k} V^\dagger)\,\rho_{i\nu}
   \nonumber \\ =&\, c_i^{\,k}\,\rho_{i\mu},
\end{align}
so $C^{i}_{\mu\nu}(A)$ is a scalar (proportional to $\delta_{\mu\nu}$) on the subspace spanned by the $\{\rho_{i\mu}\}$. For any $M\in G$, write $M=A_1 A_2 \ldots$ with $A_1$, $A_2$, $\ldots \in \mathcal{S}$. Then
\begin{align}
  W(M)\rho_{i\mu}
  =&\, \sum_{\nu} C^{i}_{\mu\nu}(M)\,\rho_{i\nu}
  \nonumber \\ =&\,
  W(A_1 A_2 \ldots)\rho_{i\mu}
  \nonumber \\ =&\,
  W(A_1)\,W(A_2)\,\ldots\, \rho_{i\mu}
  \nonumber \\ =&\,
  c_i(M)\,\rho_{i\mu},
\end{align}
and thus $C^{i}_{\mu\nu}(M)$ is a scalar for every $M\in G$.


Now we can prove that the constant $c_{i}(V') =1$ for all $V'\in {\rm SU}(q^2)$ if $|c_{i}(V')| = 1$. Here, we notice that
\begin{align}
  c_{i}(V')c_{i}(U')=&\, c_{i}(V'U') \nonumber \\=&\, c_{i}(U'V'),
\end{align} 
which means that $c_{i}(V')$ is a one‐dimensional character of the connected group ${\rm SU}(q^2)$.
Therefore, we obtain that
\begin{align}
    c_{i}(V')
    &=  c_{i}(A) c_{i}(B) c_{i}(A^{-1}) c_{i}(B^{-1})\nonumber\\
    &= 1,
\end{align}
where we wrote the special unitary matrix $V'$ as a commutator, see Eq.\ (\ref{eq:Vcomm}).
For the same reason we discussed before, we can conclude $C^i_{\mu\nu}(V)=\delta_{\mu\nu}$ for all $V\in \mathcal{U}(q^2)$. 


The finding that $W(V) \rho_{i\mu} = \rho_{i\mu}$ for all $V \in \mathcal{U}(q^2)$ if $\rho_{i\mu}$ satisfies $\langle W \rangle \rho_{i\mu} =c_i \rho_{i\mu}$ implies that $\rho$ transforms under the trivial representation of the group action defined by $W(V)$. Consequently, the dimension of the subspace spanned by such invariant states $\rho$ is determined by the multiplicity of the trivial representation in the decomposition of $W(V)$. To analyze this, we write the representation $W(V)$ as
\begin{equation}
    W(V) = T^*(V) \otimes T(V), \quad \text{where } T(V) = V \otimes V^*.
\end{equation}
Under the adjoint action $A \mapsto VAV^\dagger$, the representation $T(V)$ acting on the space of linear operators $L(\mathcal{V})$ decomposes as \cite{fulton_representation_2004}
\begin{equation}
    T(V) = T_0 \oplus T_{\mathrm{adj}},
\end{equation}
where $T_0$ is the trivial representation (spanned by the identity matrix), and $T_{\mathrm{adj}}$ corresponds to the adjoint subspace (traceless operators).
Therefore, the full representation $W(V)$ decomposes as
\begin{align}
  & (T^*_0 \oplus T^*_{\mathrm{adj}}) \otimes (T_0 \oplus T_{\mathrm{adj}}) = (T^*_0 \otimes T_0) \oplus (T^*_0 \otimes T_{\mathrm{adj}}) \nonumber\\ & \ \ \ \
  \oplus (T^*_{\mathrm{adj}} \otimes T_0) \oplus (T^*_{\mathrm{adj}} \otimes T_{\mathrm{adj}}).
\end{align}
Since only tensor products of a representation with its complex conjugate contain the trivial representation, the subrepresentations $T^*_0 \otimes T_0$ and $T^*_{\mathrm{adj}} \otimes T_{\mathrm{adj}}$ each contain the trivial representation \cite{fulton_representation_2004, hall2003lie}. Hence, the dimension of the invariant subspace (i.e., the multiplicity of eigenvalue 1) is two, corresponding exactly to the steady states identified in Sec.\ \ref{sec:4}. Therefore, we can conclude that if $\left<W\right>\rho=c\rho$ with $|c|=1$, then $\rho$ must be a superposition of the maximum random state and the trivial state. Both of these states obviously have eigenvalue $c=1$.

\subsection{Orthogonal invariant ensembles}

A parallel argument applies to orthogonal-invariant ensembles. Unlike the unitary case, the orthogonal group is not connected, so there are two disjoint components: the special orthogonal group ${\rm SO}(q^2)$ and the determinant-$-1$ component ${\rm SO}^{-}(q^2)$ \cite{hall2003lie}. For any matrix $V''$ in ${\rm SO}^{-}(q^2)$, we can write
\begin{equation}
    V'' = R V',
\end{equation}
where $R=\mathrm{diag}(-1,1,1,\ldots)$ is a reflection and $V'\in {\rm SO}(q^2)$.

As in the unitary case, if the ensemble $\mathcal{E}$ is a group, then $C^{i}_{\mu\nu}(U)=\delta_{\mu\nu}$ for all $U\in\mathcal{E}$. Moreover, $\mathcal{E}$ is closed under conjugation: $V U V^\dagger\in\mathcal{E}$ for all $V\in {\rm O}(q^2)$. If $\mathcal{E}$ is not a group, the conclusion must be adjusted slightly. Let $\{U_i\}$ denote the elements of $\mathcal{E}$ and define
\begin{equation}
  \mathcal{S}=\{\, V U_i^{\,k} V^\dagger \mid k\in\mathbb{Z},\ U_i\in\mathcal{E},\ V\in {\rm O}(q^2)\,\}.
\end{equation}
Let $G$ be the set of all finite products of elements of $\mathcal{S}$. The same closure, inverse, identity, and associativity arguments as before show that $G$ is a group. For any $A\in G$ we then have $C^i_{\mu\nu}(A)=c_i(A)\,\delta_{\mu\nu}$ (a scalar on the $\{\rho_{i\mu}\}$ subspace).
Furthermore, \(G\subset U(q^{2})\) is a subgroup that is preserved under orthogonal conjugation, i.e., normalized by \({\rm O}(q^{2})\).
In the \(W(V)\)-representation, central phases act trivially. Up to this trivial central action, there are exactly three nontrivial \({\rm O}(q^{2})\)-invariant possibilities for the image of \(G\): the special orthogonal group \({\rm SO}(q^{2})\), the full orthogonal group \({\rm O}(q^{2})\), and the image of ${\rm SU}(q)$, which is isomorphic to the projective unitary group \({\rm PU}(q)\) \cite{Grove2002,dummit2003abstract}.
Moreover, if  \(\mathcal{E}\) consists only of elements whose action on \(W\) lies in \({\rm O}(q^{2})\) (equivalently, in \({\rm U}(1)\cdot {\rm O}(q^{2})\)), then \(G\) is a normal subgroup of \({\rm O}(q^{2})\) \cite{Grove2002,dummit2003abstract}; for \(q\ge 2\), the only nontrivial normal subgroups of \({\rm O}(q^{2})\) (besides its center) are \({\rm SO}(q^{2})\) and \({\rm O}(q^{2})\).
If \(\mathcal{E}\) contains an element outside \({\rm O}(q^{2})\) (or outside \(U(1)\cdot {\rm O}(q^{2})\)), then \(G\) has the same image on \(W\) as \(U(q)\) (equivalently, as \({\rm SU}(q)\)), which is isomorphic to ${\rm PU}(q)$.
In particular, when \(G={\rm SO}(q^{2})\), every element of \(G\) is a commutator \cite{fraleigh2003first,sepanski2007compact}, which implies
\begin{equation}
  C^{i}_{\mu\nu}(V')=\delta_{\mu\nu}
\end{equation}
for all $V' \in G$. 
If $G={\rm O}(q^2)$, then for $V''=RV'$ with $V'\in {\rm SO}(q^2)$ we obtain
\begin{equation}
  C^{i}_{\mu\nu}(V'')=C^{i}_{\mu\nu}(R)=c_i(R),
\end{equation}
and since
\begin{equation}
  c_i(R)\,c_i(R)=c_i(\openone)=1,
\end{equation}
it follows that $c_i(R)=\pm1$.

For the full group ${\rm SO}(q^2)$, the representation of $W(V')$ decomposes as \cite{fulton_representation_2004, hall2003lie}
\begin{align*}
  & (T_0 \otimes T_0) \oplus (T_{\mathrm{sym}} \otimes T_{\mathrm{sym}}) \oplus (T_{\mathrm{ant}} \otimes T_{\mathrm{ant}}) \nonumber\\
  & \ \ \ \ \mbox{} \oplus (T_0 \otimes T_{\mathrm{sym}}) \oplus (T_{\mathrm{sym}} \otimes T_{0}) \oplus (T_{0} \otimes T_{\mathrm{ant}}) \nonumber\\
  & \ \ \ \ \mbox{} \oplus (T_{\mathrm{ant}} \otimes T_0) \oplus (T_{\mathrm{sym}} \otimes T_{\mathrm{ant}}) \oplus (T_{\mathrm{ant}} \otimes T_{\mathrm{sym}}).
\end{align*}
Only tensor products of a representation with its complex conjugate contain the trivial representation; since $T_0$, $T_{\mathrm{sym}}$, and $T_{\mathrm{ant}}$ are self-conjugate, the subrepresentations $T_0 \otimes T_0$, $T_{\mathrm{sym}} \otimes T_{\mathrm{sym}}$, and $T_{\mathrm{ant}} \otimes T_{\mathrm{ant}}$ each contain a trivial component. Therefore, when $q>2$, there exist three independent eigenvectors with
\[
W(V')\,\rho_i=\rho_i
\]
for all $V'\in {\rm SO}(q^2)$,
namely
\[
\rho_1=\delta_{ab}\delta_{cd},\qquad \rho_2=\delta_{ad}\delta_{bc},\qquad \rho_3=\delta_{ac}\delta_{bd}.
\]
Explicitly,
\begin{align}
W(V')_{a'b'c'd';abcd}\,\delta_{ab}\delta_{cd}&=\delta_{a'b'}\delta_{c'd'},\nonumber\\
W(V')_{a'b'c'd';abcd}\,\delta_{ad}\delta_{bc}&=\delta_{a'd'}\delta_{b'c'},\nonumber\\
W(V')_{a'b'c'd';abcd}\,\delta_{ac}\delta_{bd}&=\delta_{a'c'}\delta_{b'd'}.
\end{align}
These three vectors are common eigenvectors of $W(V)$ with eigenvalue $1$ for all $V\in {\rm O}(q^2)$ and they satisfy
\begin{equation}
  W(R)\,\rho_i=\rho_i.
\end{equation}
Hence $\langle W\rangle\,\rho_i=\rho_i$ for $i=1,2,3$ when $q>2$. These invariant eigenstates $\rho_i$ coincide with those identified in Section~\ref{sec:4}.

For $q=2$, there is an additional common eigenvector of $W(V')$ for ${\rm SO}(4)$ with eigenvalue $1$, namely $\rho_{4}=\epsilon_{abcd}$. This eigenvector does not correspond to the binary or ternary sectors and is not stable under the even–odd structure. Moreover,
\begin{equation}
  W(R)\,\rho_4=-\rho_4.
\end{equation}
If the ensemble $\mathcal{E}$ lies entirely within one of the components ${\rm SO}(4)$ or ${\rm SO}^{-}(4)$, then there are four independent eigenvectors of $\langle W\rangle$ with $|c|=1$. Specifically, if $\mathcal{E}\subset {\rm SO}(4)$, then $\langle W\rangle\,\rho_i=\rho_i$ for $i=1,2,3,4$; if $\mathcal{E}\subset {\rm SO}^{-}(4)$, then $\langle W\rangle\,\rho_i=\rho_i$ for $i=1,2,3$ and $\langle W\rangle\,\rho_4=-\rho_4$. However, if $\mathcal{E}$ contains elements from both components ${\rm SO}(4)$ and ${\rm SO}^{-}(4)$, then only $\rho_1,\rho_2,\rho_3$ remain eigenvectors of $\langle W\rangle$ with $|c|=1$. (Here we used the fact that if $\langle W\rangle\,\rho_i=c_i\rho_i$, then $W(V)\rho_i=c_i\rho_i$ for all $V\in\mathcal{E}$.)

If $\mathcal{E}$ contains elements not in $U(1).{\rm O}(q^2)$, we reach the same conclusion as the unitary case. 

\subsection{Symplectic invariant ensembles}

For the symplectic-invariant ensemble, one proceeds almost the same as the orthogonal case, with a few small differences. The constructed group $G$ is invariant under symplectic transformations. In the $W(V)$ representation, up to the trivial action, there are only two nontrivial possibilities of the image of G: ${\rm Sp}(q^2/2)$, the projective unitary group ${\rm PU}(q^2)$ \cite{Grove2002,dummit2003abstract}.

If the ensemble $\mathcal{E}$ only contains elements in the symplectic group, then there are three independent eigenvalue-1 eigenvectors of $W(V)$,
\[
\rho_1=\delta_{ab}\delta_{cd},\qquad \rho_2=\delta_{ad}\delta_{bc},\qquad \rho_3=Z_{ac}Z_{bd},
\]
where the involution matrix $Z$ is defined in Eq.\ (\ref{eq:Zx2}).
Unlike the orthogonal case, there does not exist a fourth independent eigenvalue-1 eigenvector for $q=2$, as ${\rm Sp}(2)$ is not a real matrix group, but a subgroup of ${\rm SU}(4)$ \cite{hall2003lie}. 

If the ensemble $\mathcal{E}$  contains elements outside $U(1)\cdot Sp(q^2/2)$, then there exist only two independent eigenvalue-1 eigenvector of $W(V)$,
namely, $\rho_1 $, $\rho_2$. 

\subsection {Relaxation time}

One can verify that $\langle \WW\rangle$ commutes with $\BB$, by using the Weingarten calculus result in App.~\ref{app:3}, which also guarantees the validation of the projection method. 

As an example, consider the term $C_{16} = B_{16}\,\tr\!\left(\Sigma_a\Sigma_p^\dagger\Sigma_b^*\Sigma_q^T\right)$ and investigate its action on a density matrix $\rho_{pq}$ of ternary form. The nonzero elements of such density matrices have $p=q$ or $p=q^T$, see Eq.\ (\ref{eq:chioriginal}).
If $p=q$, then
$$q^{-2}\,\delta_{a'+b',\,2p'}\,\delta_{a''b''}\,e^{\,i(a'-p')(p''-a'')}\,B_{16}$$
is obtained. Summing over all $p = (p_x, p_{x+1})$ where both indices are non-identity ($p_x, p_{x+1} \neq 0$) yields
$$\xi(a_x,b_x)\xi(a_{x+1},b_{x+1})B_{16}$$
with $\xi(a,b)=\delta_{a,b}-q^{-1}\delta_{a,\,b^{T}}e^{-i\phi_a}$.
If, instead, $p=q^{T}$, we multiply by $e^{-i\phi_p}$ and sum over all non-identity indices $p_x, p_{x+1} \neq 0$, we obtain
\begin{multline*}
  q^{-2}\,\delta_{a,\,b^{T}}e^{-i\phi_a}B_{16}\,
\bigl[(1-\delta_{a_x,0})(1-\delta_{a_{x+1},0})\\
-(q^2-1)(1-\delta_{a_x,0})\delta_{a_{x+1},0}-(q^2-1)\delta_{a_x,0}(1-\delta_{a_{x+1},0})\bigr].
\end{multline*}
 Both contributions depend only on $\delta_{a,b}$, $\delta_{a,\,b^{T}}e^{-i\phi_a}$, and on whether the indices $a_x$ and $a_{x+1}$ are zero or non-zero. Hence, they exhibit a clear ternary structure. Repeating the same analysis for the remaining terms shows that each term likewise assumes a ternary form. Because $a$ and $p$ play symmetric roles, summing over $a$ and $b$ with a fixed ternary pattern induces a ternary pattern on $p$ and $q$. Consequently, the pair transition matrix $\langle W(U) \rangle$
commutes with the pair factor of the ternary projection operator
$\BB$ and hence, by the factorization of a layer into pair
transition matrices, $\langle \WW \rangle$ commutes with the ternary projection operator $\BB$. As a result, once the Pauli strings are in the ternary or binary form, then they will be in the ternary or binary form forever.

One can also verify that $\langle \WW\rangle$ is normal based on the Weingarten calculus results in App. \ref{app:3}. Then it follows that $\langle \WW\rangle$ and $\langle \WW\rangle\langle \WW\rangle^\dagger$ share eigenvectors. Consequently, the eigenspace of $\langle \WW\rangle\langle \WW\rangle^\dagger$ corresponding to the maximum singular value coincides with the eigenspace of $\langle \WW\rangle$ with eigenvalue $c$ satisfying $|c|=1$ (as established above). Therefore, the invariant subspace of $\langle \WW\rangle$ is precisely the top-singular-value eigenspace of $\langle \WW\rangle\langle \WW\rangle^\dagger$. 
%

Since we have proved that the $\langle\WW\rangle$ commutes with the ternary projection operator $\BB$, the bound of \eqref{eq:twotimestepbound} over two time steps follows. It remains to construct an upper bound for the norm $\|\langle \WW_2 \rangle\langle \WW_1 \rangle\|'_{\infty}$.
Hereto, we may restrict to vectors with $\rho-\BB\rho\neq0$. 
We recall that the primed norm $\|\cdot\|'_2$ is the largest singular value of the matrix evaluated after excluding vectors with $\rho-\BB\rho=0$. 
We write $\sigma_2\in[0,1)$ for the largest singular value of the pair transition matrix $\langle W(U) \rangle$ smaller than $1$. An upper bound for $\|\langle \WW_2 \rangle\langle \WW_1 \rangle\|'_{\infty}$ follows from submultiplicativity,
\begin{equation}
\label{eq: norm2}
       \|\langle \WW_2 \rangle\langle \WW_1 \rangle\|'_{\infty} \le  \|\langle \WW_2 \rangle\|'_\infty\|\langle \WW_1 \rangle\|'_{\infty}.
\end{equation}
In the unitary case, since the only eigenvalue-1 states are the binary steady states, $\|\langle \WW_1 \rangle\|'_\infty$ and $\|\langle \WW_2 \rangle\|'_\infty$ are strictly less than $1$. Hence, the product can be bounded as
\begin{equation}
  \|\langle \WW_2 \rangle \|'_\infty\,\|\langle \WW_1 \rangle\|'_\infty\le \sigma_2^2 . 
  \label{eq:WWbound}
\end{equation}
For the orthogonal-invariant ensemble with $q\ge 3$ and for the symplectic-invariant ensemble, the same conclusion as in the unitary case holds. Since $0 \le \sigma_2 < 1$, it follows that in the long-time limit, the state $\rho$ approaches the binary/ternary form $\BB \rho$.

For the special case $q=2$ in the orthogonal-invariant ensemble,
$\WW$ possesses additional eigenvalue-1 eigenstates, as discussed
above. In this case, showing that $\rho(t)$ acquires a ternary form
for large $t$ requires extra care, because Eq.~\eqref{eq:WWbound}
does not apply. We recall that $\PP_j$ is the projector onto the
eigenvalue-1 eigenspace of $\langle \WW_j \rangle$, $j=1,2$, and set
$\QQ_j = \openone - \PP_j$. Since $\langle \WW \rangle$ is normal,
$\PP_j^\dagger = \PP_j$, $\PP_j \langle \WW_j \rangle = \PP_j$, and
$\|\QQ_j \langle \WW_j \rangle \rho\|_2 \le \sigma_2 \|\rho\|_2$. For the
overlap parameter we have
\begin{equation}
  \chi = \|\PP_1 \PP_2\|'^2_\infty
  = \sup_{\substack{\phi\in \mathrm{Ran}(\PP_2)\\ \|\phi\|_2=1\;\&\;
    \phi-\BB\phi\ne 0}}\langle\phi|\PP_1|\phi\rangle .
  \label{eq:chidef}
\end{equation}
For any $\rho$ we set $\alpha = \|\PP_1(\rho-\BB\rho)\|_2$ and
$|x\rangle = \langle \WW_1 \rangle (\rho - \BB\rho)$. Splitting off
the $\PP_2$ sector at the second layer and the $\PP_1$ sector at the
first, one finds
\begin{align}
  \|\langle \WW_2 \rangle\langle \WW_1 \rangle\|'^2_{\infty}
  \le& \ \sigma_2^2\!\left[\sigma_2^2
       + (1-\sigma_2^2)\,\|\PP_1\|'^2_\infty\right] \nonumber \\
     &+ (1-\sigma_2^2)\,\sup_{\substack{\|\rho\|_2=1\\
       \rho-\BB\rho \ne 0}} \langle x|\PP_2 |x\rangle ,
\end{align}
where $\|\PP_1\|'_\infty=\|\PP_2\|'_\infty=0$ if the only
eigenvalue-1 vectors in $\mathrm{Ran}(\PP_1)$ and
$\mathrm{Ran}(\PP_2)$ are ternary states, and
$\|\PP_1\|'_\infty=\|\PP_2\|'_\infty=1$ otherwise.\\
It remains to bound $\langle x|\PP_2|x\rangle$. We decompose
$|x\rangle = a\,\hat x_1 + b\,\hat x_1'$ into the orthogonal ranges
of $\PP_1$ and $\QQ_1$, with $\hat x_1$, $\hat x_1'$ normalized and
$a,b\geq0$ without loss of generality. The identities above give
$a=\alpha$ and
$b\leq\sigma_2\sqrt{1-\alpha^2}$.

Since $|x\rangle$ lies in the non-ternary sector,
\begin{equation}
  \langle x|\PP_2 |x\rangle =
  \sup_{\substack{\phi\in \mathrm{Ran}(\PP_2)\\\|\phi\|_2=1\;\&\;
  \phi-\BB \phi\ne 0}}\big|\langle\phi|x\rangle\big|^2\,
  \|\PP_2\|'^2_\infty .
\end{equation}
For any admissible $\phi$, define
$c=\langle\hat x_1|\phi\rangle$ and
$d=\langle\hat x_1'|\phi\rangle$. Since
$\hat x_1\in\mathrm{Ran}(\PP_1)$,
\begin{equation}
 |c|^2\leq\langle\phi|\PP_1|\phi\rangle\leq\chi,
\end{equation}
while normalization gives $|c|^2+|d|^2\leq1$. Hence
\begin{equation}
 |\langle\phi|x\rangle|
 \leq a|c|+b|d|
 \leq a|c|+b\sqrt{1-|c|^2}.
\end{equation}
Writing $|c|=\cos\theta$, we therefore obtain
\begin{equation}
   \langle x|\PP_2 |x\rangle \le \max_{\cos^2\!\theta \le \chi}
   \big(a\cos \theta +b\sin \theta\big)^2\,\|\PP_2\|'^2_\infty.
\end{equation} 
The maximum occurs either at
the unconstrained extremum or at the boundary, hence
\begin{equation}
 \langle x|\PP_2 |x\rangle \le
 \begin{cases}
 (a^2+b^2)\,\|\PP_2\|'^2_\infty,
 & \text{if } \dfrac{a^2}{a^2+b^2}\le \chi,\\[6pt]
 (a\sqrt{\chi}+b\sqrt{1-\chi})^2\,\|\PP_2\|'^2_\infty,
 & \text{if } \dfrac{a^2}{a^2+b^2}> \chi.
\end{cases}
\end{equation}
Inserting $a=\alpha$ and $b^2\le\sigma_2^2(1-\alpha^2)$, and writing
\begin{align}
\Xi =&\, \alpha\sqrt{\chi}+\sigma_2\sqrt{1-\alpha^2}\sqrt{1-\chi}, \\
\eta =&\, \frac{a^2}{a^2+b^2}
 = \frac{\|\PP_1(\rho-\BB\rho)\|_2^2}{\|\PP_1(\rho-\BB\rho)\|_2^2
   +\|\QQ_1\langle \WW_1 \rangle(\rho-\BB\rho)\|_2^2},
\end{align}
we obtain the inequalities
\begin{align}
  \lefteqn{\| \PP_2 \langle \WW_1 \rangle(\rho-\BB\rho)\|_2^2}
  ~~~~~~~~ \nonumber \\ \le&\,
\begin{cases}
    (\alpha^2+\sigma_2^2(1-\alpha^2))\,\|\PP_2\|'^2_\infty,
    & \text{if } \eta \le \chi, \\[1ex]
    \Xi^2\,\|\PP_2\|'^2_\infty, & \text{if } \eta > \chi .
\end{cases}
\end{align}
The two cases are bounded separately. In the first case,
$\eta\le\chi$ is equivalent to
$b^2 \ge \frac{1-\chi}{\chi}\,\alpha^2$, which together with
$b^2\le\sigma_2^2(1-\alpha^2)$ restricts $\alpha$ to
\begin{equation}
    \alpha^2\le \frac{\sigma_2^2\chi}{1-\chi+\sigma_2^2\chi},
\end{equation}
so that $\alpha^2+\sigma_2^2(1-\alpha^2)\le
\sigma_2^2/(1-\chi+\sigma_2^2\chi)$. In the second case, $\Xi$ is
maximal at
\begin{equation}
\alpha^2=\frac{\chi}{\chi+\sigma_2^2(1-\chi)}
 > \frac{\sigma_2^2\chi}{1-\chi+\sigma_2^2\chi},
\end{equation}
where $\Xi^2 = \chi+(1-\chi)\sigma_2^2$. Hence the inequality
simplifies to
\begin{align}
\| \PP_2 \langle \WW_1 \rangle(\rho-\BB\rho)\|_2^2
 &\le \begin{multlined}[t]
    \max \bigg\{ \frac{\sigma_2^2}{1-\chi+\sigma_2^2\chi}\,
    \|\PP_2\|'^2_\infty, \\
    \big(\chi+(1-\chi)\sigma_2^2\big)\,\|\PP_2\|'^2_\infty \bigg\}
    \end{multlined} \nonumber \\
    &\le \big(\chi+(1-\chi)\sigma_2^2\big)\,\|\PP_2\|'^2_\infty ,
\end{align}
where the last step uses
$\big(\chi+(1-\chi)\sigma_2^2\big)\big(1-\chi+\sigma_2^2\chi\big)
-\sigma_2^2 = \chi(1-\chi)(1-\sigma_2^2)^2 \ge 0$.

Combining these results gives Eq.~\eqref{eq:lambdaWestimate}.

The parameter $\chi$ can be calculated numerically and is found to be less than one. This ensures that $\|\langle \WW_2 \rangle\langle \WW_1 \rangle\|'^2_{\infty} < 1$, a result that is independent of the system size. While $\|\PP_1\|'_\infty = \|\PP_2\|'_\infty = 1$ for any $\chi > 0$, both norms vanish in the limit $\chi = 0$. Consequently, at $\chi = 0$, the vanishing of $\|\PP_2\|'_\infty$ simplifies the bound to $\|\langle \WW_2 \rangle\langle \WW_1 \rangle\|'^2_{\infty} \le \sigma_2^4$, thereby recovering the unitary limit.

\section{Haar orthogonal circuit for general qudit size $q$}
\label{app:7}
For a Haar orthogonal circuit of a general qudit size $q$, the steady state is defined in terms of the function
\begin{equation}
\label{eq:weight}
  \rho^{\infty}_{p,s_p}(t)
  = \sum_{q}\gamma_p(t)\gamma^*_q(t)
    \prod_x\chi_{s_{p_x}}(p_x,q_x),
\end{equation}
where each $s_{p_x} \in \{+1, -1\}$ and $\chi_{s_{p_x}}(p_x,q_x)$ is defined in Eq.\ (\ref{eq:chioriginal}). The function $\rho_{p;s_p}(t)$ follows a closed Markovian evolution equation,
\begin{align}
  \rho_{p,s_p}(t)
  =&\,
  \sum_{a,s_a}\
  \rho_{a,s_a}(t-1)
  \langle |W_{ap,s_as_p}|^2 \rangle.
\end{align}
The transition probabilities $\langle |W_{ap}|^2 \rangle$ in this stochastic model factorize into pair contributions from the two-qudit gates, with the specific layer of gates depending on the parity of the timestep $t$
\begin{align}
  \langle |W_{ap,s_as_p}|^2 \rangle
  =&\,
  \prod_{x\, {\rm even}}
  \left\{ \begin{array}{ll}
    \langle |{\cal W}_{ap,s_as_p;x,x+1}|^2 \rangle & \mbox{for $t$ even}, \\
    \langle |{\cal W}_{ap,s_as_p;x-1,x}|^2 \rangle & \mbox{for $t$ odd}.
  \end{array} \right.
\end{align}
The pair-transition probability $ \langle |W_{ap;s_as_p;x,y}|^2 \rangle$ for the Haar orthogonal circuit at a generic dimension $q$ can be expressed in the form of Eq. \eqref{eq:Wgeneral}. For the Haar-random case, the coefficients are $\mathcal{P}_2=\mathcal{P}_3=0$, and
\begin{equation}
  \mathcal P_1 = \frac{ \eta_{s_p}(p)}{q^2-1} \times \begin{cases}
  \frac{2}{q^2+2} & \mbox{if $s_{p_{x}} s_{p_{x+1}} = s_{a_{x}} s_{a_{x+1}} = 1$}, \\
  \frac{2}{q^2} & \mbox{if $s_{p_{x}} s_{p_{x+1}} = s_{a_{x}} s_{a_{x+1}} = -1$}, \\
  0 & \mbox{if $s_{p_{x}} s_{p_{x+1}} \neq s_{a_{x}} s_{a_{x+1}}$},
  \end{cases}  
\end{equation}
where the function $\eta_{s_p}(p)$ is defined as
\begin{equation}
 \eta_{s_p}(p)=\frac{(1+s_{p_x}\Delta(p_x))(1+s_{p_{x+1}}\Delta(p_{x+1}))}{4},  
\end{equation}
with
\begin{equation}
\Delta(p_x) =
\begin{cases}
    \delta_{p'_{x},0} , & q \text{ odd},\\[4pt]
    \delta_{p'_{x},0}
    + \delta_{p'_{x},\frac{q}{2}}(-1)^{p''_{x}},
    & q \text{ even}.
\end{cases}
\end{equation}

\begin{figure*}[t]
\centering
\includegraphics[scale=0.48, trim=0.4 0 
0 0, clip]{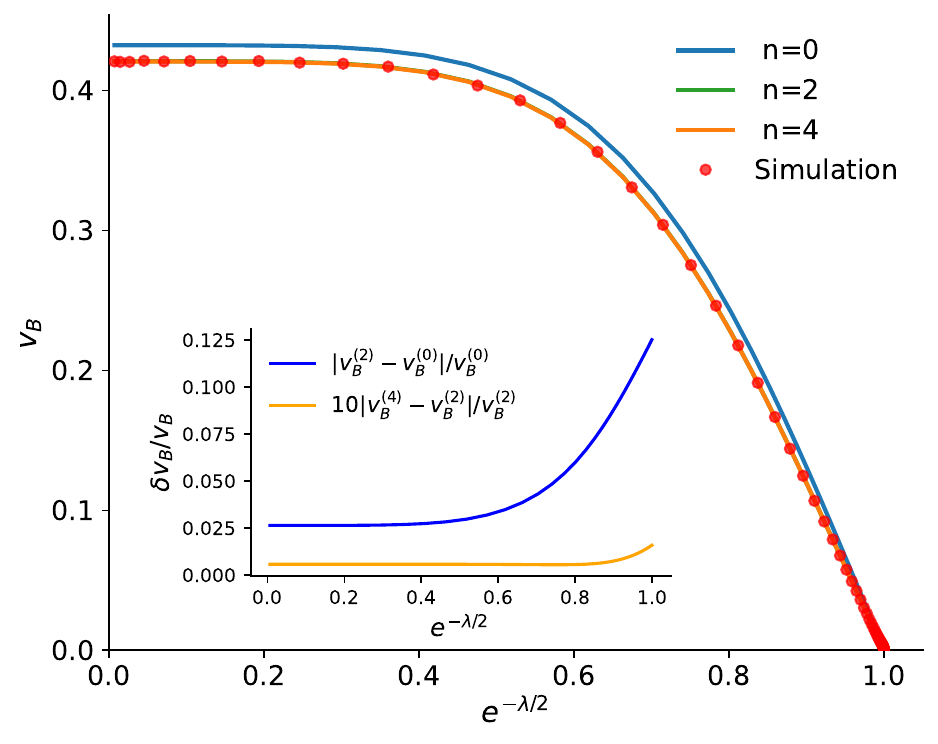}  
\vskip 0.5cm
\caption{\justifying \small
 Butterfly velocity for Brownian symplectic circuits ($q=2$), plotted analogously to Fig.~\ref{fig:BrownianV}. The simulation parameters ($L=320$, $t=200$) are the same as in Fig.\ \ref{fig:BrownianV}, while the ensemble average is taken over $N = 2 \times 10^4$ realizations. Note that the analytical results for $n=2$ and $4$ rapidly converge to the numerical data. Inset: Relative deviations between approximations of orders $n=4, 2$ and $n=0$. }
\label{fig:symplecticvb}
\end{figure*}

\begin{figure*}[t]
\centering
\includegraphics[scale=0.48, trim=0 0 
0 0, clip]{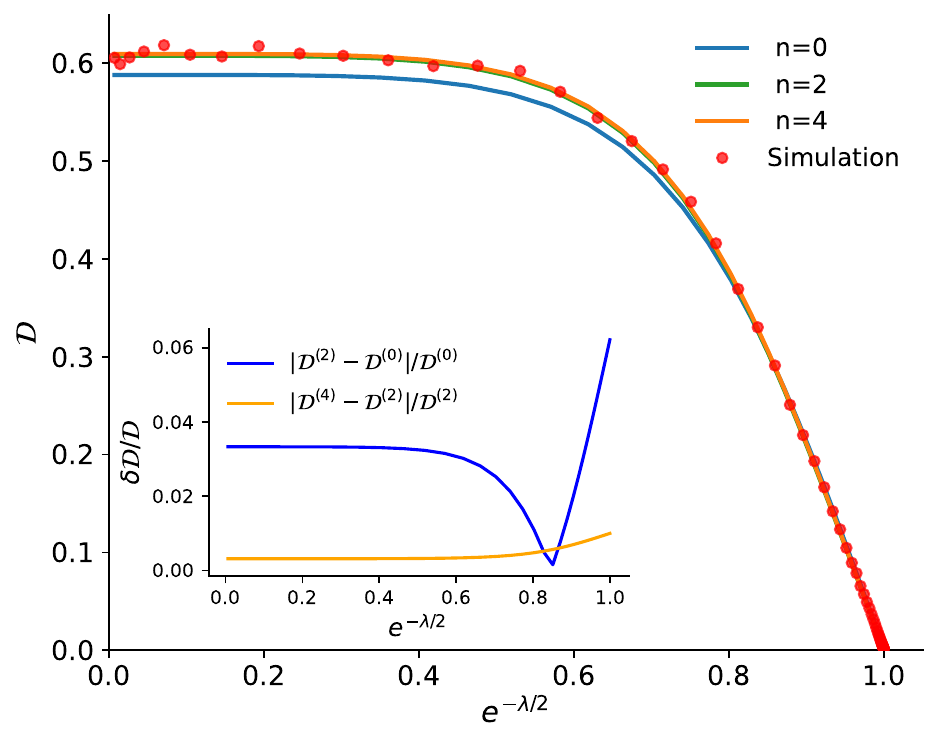} 
\vskip 0.5cm
\caption{\justifying \small
Analogous to Fig.~\ref{fig:symplecticvb}, but for the diffusion constant $\mathcal{D}$ in the Brownian symplectic ensemble. Deviations in the numerical data stem from finite-sample statistical noise. The analytical predictions for $n=2$ and $4$ coincide, demonstrating the rapid convergence of the expansion.}
\label{fig:symplecticD}
\end{figure*}

\section{Brownian-motion circuits}
\label{app:8}

To illustrate our findings, we use the Brownian-motion special orthogonal ensemble as an example, see Eqs.\ (\ref{eq:UBrownian}) and (\ref{eq:HBrownian}). In this ensemble $\lambda$ is a parameter that interpolates between the trivial circuit ($\lambda = 0$) and the Haar-random circuit ($\lambda \to \infty$).

Based on the result of Ref.~\cite{TanBrownian}, we then find the following expressions for the moments of the orthogonal Brownian-motion two-qudit evolution operators:
\begin{widetext}
\begin{align}
  \label{eq:Rapp}
  \mathcal{R}_{1,1;1,1} =&\, \frac{3}{q^8}
  - 3 \frac{q^2-1}{q^8} e^{-\lambda}
  \left[e^{\frac{2 \lambda}{q^2}} q^2
  - (q^2+2) \right] 
  + \frac{1}{24 q^8} e^{-2 \lambda}
  \left[ 
  e^{\frac{8 \lambda}{q^2}} q^2 (q^2-1)(q^2-2)(q^2-3)
  \right.  \nonumber \\ &\, \left. \ \ \ \ \mbox{}
  + 9 e^{\frac{4 \lambda}{q^2}} q^2 (q^2-1)(q^2+2)(q^2-3)
  + 4 e^{\frac{2 \lambda}{q^2}} q^2 (q^2+2)(q^2+1)(q^2-3)
  \right.  \nonumber \\ &\, \left. \ \ \ \ \mbox{}
  + 9 (q^2-1)(q^2-2)(q^2+1)(q^2+4)
  + e^{-\frac{4 \lambda}{q^2}} q^2 (q^2-1)(q^2+1)(q^2+6) \right],  
  \\
  \mathcal{R}_{1,1;2} =&\, \frac{(q^2-1)(q^2+2)}{q^6} e^{-\lambda}
  - \frac{q^2-1}{24 q^6} e^{-2 \lambda}
  \left[
  e^{\frac{8 \lambda}{q^2}} q^2(q^2-2)(q^2-3)
  \right.  \nonumber \\ &\, \left. \ \ \ \ \mbox{}
  + 3 e^{\frac{4 \lambda}{q^2}} q^2(q^2+2)(q^2-3)
  - 3 (q^2-2)(q^2+1)(q^2+4)
  - e^{-\frac{4 \lambda}{q^2}} q^2(q^2+1)(q^2+6) \right], \\
  \mathcal{R}_{2;2} =&\, \frac{3}{q^4} 
  - \frac{1}{q^4} e^{-\lambda} \left[
    e^{\frac{2 \lambda}{q^2}} q^2(q^2-1) - (q^2-2)(q^2+1) 
  \right]
  + \frac{1}{24 q^4} e^{-2 \lambda} \left[
  e^{\frac{8 \lambda}{q^2}} q^2(q^2-1)(q^2-2)(q^2-3)
  \right.  \nonumber \\ &\, \left. \ \ \ \ \mbox{}
  - 3 e^{\frac{4 \lambda}{q^2}} q^2(q^2-1)(q^2+2)(q^2-3)
  + 4  e^{\frac{2\lambda}{q^2}} q^2 (q^2+1)(q^2+2)(q^2-3)
  \right.  \nonumber \\ &\, \left. \ \ \ \ \mbox{}
  - 3 (q^2-1)(q^2+1)(q^2-4)(q^2+4)
  + e^{-\frac{4 \lambda}{q^2}} q^2 (q^2-1)(q^2+1)(q^2+6)
  \right],
  \\
  \mathcal{R}_{1;3} =&\, - \frac{1}{24 q^2} e^{-2 \lambda}
  \left[
  e^{\frac{8 \lambda}{q^2}} (q^2-1)(q^2-2)(q^2-3)
  - e^{\frac{2\lambda}{q^2}} (q^2+1)(q^2+2)(q^2-3)
   \right.  \nonumber \\ &\, \left. \ \ \ \ \mbox{}
 + e^{-\frac{4 \lambda}{q^2}} (q^2-1)(q^2+1)(q^2+6) \right] ,\\
  \mathcal{R}_{4} =&\,  - \frac{q^2-1}{24 q^2} e^{-2 \lambda} 
  \left[ e^{\frac{8 \lambda}{ q^2}}  q^2 (q^2-2)(q^2-3)
  - 3 e^{\frac{4 \lambda}{ q^2}}  q^2 (q^2+2)(q^2-3)
  \right. \nonumber\\ & \left. \ \ \ \
  + 3 (q^2+1)(q^4+4)(q^2-2)
  - e^{\frac{-4 \lambda}{ q^2}} q^2(q^2+1)(q^2+6)
 \right],  \\
  \mathcal{R}_{1;1} =&\, \frac{1}{q^4} + \frac{q^2-1}{2 q^4} e^{-\lambda}
  \left(q^2 e^{\frac{2 \lambda}{ q^2}} + 2+ q^2 \right),  \\
  \mathcal{R}_{2} =&\, \frac{1}{q^2} - \frac{q^2-1}{2 q^2} e^{-\lambda}
  \left(q^2e^{\frac{2 \lambda}{ q^2}} - (q^2 + 2) \right).
  \label{eq:R11App}
\end{align}
\end{widetext}

For the moments of the symplectic Brownian-motion two-qudit operators, for which Eq.\ (\ref{eq:HBrownian}) is replaced by
\begin{equation}
  \label{eq:HBrownian2}
  \langle H_{ij}(t) H_{kl}(t') \rangle =
  \frac{\lambda}{q^2} (
  \delta_{il} \delta_{jk} - \Omega_{ik} \Omega_{jl}
) \delta(t-t'),
\end{equation}
where the antisymmetric $q^2 \times q^2$ matrix $\Omega$ is defined in Eq.\ (\ref{eq:OmegaxxDef}). In this case, one has to replace $q^2$ by $-q^2$ in Eqs.\ (\ref{eq:Rapp})--(\ref{eq:R11App}). Results for the butterfly velocity and diffusion constant in the Brownian symplectic circuit are shown in Figs.~\ref{fig:symplecticvb} and \ref{fig:symplecticD}.
\bibliography{symmetry}
\end{document}